\documentstyle[11pt,aaspp4]{article}

\lefthead{Wyatt et al.}
\righthead{Circumstellar Disk Asymmetries}

\begin{document}

\title{How Observations of Circumstellar Disk Asymmetries Can Reveal
  Hidden Planets: Pericenter Glow and its Application to the HR 4796 Disk}

\author{M. C. Wyatt, S. F. Dermott, C. M. Telesco, R. S. Fisher, \\
  K. Grogan, E. K. Holmes and R. K. Pi\~{n}a}
\affil{Department of Astronomy, University of Florida, Gainesville,
  FL 32611}

\begin{abstract}
Recent images of the disks of dust around the young stars
HR 4796A (\cite{jfht98}; \cite{krwb98}; \cite{ssbk99}; \cite{tfpk99}) and
Fomalhaut (\cite{hgzw98}) show, in each case, a double-lobed feature
that may be asymmetric (one lobe may be brighter than the other).
A symmetric double-lobed structure is that expected from a disk of
dust with a central hole that is observed nearly edge-on
(i.e., close to the plane of the disk).
This paper shows how the gravitational influence of a second body in the
system with an eccentric orbit would cause a brightness asymmetry
in such a disk by imposing a "forced eccentricity" on the orbits
of the constituent dust particles, thus shifting the center of symmetry
of the disk away from the star and causing the dust near the
forced pericenter of the perturbed disk to glow.
Dynamic modeling of the HR 4796 disk shows that its
$\sim 5$\% brightness asymmetry could be the result of a forced
eccentricity as small as 0.02 imposed on the disk by either the binary
companion HR 4796B, or by an unseen planet close to the inner edge of the
disk.
Since it is likely that a forced eccentricity of 0.01 or higher would be
imposed on a disk in a system in which there are planets, but no binary
companion, the corresponding asymmetry in the disk's structure could serve
as a sensitive indicator of these planets that might otherwise remain
undetected.
\end{abstract}


\section{Introduction}
\label{sec-intro}
A new generation of astronomical instrumentations is now making it possible
to image the thermal emission from the disks of dust that surround some stars.
The disks that have been imaged around main sequence stars (i.e., the
Vega-type, or debris, disks) have aroused considerable interest because
it is hoped that interpretation of their structure will provide valuable
information about the evolution of planetary systems, possibly even
leading to the indirect detection of planets hiding in the disks.
One such disk is that around the A0V star HR 4796A.
Mid-IR (10-20 $\mu$m) images of the HR 4796 disk
(\cite{jfht98}; \cite{krwb98}; \cite{tfpk99}) show that its emission is
concentrated in two lobes, one on either side of the star, indicating that
the disk is being observed nearly edge-on, and that its inner region is
almost completely devoid of dust.
The same double-lobed feature is seen in NICMOS (1.1 $\mu$m) images
of the disk's scattered light (\cite{ssbk99}).
The size of the disk's central cavity, which was previously inferred from
the star's spectral energy distribution (\cite{jzbs93}),
is approximately 40 AU in radius from the star, about the same size
as the solar planetary system.
Since the age of the HR 4796 system, $t_{sys} \approx 10$ Myr
(\cite{shb95}; \cite{jmwt98}) places it at a stage in its evolution
when the formation of any planets is expected to be almost complete
(e.g., \cite{liss93}), many authors have speculated that the central
cavity could be indicative of planetary formation in this inner region
(\cite{jgwm95}; \cite{jfht98}; \cite{krwb98}; \cite{jmwt98}).
A similar double-lobed feature has been observed in the disk around the
star Fomalhaut, with a similar explanation proposed for its origin
(\cite{hgzw98}).
Some authors, however, remain skeptical about the existence of planets
in these systems (e.g., \cite{kala98}).

Observations of the HR 4796 disk show a further interesting feature:
the disk's lobes appear to be of unequal brightness, although this observed
asymmetry is of low statistical significance ($\sim 1.8\sigma$).
Mid-IR observations of the disk are described in a companion paper by
Telesco et al. (1999, hereafter T99), and their IHW18
(18.2 $\mu$m) waveband observation (their Fig.~4b) is reproduced
in this paper in Fig.~\ref{fig8}a.
Their observation suggests that the NE lobe (on the left of the image in
Fig.~\ref{fig8}a) is $\sim 5$\% brighter than the SW lobe.
This lobe asymmetry may also be apparent in the NICMOS (1.1 $\mu$m)
images of the disk (\cite{ssbk99}), and in the $\sim 20$ $\mu$m images
of Koerner et al. (1998).
The brightnesses of the Fomalhaut disk's lobes also appear to be
asymmetric (\cite{hgzw98}).
There are many possible explanations for the HR 4796 disk's lobe
asymmetry.
This paper describes a model of the T99 observation that provides one
possible dynamical explanation for the asymmetry:
that it is the long-term consequence of the gravitational
perturbations of one or more massive bodies on the disk
(i.e., the consequence of the system's "secular perturbations").

Quite apart from any speculation about a nascent planetary system in
the inner region of the HR 4796 disk, we know that the disk must have
been gravitationally perturbed, since HR 4796 is a visual binary
system.
The M dwarf star HR 4796B, located at a projected distance of $517$ AU
(\cite{jmwt98}), is the common proper motion companion
of HR 4796A (\cite{jzbs93}).
An understanding of how asymmetries in the observed structure of the
zodiacal cloud, the tenuous disk of dust in the solar system, are linked to the
solar system's planets (\cite{dghk99}), shows that if there is at least one
massive perturber in the HR 4796 system that is on an eccentric orbit,
then the system's secular perturbations would have caused the
disk's center of symmetry to be offset from the star (\cite{dghw98}).
This offset would mean that the material in one of the disk's observed
lobes is closer to the star than that in the other lobe;
consequently this lobe would be hotter and brighter.
The aim of this paper is to ascertain how large the perturbations would
have to be to cause the observed 5\% asymmetry,
and to discuss whether perturbations of this magnitude are physically
realistic, or even to be expected, in this system.

Since this paper is based on the interpretation of a circumstellar disk
observation, it starts in \S \ref{sec-ocd} with a consideration of how
information about a disk is stored in an observation.
\S \ref{sec-ddc} gives a comprehensive discussion of the physical
processes that govern the dynamical evolution of a disk's particles,
and sets up a broad theoretical framework (the "dynamic disk")
with which to investigate a disk's structure.
\S \ref{sec-spd} then shows how secular perturbations cause a disk's
structure to have offset and warp asymmetries, and provides evidence
that these asymmetries have been observed in the structure of the zodiacal
cloud.
\S \ref{sec-hr4796} uses the theoretical discussions of \S\S
\ref{sec-ocd}-\ref{sec-spd} to create an offset model for the HR 4796 disk
that matches the 18.2 $\mu$m brightness distribution observed by T99.
The interpretation of this model is discussed in \S \ref{sec-disc}.

\section{The Observable Circumstellar Disk}
\label{sec-ocd}
The observed brightness of a circumstellar disk comes from two sources:
starlight that has been absorbed by the disk particles and re-emitted as
thermal radiation (primarily at mid-IR, far-IR, and submillimeter
wavelengths at $\lambda > 5$ $\mu$m), and starlight that has been
scattered by the disk particles (primarily at optical and near-IR
wavelengths at $\lambda < 2$ $\mu$m).
This paper only discusses a disk's thermal emission.

\subsection{Thermal Emission of a Single Disk Particle}
\label{ssec-te}
A particle of diameter $D$, that is at a distance $r$ from a star, is heated
by the stellar radiation to a temperature $T$ that can be calculated from
the equilibrium between the energy that the particle absorbs and that
which it re-emits as thermal radiation.
This temperature depends on the particle's optical properties (\cite{gust94}):
\begin{equation}
  T(D,r) = [\langle Q_{abs} \rangle_{T_\star} /
            \langle Q_{abs} \rangle_{T(D,r)}]^{1/4}*T_{bb}, \label{eq:tdr}
\end{equation}
where this equation must be solved iteratively, since the particle's temperature
appears on both sides of the equation,
$\langle Q_{abs} \rangle_{T_\star}$ and $\langle Q_{abs} \rangle_{T(D,r)}$
are the particle's absorption efficiency averaged over the stellar
spectrum (which can be approximated as that of a black body radiating at
the star's effective temperature $T_\star$)
and the spectrum of a black body radiating at a temperature $T$,
and $T_{bb}$ is the equilibrium temperature of the particle if it were a
black body:
\begin{equation}
  T_{bb} = 278.3\sqrt{a_\oplus/r}*(4\sigma/A)(L_\star/L_\odot)^{1/4},
  \label{eq:tbb}
\end{equation}
where $T_{bb}$ is given in K, $\sigma/A$ is the ratio of the particle's
cross-sectional area to its surface area (e.g., spherical particles have
$\sigma = \pi D^2/4$ and $A = \pi D^2$, giving $\sigma/A = 1/4$),
and $L_\star$ and $L_\odot$ are the luminosities of the star and the
Sun.

If this particle is at a distance $R_\oplus$ from the Earth, the
contribution of its thermal emission to the flux density at a wavelength,
$\lambda$, received at the Earth is given by:
\begin{equation}
  F_\nu(\lambda,D,r) = Q_{abs}(\lambda,D)B_\nu[\lambda,T(D,r)]\Omega(D),
  \label{eq:flux}
\end{equation}
where $B_\nu$ is the Planck function, and $\Omega = \sigma/R_\oplus^2$ is
the solid angle subtended at the Earth by the cross-sectional area of the
particle.

\subsection{Definition of Disk Structure}
\label{ssec-dds}
A circumstellar disk consists of particles with a range of sizes,
compositions and morphologies.
Throughout this paper, however, disk particles are assumed to have the same
composition and morphology, and a particle's size is characterized by its
diameter, $D$.
Disk particles span a range of sizes from $D_{min}$, the smallest
(probably submicron-sized) particles sustainable for a given disk,
up to $D_{max}$, the largest (probably kilometer-sized) members of
the disk that were formed from the proto-planetary disk.
The spatial distribution of these particles can be defined by
$n(D,r,\theta,\phi)$, where $n(D,r,\theta,\phi)dD$ is the volume density
(number per unit volume) of particles in the size range
$D \pm dD/2$ at a location in the disk defined by $r$, the radial distance
from the star, $\theta$, the longitude relative to an arbitrary direction,
and $\phi$, the latitude relative to an arbitrary reference plane.
However, since it is a particle's cross-sectional area that is apparent in an
observation (eq.~[\ref{eq:flux}]), a disk's observable structure is better
defined in terms of $\sigma(D,r,\theta,\phi) = n(D,r,\theta,\phi)\sigma$,
the cross-sectional area per unit volume per unit diameter.

The definition of a disk's structure can be simplified by assuming
the size distribution of its particles to be independent of
$\theta$ and $\phi$:
\begin{equation}
  \sigma(D,r,\theta,\phi) = \bar{\sigma}(D,r)\sigma(r,\theta,\phi),
  \label{eq:sig2}
\end{equation}
where $\bar{\sigma}(D,r)dD$ is the proportion of the total cross-sectional
area of the disk at $r$ that is in particles in the size range $D \pm dD/2$,
and
\begin{equation}
  \sigma(r,\theta,\phi) = \int_{D_{min}}^{D_{max}} \sigma(D,r,\theta,\phi) dD
\end{equation}
is the spatial distribution of cross-sectional area of particles of all
sizes in the disk.

\subsection{Line of Sight Brightness of a Disk Observation}
\label{ssec-los}
To determine the observed brightness of a circumstellar disk in, for example,
one pixel of an image of the disk, two components of the observation
must be defined:
the vector, $\mathbf{R}$, which extends from the observer to the disk, and
which describes how the line of sight intersects the disk in terms of
$r$, $\theta$, and $\phi$;
and the solid angle of the observation, $\Omega_{obs}$, where
$\Omega_{obs} = d_{pix}^2$ for pixels of width $d_{pix}$ radians.

Consider a volume element along this line of sight that is at a location in
the disk defined by $r,\theta,\phi$, and that has a length $\mathbf{dR}$;
the element volume is $dV = \Omega_{obs}R_\oplus^2\mathbf{dR}$.
The contribution of the thermal emission of the particles in this element
to the disk's brightness in the observation is given by:
\begin{eqnarray}
  dF_\nu(\lambda,r,\theta,\phi)/\Omega_{obs} & = &
    \int_{D_ {min}}^{D_{max}} Q_{abs}(\lambda,D)B_\nu[\lambda,T(D,r)]
    \sigma(D,r,\theta,\phi)dD \mathbf{dR}, 
    \label{eq:brightness} \\
    & = & P(\lambda,r)\sigma(r,\theta,\phi)\mathbf{dR}, \label{eq:fnu}
\end{eqnarray}
where equation (\ref{eq:fnu}) uses the simplification for the disk structure
given by equation (\ref{eq:sig2}), and
\begin{equation}
  P(\lambda,r) = \int_{D_ {min}}^{D_{max}} Q_{abs}(\lambda,D)
    B_\nu[\lambda,T(D,r)]\bar{\sigma}(D,r)dD. \label{eq:p}
\end{equation}
Thus, the brightness of this element is not affected by the solid angle
of the observation, neither is it affected by the distance of the element
from the Earth.

Equation (\ref{eq:p}) can also be written as:
\begin{equation}
  P(\lambda,r) = \langle Q_{abs}(\lambda,D)B_\nu[\lambda,T(D,r)]
    \rangle_{\sigma(D,r)}; \label{eq:p2}
\end{equation}
i.e., $P(\lambda,r)$ is the combination of the particles' optical properties
given by $Q_{abs}(\lambda,D)B_\nu[T(D,r),\lambda]$ averaged over
the disk's cross-sectional area distribution, $\sigma(D,r)$.
Thus, $P(\lambda,r)$ can also be given by:
\begin{equation}
  P(\lambda,r) = Q_{abs}(\lambda,D_{typ})B_\nu[\lambda,T(D_{typ},r)],
    \label{eq:pdtyp}
\end{equation}
where $D_{typ}$ is the size of disk particle that characterizes
(and hence dominates) the disk's emission.
This characteristic particle size could be different in different
wavebands, as well as at different distances from the star, but always
lies in the range $D_{min} < D_{typ} < D_{max}$;
its value can be found by considering the relative contribution
of particles of different sizes to $P(\lambda,r)$.
However, unless the optical properties of the particles prevent it, the
particles that dominate a disk's emission are also those that
contribute most to its cross-sectional area, i.e., they are
those that dominate the disk's structure, $\sigma(r,\theta,\phi)$.
This is usually the case for mid-IR ($\lambda = 10-20$ $\mu$m)
observations, such as those of T99 (see e.g., \S \ref{ssec-mpp}).

The total brightness of the disk in this observation is the integral of
equation (\ref{eq:brightness}) over $\mathbf{R}$.
Thus, a disk observation is composed of three parts:
the disk's structure, the optical properties of the disk particles,
and the orientation of the disk to the line of sight.

\subsection{Real Circumstellar Disk Images}
\label{ssec-im}
An image of the disk is made up of many pixels, each of which has a
different line of sight vector, $\mathbf{R}$, and corresponding brightness.
The image at the detector has been convolved with the observational
point spread function (PSF), a combination of the seeing conditions and the
telescope and instrumental optics, which, in the diffraction-limited case,
for an ideal instrument, can be approximated as gaussian smoothing with
$\textrm{FWHM} = \lambda/D$, where this $D$ is the diameter of
the telescope.
The observed image also contains photospheric emission from the point-like
star, as well as random noise fluctuations.
A useful image of the disk can be recovered by subtracting the
image of the star, but the accuracy of this subtraction depends
on how well both the PSF and the stellar flux density are known.
Further smoothing of the image to increase the signal-to-noise may also
prove useful.
 
\section{The Dynamic Disk}
\label{sec-ddc}
This section outlines the theoretical framework upon which later
discussion of circumstellar disk structure (such as how secular
perturbations affect this structure) is based.
A disk is a dynamic entity, the constituent particles of which
are undergoing constant dynamical and physical evolution;
\S \ref{ssec-pp} gives an extensive discussion of the physical
processes acting on disk particles.
\S \ref{ssec-ddpc} then summarizes our understanding of the dynamic disk
by showing how disk particles can be categorized according to the dominant
physical processes affecting their evolution.

\subsection{Physical Processes}
\label{ssec-pp}
If we are to make generalizations about the physical processes relevant to
circumstellar disk evolution, then the zodiacal cloud is the best
example of a circumstellar disk on which to base this understanding,
since its properties have been determined with a certain degree of
confidence.
This confidence stems from the wealth of observational, theoretical,
and physical evidence describing its present state, its evolutionary
history, and the physical environment of the system it is in.
The physical processes that are described in this section are those
thought to have dominated the evolution of the zodiacal cloud,
supposedly since the Sun reached the main sequence
(e.g., \cite{lg90}; \cite{gust94}).
These processes should serve as an adequate basis for an
understanding of the evolution of the circumstellar disks around
other main sequence stars.

There is, however, one obvious distinction between the zodiacal cloud
and exosolar dust disks.
The emission observed from the zodiacal cloud is dominated by that
from dust in the inner solar system (within 5 AU of the Sun), which
has its origins in the asteroid belt (\cite{dnbh84}; \cite{gdjx97}) and
the short period comets (\cite{slhl86}).
In contrast, the emission observed from exosolar dust disks is
dominated by that originating from dust in regions analogous to the
Kuiper belt in the solar system, i.e., $>30$ AU from the star
(e.g., \cite{bp93}).
Relatively little is known about the Kuiper belt, but, like the inner
solar system, it appears to be populated with many asteroid- or
comet-like objects that are probably the remnants of the solar
system's planetary formation phase (e.g., \cite{jewi99}).

\subsubsection{Gravity}
\label{sssec-grav}
The dominant force acting on all but the smallest disk particles
is the gravitational attraction of the star:
\begin{equation}
  F_{grav} = GM_\star m/r^2, \label{eq:fgrav}
\end{equation}
where $G$ is the gravitational constant, $M_\star$ is the mass of the
star, $m$ is the mass of the particle, and $r$ is the distance of the
particle from the star.
All material is assumed to orbit the star on Keplerian (elliptical)
orbits, with other forces acting as perturbations to these orbits.
The orbit of a particle is defined by the following orbital elements:
the semimajor axis, $a$, and eccentricity, $e$, that define the radial
extent and shape of the orbit;
the inclination, $I$, and longitude of ascending node, $\Omega$, that
define the plane of the orbit (relative to an arbitrary reference plane);
and the longitude of pericenter, $\tilde{\omega}$, that defines the
orientation of the orbit within the orbital plane (relative to an arbitrary
reference direction).
The orbital period of the particle is given by:
\begin{equation}
  t_{per} = \sqrt{(a/a_\oplus)^3(M_\odot/M_\star)},
  \label{eq:tper}
\end{equation}
where $t_{per}$ is given in years, $a_\oplus = 1$ AU is the semimajor
axis of the Earth's orbit, and $M_\odot$ is the mass of the Sun.
At any instant, the location of the particle in its orbit is defined by the
true anomaly, $f$, where $f=0^\circ$ and $180^\circ$ at the pericenter and
apocenter, respectively, and its distance from the star, $r$, and its
velocity, $v$, are defined by:
\begin{eqnarray}
  r & = & a(1-e^2)/(1+e\cos{f}) \label{eq:rgrav}, \\
  v & = & \sqrt{GM_\star(2/r-1/a)}. \label{eq:vgrav}
\end{eqnarray}

Since at any given time a particle could be at any point along its orbit,
its contribution to the distribution of material in a disk can be
described by the elliptical ring that contains the mass of the particle
spread out along its orbit, the line density of which varies inversely
with the particle's velocity (eq.~[\ref{eq:vgrav}]).
Each disk particle has an orbit defined by a different set of orbital
elements, with a contribution to the spatial distribution of material in
the disk that can be described by a corresponding elliptical ring.
Thus, a disk's structure can be defined by the distribution of orbital
elements of its constituent particles, $n(D,a,e,I,\Omega,\tilde{\omega})$,
where $n(D,a,e,I,\Omega,\tilde{\omega})dDdadedId\Omega d\tilde{\omega}$
is the number of disk particles with sizes in the range $D \pm dD/2$,
and orbital elements in the range $a \pm da/2, e \pm de/2, I \pm dI/2$,
$\Omega \pm d\Omega/2, \tilde{\omega} \pm d\tilde{\omega}/2$.
A disk's orbital element distribution can be quantified in terms of
the evolutionary history of the system and the physical processes
acting on the disk's particles.
Using techniques such as those described in \S \ref{sssec-simul},
the resulting structure can then be compared with the disk's observable
structure, $\sigma(D,r,\theta,\phi)$, to link a disk observation with
the physics of the particles in that disk.
It is often convenient to discuss the dependence of the orbital
element distribution on the different parameters separately;
e.g., the distribution of semimajor axes, $n(a)$, is defined such that
$n(a)da$ is the total number of particles with orbital semimajor axes
in the range $a \pm da/2$.

\subsubsection{Collisions}
\label{sssec-coll}
A typical disk particle is created by the break-up of a
larger "parent" body, either as the result of a collision with another
body, or simply by its disintegration.
This parent body could have been created by the break-up of an
even larger body, and the particle itself will most likely end up as a
parent body for particles smaller than itself.
This "collisional cascade" spans the complete size range of disk
material, and the particles that share a common ancestor are said to
constitute a "family" of particles.

The size distribution that results from this collisional cascade can be
found from theoretical arguments (\cite{dohn69}):
\begin{equation}
  n(D) \propto D^{2-3q}, \label{eq:nd}
\end{equation}
where $q=11/6$;
this distribution is expected to hold for disk particles that are large
enough not to be affected by radiation forces (\S \ref{sssec-rf}).
A disk with this distribution has its mass, $m(D) = n(D)m$,
concentrated in its largest particles, while its cross-sectional area,
$\sigma(D) = n(D)\sigma$, is concentrated in its smallest particles.
Collisions in such a disk are mostly non-catastrophic (see
eqs.~[\ref{eq:dimp}]-[\ref{eq:fcc}]), and a particle in this disk is most
likely to be broken up by a particle that has just enough mass
(and hence energy) to do so.
This in turn means that collisional fragments have velocities, and hence
orbital elements, that are almost identical to those of the original
particle;
i.e., in the absence of other forces, all members of the same family
have identical orbits.
Due to the interaction of the competing physical processes, the size
distribution of disk particles that are affected by radiation forces
is only really understood qualitatively (\S\S \ref{sssec-rf} and
\ref{sssec-pcsig});
their distribution is particularly important, since, in general,
a disk's cross-sectional area (and hence its observable structure)
is concentrated in these smaller particles.

The importance of collisions in determining a particle's evolution
depends on its collisional lifetime, which is discussed in Appendix
\ref{app-tcoll}.
The collisional lifetime of the particles that constitute most of
a disk's cross-sectional area (i.e., those that are expected to characterize
the disk's mid-IR emission, \S \ref{ssec-los}), can be approximated by
(eqs.~[\ref{eq:tcol5}] and [\ref{eq:fccdtyp}]):
\begin{equation}
  t_{coll}(D_{typ},r) = t_{per}(r)/4\pi\tau_{eff}(r),
    \label{eq:tcol6}
\end{equation}
where $t_{per}(r)$ is the average orbital period of particles at $r$
(eq.~[\ref{eq:tper}] with $a$ replaced by $r$), and $\tau_{eff}(r)$
is the disk's effective face-on optical depth (eq.~[\ref{eq:taueffdefn}]),
which would be equal to the disk's true optical depth if its particles
had unity extinction efficiency.
The collisional lifetime of particles with $D > D_{typ}$ can be
considerably longer than that of equation (\ref{eq:tcol6})
(see e.g., eq.~[\ref{eq:tcol7}]), and a disk's largest particles,
those for which $t_{coll}(D,r) > t_{sys}$, may not have suffered
any catastrophic collisions since they were first created;
such particles are primordial particles.
The cascades of very young disks may still contain a significant proportion
of primordial particles;
i.e., their cascades may not be fully evolved.

The collisional cascade theory is well-supported by evidence from the
zodiacal cloud.
The size distribution of the largest ($D>3$ km) members of the zodiacal
cloud's collisional cascade, the observable asteroids, is well-approximated
by equation (\ref{eq:nd}) (\cite{dd97}; Durda, Greenberg, \& Jedicke 1998);
the distribution of the very largest ($D > 30$ km) asteroids deviates from
this distribution, however, because of the transition from strength-scaling
to gravity-scaling for asteroids larger than $\sim 150$ m (\cite{dgj98}).
The size distribution of the zodiacal cloud's medium-sized
(1 mm $< D < 3$ km) members is also expected to follow equation
(\ref{eq:nd}) (\cite{dd97}), but there is no observational proof of this,
since these members are too faint to be seen individually, and too few to
be studied collectively (\cite{lg90}).
There is, however, proof that the zodiacal cloud's collisional cascade
extends from its largest members down to its smallest dust particles:
the shapes of the "dust band" thermal emission features
(\cite{lbgg84}) correspond to those expected from the small
($1-1000$ $\mu$m) particles resulting from the break-up, some time ago,
of a few very large asteroids, the largest fragments of which are still
observable as the asteroids in the Themis, Koronis, and, possibly, Eos
families (\cite{dnbh84}; \cite{gdjx97}).
The size distribution of the zodiacal cloud's smallest ($D < 1$ mm)
dust particles (e.g., \cite{lg90}; \cite{lb93}) can be explained
qualitatively (e.g., \cite{gzfg85}; see also \S \ref{sssec-rf}).

Analysis of the collision rates of objects in the Kuiper belt
(\cite{ster95}) shows that a collisional cascade should exist here too;
there is also evidence to suggest that the Kuiper belt was once more
massive than it is today (\cite{jewi99}), meaning that in the past
collisions would have played a much larger role in determining its
structure than they do today, maybe even causing the supposed mass loss
(\cite{sc97}).
The size distribution of the observed Kuiper belt objects appears to
be slightly steeper than that in the inner solar system
($q > 11/6$, \cite{jewi99}), while observations have been unable, as yet,
to determine its dust distribution (\cite{bds95}; \cite{gakg97}), 

\subsubsection{Radiation Forces, $\beta$}
\label{sssec-rf}
For most of the collisional cascade, gravity can be considered to be
the only significant force acting on disk particles.
The smallest particles, however, are significantly affected by their
interaction with the photons from the star.

\paragraph{Radiation Pressure}
\label{ssssec-rp}
Radiation pressure is the component of the radiation force that points
radially away from the star.
It is inversely proportional to the square of a particle's distance from
the star, and is defined for different particles by its ratio to the
gravitational force of equation (\ref{eq:fgrav}) (\cite{gust94}):
\begin{equation}
  \beta(D)  =  F_{rad}/F_{grav} =  C_r (\sigma/m) \langle Q_{pr}
    \rangle_{T_\star}(L_\star/L_\odot)(M_\odot/M_\star),  \label{eq:beta}
\end{equation}
where $C_r = 7.65\times 10^{-4}$ kg/m$^2$, $\sigma/m$ is the ratio of
the particle's cross-sectional area to its mass (e.g.,
$\sigma/m = 1.5/\rho D$ for spherical particles of density $\rho$), and
$\langle Q_{pr} \rangle_{T_\star} = \int Q_{pr}(D,\lambda)F_\lambda
d\lambda / \int F_\lambda d\lambda$ is the particle's radiation
pressure efficiency\footnote{A particle's radiation pressure
efficiency is related to its absorption and scattering efficiencies by
$Q_{pr} = Q_{abs} + Q_{sca}(1-\langle \cos{\theta}\rangle)$, where
$\langle \cos{\theta}\rangle$ accounts for the asymmetry of the
scattered radiation.} averaged over the stellar spectrum, $F_\lambda$.

An approximation for large particles is that
$\langle Q_{pr} \rangle_{T_\star} \approx 1$;
thus, large spherical particles have: 
\begin{equation}
  \beta(D) \approx (1150/\rho D)(L_\star/L_\odot)(M_\odot/M_\star),
  \label{eq:beta2}
\end{equation}
where $\rho$ is measured in kg/m$^3$, and $D$ in $\mu$m.
This approximation is valid for particles in the solar system with
$D>20$ $\mu$m (\cite{gust94}).
For particles in exosolar systems, this limit scales with the wavelength
at which the star emits most of its energy,
$\lambda_\star \propto 1/T_\star$;
i.e., equation (\ref{eq:beta2}) is valid for spherical particles with
$D > 20(T_\odot/T_\star)$ $\mu$m, where $T_\odot = 5785$ K is the effective
temperature of the Sun.
Since $\beta \propto 1/D$, this means that the smaller a particle, the
larger its $\beta$;
this holds down to micron-sized particles, smaller than which $\beta$
decreases to a level that is independent of the particle's size
(\cite{gust94}).

The effect of radiation pressure is equivalent to reducing the mass of the
star by a factor $1-\beta$.
This means that a particle for which $\beta \neq 0$ moves slower
around the same orbit by a factor of $\sqrt{1-\beta}$ than one for which
$\beta = 0$ (eq.~[\ref{eq:vgrav}]).
It also means that daughter fragments created by the break-up of
a parent body move on orbits that can differ substantially from that
of the parent.
The reason for this, is that while the positions and velocities of a
parent and its daughter fragments are the same at the moment of break-up
(apart from a small velocity dispersion), their $\beta$ are different,
and so the daughter fragments move in effective potentials that are different
from that the parent moved in.
Daughter fragments created in the break-up of a parent particle that had
$\beta = 0$, and for which orbital elements at the time of the
collision were $a,e,I,\Omega,\tilde{\omega}$, and $f$, move in the
same orbital plane as the parent, $I^{'} = I$ and
$\Omega^{'} = \Omega$, but on orbits with semimajor axes, $a^{'}$,
eccentricities, $e^{'}$, and pericenter orientations,
$\tilde{\omega}^{'}$, that are given by (Burns, Lamy, \& Soter 1979):
\begin{eqnarray}
  a^{'} & = & a(1-\beta) / \left[1 - 2\beta(1+e\cos{f})/(1-e^2)\right],
  \label{eq:aprime} \\
  e^{'} & = & (1-\beta)^{-1}\sqrt{e^2 + 2\beta e\cos{f} + \beta^2},
  \label{eq:eprime} \\
  \tilde{\omega}^{'} - \tilde{\omega} & = & f - f^{'} =
  \arctan{\left[ \beta \sin{f}/(\beta\cos{f} + e) \right]}. \label{eq:wprime}
\end{eqnarray}

\begin{figure}
\epsscale{0.45}
  \begin{center}
    \begin{tabular}{c}
      \epsscale{0.8} \plotone{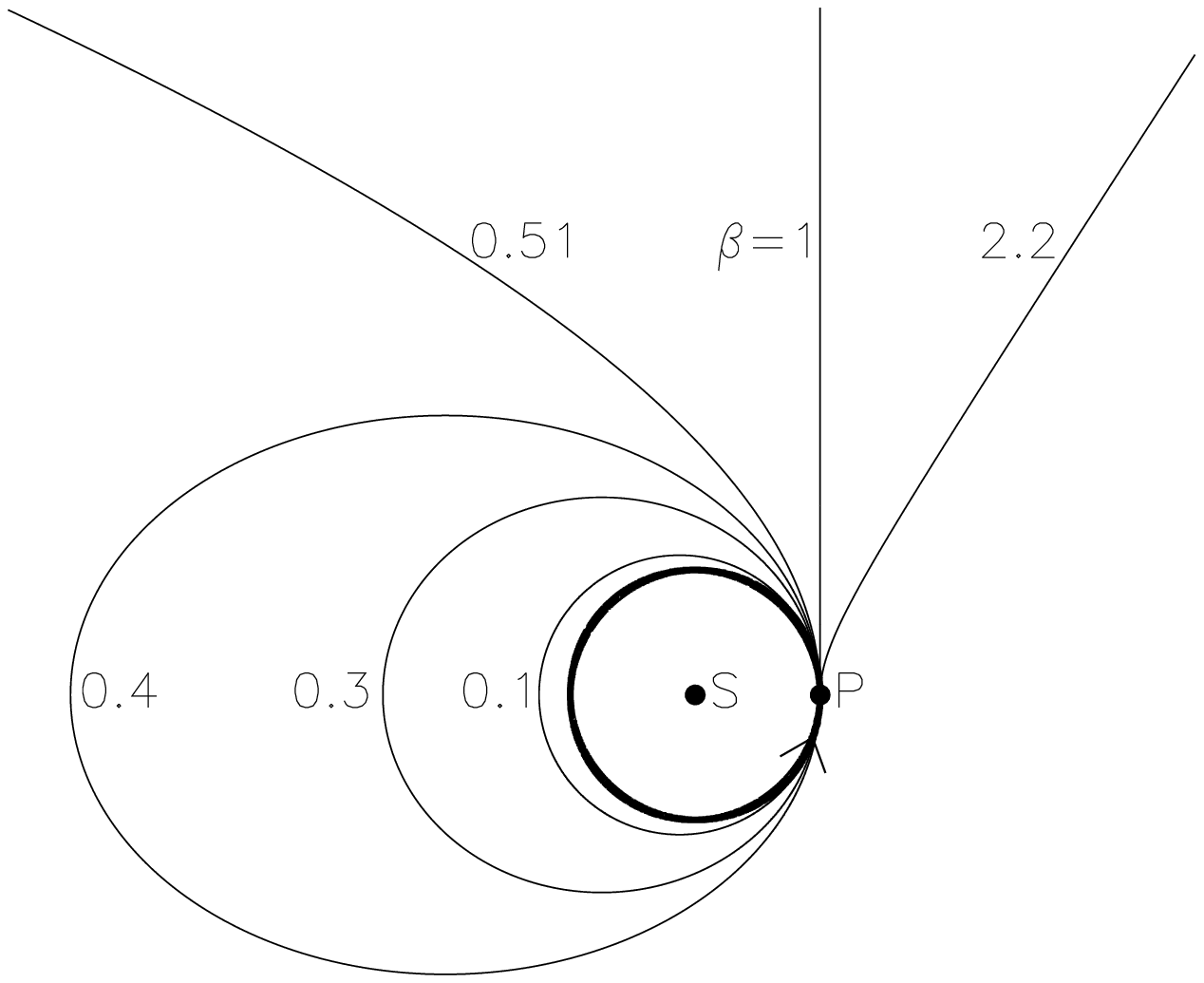}
    \end{tabular}
  \end{center}
  \caption{Figure showing the new orbits of the fragments of a collision
  in which a large parent particle "P", that was on a circular orbit
  around a star "S", was broken up.
  Fragments of different sizes have different radiation pressure forces,
  characterized by a particle's $\beta$, acting on them, and so have
  different orbits:
  those with $\beta < 0.1$, the "large" particles, have orbits that are
  close to that of the parent;
  those with $0.1 < \beta < 0.5$, the "$\beta$ critical" particles, have
  orbits that have the same pericenter distance as the parent,
  but larger apocenter distances;
  and those with $\beta > 0.5$, the "$\beta$ meteoroids", have hyperbolic
  orbits.
  Since when they are created, the velocity vector of all fragments is
  perpendicular to the stellar direction, this is the point of their orbit's
  closest approach to the star.
  The orbits of fragments with $\beta = $0, 0.1, 0.3, 0.4, 0.51, 1.0, and 2.2
  are shown here.
  The thick circular line denotes both the orbit of the parent particle
  and that of collisional fragments with $\beta = 0$.
  All particles are orbiting the star counterclockwise.}
  \label{fig1}
\end{figure}

\placefigure{fig1}

Analysis of equations (\ref{eq:aprime})-(\ref{eq:wprime}) shows that
the orbits of the largest fragments, those for which $\beta < 0.1$,
are similar to that of the parent.
On the other hand, the smallest fragments, those for which
$\beta > 0.5(1-e^2)/(1+e\cos{f})$, have hyperbolic orbits ($e^{'} > 1$);
these particles are known as "$\beta$ meteoroids"\footnote{Note that
particles with $\beta > 1$ are $\beta$ meteoroids even
if they were not created collisionally, since they "see" a negative
mass star.}.
Since $\beta$ meteoroids are lost from the system on the timescale
of the orbital period of the parent (eq.~[\ref{eq:tper}]), the diameter of
particle for which $\beta > 0.5$ essentially defines the lower end of
the collisional cascade.
However, there may also be a population of submicron particles that have
$\beta < 0.5$ (\cite{gust94}).
The intermediate-sized fragments, those for which $0.1 < \beta < 0.5$,
that we call "$\beta$ critical" particles, have orbits that differ
substantially from that of the parent.
However, the point of closest approach to the star of the orbits of
all daughter fragments, irrespective of their size, is the same as that of
the parent:
combining equations (\ref{eq:aprime}) and (\ref{eq:eprime}) gives the
pericenter distance of daughter fragments, $r_p^{'} = a^{'}(1-e^{'})$, as
\begin{equation}
  r_p^{'}/r_p = 1 + e(1-\cos{f}) + O(e^2). \label{eq:rp}
\end{equation}
The orbits of collisional fragments with different $\beta$ from a
parent particle that was on a circular orbit are shown in Fig.~\ref{fig1}.

\paragraph{Poynting-Robertson (P-R) Light Drag}
\label{ssssec-pr}
The component of the radiation force tangential to a particle's
orbit is called the P-R drag force.
This force is also proportional to $\beta$.
It results in an evolutionary decrease in both the semimajor
axis and eccentricity of the particle's orbit (\cite{bls79}):
\begin{eqnarray}
  \dot{a}_{pr} & = & -(\alpha/a)(2+3e^2)/(1-e^2)^{3/2}
              = -2\alpha/a + O(e^2) \label{eq:adot} \\
  \dot{e}_{pr} & = & -(\alpha/a^2)2.5e/(1-e^2)^{1/2}
              = -2.5\alpha e/a^2 + O(e^2) \label{eq:edot},
\end{eqnarray}
where $\alpha = 6.24 \times 10^{-4} (M_\star/M_\odot)\beta$
AU$^2$/year.
P-R drag does not change the plane of the particle's orbit,
$\dot{I}_{pr} = \dot{\Omega}_{pr} = 0$;
neither does it affect the orientation of the particle's pericenter,
$\dot{\tilde{\omega}}_{pr} = 0$.
For a particle with zero eccentricity, equation (\ref{eq:adot})
can be solved to find the time it takes for the particle to spiral in from a
radial distance of $r_1$ to $r_2$:
\begin{equation}
  t_{pr} = 400(M_\odot/M_\star)[(r_1/a_\oplus)^2-(r_2/a_\oplus)^2]/\beta
  \label{eq:tpr},
\end{equation}
where $t_{pr}$ is given in years.

Consider the daughter fragments created in the break-up of a parent body that
was on an orbit at a distance $r$ from the star.
The largest fragments are broken up by collisions before their orbits have
suffered any significant P-R drag evolution, while the smaller fragments,
for which the P-R drag evolution is faster, can reach the star without having
encountered another particle (at which point they evaporate).
Particles for which P-R drag significantly affects their orbits in their
lifetime can be estimated as those for which their collisional lifetime
(eq.~[\ref{eq:tcol6}]; the use of this equation is justified in the next
paragraph) is longer than their P-R drag lifetime (eq.~[\ref{eq:tpr}]
with $r_2=0$), i.e., those for which $\beta > \beta_{pr}$, where
\begin{equation}
  \beta_{pr} = 5000\tau_{eff}(r)\sqrt{(M_\odot/M_\star)
    (r/a_\oplus)}; \label{eq:btr}
\end{equation}
for large spherical particles, this is also those for which $D < D_{pr}$,
where
\begin{equation}
  D_{pr} = [0.23/\rho \tau_{eff}(r)](L_\star/L_\odot)\sqrt{(M_\odot/M_\star)
                      (a_\oplus/r)}, \label{eq:dtr}
\end{equation}
$\rho$ is measured in kg/m$^3$, and $D_{pr}$ in $\mu$m.

Consider the daughter fragments created in the break-up of an endless supply
of parent bodies that are on orbits with the same semimajor axis, $a_s$.
Ignoring collisional processes, the fragments with orbits that are
affected by P-R drag, those with $\beta > \beta_{pr}$, have their
semimajor axes distributed from $a=a_s$ to $a=0$ according to
(eq.~[\ref{eq:adot}]):
\begin{equation}
  n(a) \propto 1/\dot{a}_{pr} \propto a; \label{eq:napr}
\end{equation}
this corresponds to a volume density distribution that is roughly
inversely proportional to distance from the star\footnote{If the
particles had circular orbits, equation (\ref{eq:napr}) means
a spherical shell of width $dr$, the volume of which is $\propto r^2dr$,
would contain a number of particles that is $\propto rdr$
(see e.g., \cite{gomt97}).}.
If the collisional processes leading to the size distribution of the
parent bodies, $n_s(D)$, still holds for the production of the P-R drag
affected particles, then their size distribution is given by:
\begin{equation}
  n(D) \propto n_s(D)/\dot{a}_{pr} \propto n_s(D)D.
  \label{eq:ndpr}
\end{equation}
If $n_s(D)$ can be given by equation (\ref{eq:nd}) with $q = 11/6$,
the cross-sectional area of a disk's P-R drag affected particles
is concentrated in the largest of these particles, while that
of its unaffected particles is concentrated in the smallest of these
particles;
i.e., most of a disk's cross-sectional area is expected to be concentrated
in particles with $D_{typ} \approx D_{pr}$, justifying the use of equation
(\ref{eq:tcol6}) for the collisional lifetime of these particles.

Observations of the zodiacal cloud at 1 AU show that its effective
optical depth here is $\tau_{eff} = O(10^{-7})$.
Since these particles originated in the asteroid belt at $\sim 3$ AU,
arriving at 1 AU due to the P-R drag evolution of their orbits,
the zodiacal cloud's volume density should vary $\propto 1/r$,
and its effective optical depth at 3 AU should be similar to that at 1 AU.
Assuming zodiacal cloud particles to have a density $\sim 2500$
kg/m$^3$ (\cite{lg90}), the cross-sectional area of material in the
asteroid belt should be concentrated in particles with
$D_{pr} = O$(500 $\mu$m) (eq.~[\ref{eq:dtr}]), for which both the
collisional lifetime, and the P-R drag lifetime, is $\sim 4$ Myr.
The cross-sectional area of material at 1 AU is expected to be
concentrated in particles smaller than that in the asteroid
belt, since many of the larger particles should have been
broken up by collisions before they reach the inner solar system;
this is in agreement with observations that show the
cross-sectional area distribution at 1 AU to peak for particles
with $D = 100-200$ $\mu$m (\cite{lg90}; \cite{lb93}).
Also, equation (\ref{eq:tcol7}) with $D_{typ} = 500$ $\mu$m,
$D_{cc}(D)/D = (10^{-4})^{1/3}$, and $q = 11/6$, predicts that the
collisional lifetime of large bodies in the asteroid belt should be:
\begin{equation}
  t_{coll} \approx 10^9\sqrt{D}, \label{eq:tcollast}
\end{equation}
where $t_{coll}$ is given in years, and $D$ in km.
Since the solar system is $\sim 4.5 \times 10^9$ years old,
this implies that asteroids larger than $\sim 20$ km should be
primordial asteroids;
this is in agreement with more accurate models of the observed size
distribution of these asteroids (\cite{dgj98}).

\subsection{Division of a Disk into Particle Categories}
\label{ssec-ddpc}
Disk particles of different sizes can be categorized according to the
dominant physical processes affecting their evolution.
Particles in the different categories have different lives;
i.e., the way they are created, their dynamical evolution,
and the way they are eventually destroyed, are all different.
Each of a disk's categories has a different spatial distribution;
which of these categories dominates the disk's observable structure
depends on the relative contribution of each to the disk's
cross-sectional area (see \S \ref{ssec-los}).

\subsubsection{Category Definitions}
\label{sssec-dpc}
A disk's largest particles, those with $\beta < 0.1$, have orbital
elements that are initially the same as, or at least very similar to,
those of their parents.
Of these large particles, only those with $\beta < \beta_{pr}$
suffer no significant P-R drag evolution to their orbits in their
lifetime.
These truly large particles continue on the same orbits as those
of their ancestors until they collide with a particle large enough to
cause a catastrophic collision;
the resulting collisional fragments populate the collisional cascade.
The spatial distribution of these "large" particles in a disk is
its base distribution.
The spatial distributions of a disk's smaller particles can only be
understood in terms of how they differ from the disk's base distribution.

Disk particles with $\beta_{pr} < \beta < 0.1$, spiral in from their
parent's orbits due to P-R drag, so that they are closer to the star than
their parents by the time of their demise (which could be caused
either by collisions or by evaporation close to the star);
the spatial distribution of these "P-R drag affected" particles in
a disk differs from the disk's base distribution in that it extends
closer in to the star.
The orbits of particles with $0.1 < \beta < 0.5$ also undergo
significant P-R drag evolution before their demise, but their original
orbits are already different from those of their parents;
the spatial distribution of these "$\beta$ critical" particles
in a disk extends both further out, and further in, from the disk's base
distribution.
Particles with $\beta > 0.5$ leave their parents on hyperbolic
orbits, and so are quickly lost from the system;
the spatial distribution of these "$\beta$ meteoroids" in a disk
extends further out, but not further in, from the disk's base distribution. 

Thus, a disk comprises four particle categories, each of which
has a different spatial distribution, although all are inextricably
linked to that of the large particles through the collisional
cascade.
For disks that have $\beta_{pr} > 0.5$, however,
i.e., those with
\begin{equation}
  \tau_{eff}(r) > 10^{-4}\sqrt{(M_\star/M_\odot)(a_\oplus/r)},
  \label{eq:taueffnopr}
\end{equation}
there is no significant P-R drag evolution of any of its constituent
particles.
Such disks comprise just three categories, since the P-R drag affected
category is empty.

\subsubsection{Category Cross-Sectional Area}
\label{sssec-pcsig}
The size distributions of a disk's large particles, and its P-R drag
affected particles, were discussed in \S\S \ref{sssec-coll} and
\ref{sssec-rf}.
These discussions imply that the cross-sectional area of a disk
in which there is a population of P-R drag affected particles
(see eq.~[\ref{eq:taueffnopr}]) is dominated by particles with
$D \approx D_{pr}$ (eq.~[\ref{eq:dtr}]).
As a first-cut approximation, the size distribution of a disk in which
there are no P-R drag affected particles follows equation
(\ref{eq:nd}) from $D_{max}$ down to $D_{min} = D(\beta=0.5)$;
however, since a particle's catastrophic collision rate is affected by
the size distribution of particles smaller than itself (\cite{dd97};
eq.~[\ref{eq:rcoll}]), this distribution cannot be expected to
hold all the way down to $D(\beta=0.5)$.
This means that the cross-sectional area of such a disk is concentrated
in its smallest particles, and the contribution of $\beta$ critical
particles to the disk's total cross-sectional area is given by 
\begin{equation}
  d\sigma/\sigma_{tot} = [D^{5-3q}]_{D(\beta=0.1)}^{D(\beta=0.5)} /
                      [D^{5-3q}]_{D_{max}}^{D(\beta=0.5)};
\end{equation}
e.g., if $q=11/6$, and $\beta \propto 1/D$, this means that half of the
disk's cross-sectional area comes from its $\beta$ critical particles.

Since $\beta$ meteoroids have hyperbolic orbits, they are expected to
contribute little to a disk's cross-sectional area unless they are
produced at a high enough rate to replenish their rapid loss from the
system.
This could be the case if the disk was very dense, since material would
pass quickly through the cascade;
such a disk would undergo considerable mass loss.
An estimate of how dense the disk would have to be for this to be the
case depends on the assumptions made about the physics of collisions
between small particles.
For heuristic purposes, it is assumed here that the total cross-sectional
area of $\beta$ meteoroids created by the collisional break-up of a parent
body is comparable to that of the parent itself;
this is probably an underestimate if the collision is destructive,
but an overestimate if the collision is erosive.
If this were the case, then the disk's $\beta$ meteoroids would dominate
a disk's cross-sectional area only if their lifetime,
which is of the order of the orbital period of their parents,
is longer than the lifetime of these parents,
which can be approximated by equation (\ref{eq:tcol6}),
i.e., only if
\begin{equation}
  \tau_{eff}(r) > 0.1. \label{eq:taueffnobm}
\end{equation}

In conclusion, from a theoretical stand-point, there are few solid
assumptions that can be made about a disk's size distribution,
other than that the denser a disk is, the smaller the diameter of
particles that its cross-sectional area is concentrated in.

\subsection{The Perturbed Dynamic Disk}
\label{ssec-pf}
In addition to the physical processes described in \S \ref{ssec-pp}, the
particles of the dynamic disk described in this section are affected by
a number of perturbing processes;
these produce subtle, but perhaps observable, changes in a disk's
structure.
The dominant perturbing processes in the zodiacal cloud are the
secular and resonant gravitational perturbations of the planets.
Secular perturbations are discussed in \S \ref{sec-spd}.
Resonant perturbations give rise to planetary resonant rings:
a planet's resonant ring is an asymmetric circumstellar ring of material
that co-orbits with the planet as a result of the resonant trapping of
particles into the planet's exterior mean motion resonances (not to be
confused with circumplanetary dust rings).
There is both observational and theoretical evidence for the existence
of the Earth's resonant ring (\cite{djxg94}; \cite{rfwh95});
many of the observed Kuiper belt objects are trapped in 2:3 resonance
with Neptune (\cite{jewi99}), thus forming Neptune's resonant ring
(\cite{malh95}).

Other possible perturbing processes include:
stellar wind forces, that, at least for dust in the solar system, effectively
increase the value of $\beta$ for P-R drag (e.g., \cite{lg90});
Lorentz forces acting on charged particles, that, while negligible for
particles in the inner solar system, are increasingly important for
particles at distances further from the Sun (e.g., \cite{km98});
interactions with dust from the interstellar medium (e.g., \cite{ac97});
the sublimation of icy dust grains, which is one of the mechanisms that has
been suggested as the cause of the inner hole in the HR 4796 disk
(\cite{jmwt98});
and the self gravity of a massive disk, which could have played an important
role in determining the evolution of the primordial Kuiper belt
(\cite{wh98}).

\section{Structure of a Secularly Perturbed Disk}
\label{sec-spd}
The orbit of a particle in a circumstellar disk that is in a system
in which there are one or more massive perturbers is inevitably
affected by the gravitational perturbations of these bodies.
The consequent evolution of the particle's orbit can be used to
obtain a quantitative understanding of the effect of these perturbations
on the structure of the disk.

\subsection{Secular Perturbation Theory}
\label{ssec-spt}

\subsubsection{Perturbation Equations}
\label{sssec-pe}
The gravitational forces from a planetary system that act to perturb the
orbit of a particle in the system can be decomposed into the sum of many
terms that are described by the particle's disturbing function, $R$.
The long-term average of these forces are the system's secular
perturbations, and the terms of the disturbing function that contribute
to these secular perturbations, $R_{sec}$, can be identified as those that
do not depend on the mean longitudes of either the planets or the particle
(the other forces having periodic variations).

Consider a particle that is orbiting a star of mass $M_\star$, that also
has $N_{pl}$ massive, perturbing, bodies orbiting it.
This particle has a radiation pressure force acting on it represented by
$\beta$, and its orbit is described by the elements $a$, $e$, $I$,
$\Omega$ and $\tilde{\omega}$.
To second order in eccentricities and inclinations, the secular terms in the
particle's disturbing function are given by
(\cite{bc61}; \cite{dnbh85}; \cite{dn86}; \cite{md99}):
\begin{equation}
  R_{sec} =  na^2 \left[ \frac{1}{2}A(e^2-I^2)
    + \sum_{j=1}^{N_{pl}} \left[A_j e e_j cos(\tilde{\omega}-\tilde{\omega}_j)
    + B_j I I_j cos(\Omega-\Omega_j) \right] \right],
\label{eq:rsec}
\end{equation}
where $n = (2\pi/t_{year})\sqrt{(M_\star/M_\odot)(1-\beta)(a_\oplus/a)^3}$
is the mean motion of the particle in rad/s,
$t_{year} = 2\pi / \sqrt{GM_\odot/a_\oplus^3} = 3.156\times 10^7$ s is one year
measured in seconds, and
\begin{mathletters}
  \begin{eqnarray}
    A   & = & + \frac{n}{4(1-\beta)}
                \sum_{j=1}^{N_{pl}}
	        \left( \frac{M_{j}}{M_\star}     \right)
	        \alpha_j\overline{\alpha}_j b_{3/2}^{1}(\alpha_j),
    \label{eq:a} \\
    A_j & = & - \frac{n}{4(1-\beta)}
	        \left( \frac{M_{j}}{M_\star}     \right)
	        \alpha_j\overline{\alpha}_j b_{3/2}^{2}(\alpha_j),
    \label{eq:aj} \\
    B_j & = & + \frac{n}{4(1-\beta)}
	        \left( \frac{M_{j}}{M_\star}     \right)
	        \alpha_j\overline{\alpha}_j b_{3/2}^{1}(\alpha_j),
    \label{eq:bj}
  \end{eqnarray}
\end{mathletters}
where $\alpha_j = a_j/a$ and $\overline{\alpha}_j = 1$ for $a_j<a$, and 
$\alpha_j = \overline{\alpha}_j = a/a_j$ for $a_j>a$,
and $b_{3/2}^s(\alpha_j) = (\pi)^{-1}\int_0^{2\pi}(1-2\alpha_j\cos{\psi}+
\alpha_j^2)^{-3/2}\cos{s\psi}d\psi$ are the Laplace coefficients ($s=1,2$).
$A$, $A_j$ and $B_j$ are in units of rad/s, and
$R_{sec}$ is in units of m$^2$/s$^2$.

The effect of these perturbations on the orbital elements of the
particle can be found using Lagrange's planetary equations
(\cite{bc61}; \cite{md99}).
The semimajor axis of the particle remains constant, 
$\dot{a}_{sec} = 0$, while the variations of its eccentricity
and inclination are best described when coupled with the
variations of its longitude of pericenter and ascending node
using the variables defined by its complex eccentricity, $z$,
and complex inclination, $y$:
\begin{mathletters}
  \begin{eqnarray}
    z & = & e*\exp{i\tilde{\omega}}, \label{eq:zdefn} \\
    y & = & I*\exp{i\Omega}, \label{eq:ydefn}
  \end{eqnarray}
\end{mathletters}
where $i^2 = -1$.
Using these variables Lagrange's planetary equations give the orbital
element variations due to secular perturbations as:
\begin{mathletters}
  \begin{eqnarray}
    \dot{z}_{sec} & = & +iAz + i\sum_{j=1}^{N_{pl}}A_j z_j, \label{eq:zdot} \\
    \dot{y}_{sec} & = & -iAy + i\sum_{j=1}^{N_{pl}}B_j y_j, \label{eq:ydot}
  \end{eqnarray}
\end{mathletters}
where $z_j$ and $y_j$ are the complex eccentricities and inclinations
of the perturbers, which have a slow temporal variation due to the
secular perturbations of the perturbers on each other (\cite{bc61};
\cite{md99}):
\begin{mathletters}
  \begin{eqnarray}
    z_j(t) & = & \sum_{k=1}^{N_{pl}}e_{jk}\ast\exp{i(g_kt + \beta_k)},
      \label{eq:zj} \\
    y_j(t) & = & \sum_{k=1}^{N_{pl}}I_{jk}\ast\exp{i(f_kt + \gamma_k)},
      \label{eq:yj}
  \end{eqnarray}
\end{mathletters}
where $g_k$ and $f_k$ are the eigenfrequencies of the perturber system,
the coefficients $e_{jk}$ and $I_{jk}$ are the corresponding
eigenvectors, and $\beta_k$ and $\gamma_k$ are constants found from the
initial conditions of the perturber system.

\subsubsection{Solution to Perturbation Equations}
\label{sssec-lp}
Ignoring the evolution of a particle's orbital elements due to P-R drag,
equations (\ref{eq:zdot}) and (\ref{eq:ydot}) can be solved to give
the secular evolution of the particle's instantaneous complex
eccentricity and inclination (a.k.a.~the particle's osculating
elements).
This secular evolution is decomposed into two distinct time-varying
elements --- the "forced", subscript $f$, and "proper", subscript $p$,
elements --- that are added vectorially in the complex planes
(see e.g., Fig.~\ref{fig2}a):
\begin{mathletters}
  \begin{eqnarray}
    z(t) = z_f(t) + z_p(t) & = & \sum_{k=1}^{N_{pl}} \left[
             \frac{ \sum_{j=1}^{N_{pl}}A_j e_{jk}}{g_k-A} \right]
             \ast\exp{i(g_kt + \beta_k)} +
             e_p\ast\exp{i(+At+\beta_0)},  \label{eq:z} \\
    y(t) = y_f(t) + y_p(t) & = & \sum_{k=1}^{N_{pl}} \left[
             \frac{ \sum_{j=1}^{N_{pl}}B_j I_{jk}}{f_k+A} \right]
             \ast\exp{i(f_kt + \gamma_k)} +
             I_p\ast\exp{i(-At+\gamma_0)},  \label{eq:y}	    
  \end{eqnarray}
\end{mathletters}
where $e_p$, $\beta_0$, and $I_p$, $\gamma_0$ are
determined by the particle's initial conditions.

These equations have simple physical and geometrical interpretations.
A particle's forced elements, $z_f$ and $y_f$, depend only on the
orbits of the perturbers in the system (that have a slow secular
evolution, eqs.~[\ref{eq:zj}] and [\ref{eq:yj}]), as well as on the
particle's semimajor axis (which has no secular evolution).
Thus, at a time $t_0$, a particle that is on an orbit with a
semimajor axis $a$, has forced elements imposed on its orbit by
the perturbers in the system that are defined by $z_f(a,t_0)$ and
$y_f(a,t_0)$.
The contribution of the particle's proper elements to its osculating
elements, $z(t_0)$ and $y(t_0)$, is then given by:
$z_p(t_0) = z(t_0) - z_f(a,t_0)$ and $y_p(t_0) = y(t_0) - y_f(a,t_0)$;
thus defining the particle's proper eccentricity, $e_p$, and proper
inclination, $I_p$, which are its fundamental orbital elements
(i.e., those that the particle would have if there were no perturbers in
the system), as well as the orientation parameters $\beta_0$ and
$\gamma_0$.
Since both the forced elements, and the osculating elements, of
collisional fragments are the same as those of their parent
(apart from fragments with $\beta > 0.1$), particles from the same
family have the same proper elements, $e_p$ and $I_p$.

\placefigure{fig2}

The evolution of a particle's proper elements is straight-forward
--- they precess around circles of fixed radius, $e_p$ and $I_p$, at a
constant rate, $A$, counterclockwise for $z_p$, clockwise for
$y_p$.
The secular precession timescale depends only on the semimajor
axis of the particle's orbit:
\begin{equation}
  t_{sec} = 2\pi /At_{year},
  \label{eq:tsec}
\end{equation}
where $t_{sec}$ is given in years, and $A$ is given in
equation (\ref{eq:a});
secular perturbations produce long period variations in a particle's
orbital elements (e.g., $t_{sec} = O$(0.1 Myrs) in the asteroid belt).
The centers of the circles that the proper elements precess
around are the forced elements (see e.g., Fig.~\ref{fig2}a).
Actually the forced elements vary on timescales that are
comparable to the precession timescale (eq.[\ref{eq:tsec}]);
thus, it might appear ambitious to talk of the precession of a particle's
osculating elements around circles when its real evolutionary track in
the complex eccentricity and complex inclination planes may not be
circular at all.
The reason it is presented as such is that at any given time, all of the
particles at the same semimajor axis precess (at the same rate)
around the same forced elements on circles of different radii, and this
has consequences for the global distribution of orbital elements
(see \S \ref{sssec-opsfam}).

There are two things that are worth mentioning now about a particle's
forced elements.
If there is just one perturber in the system, $N_{pl} = 1$, its complex
eccentricity and complex inclination do not undergo any secular
evolution, and the forced elements imposed on a particle in the system
are not only constant in time, but also independent of the mass of the
perturber:
\begin{mathletters}
  \begin{eqnarray}
    z_f & = & \left[ b_{3/2}^{2}(\alpha_j) / b_{3/2}^{1}(\alpha_j) \right]
              e_j\ast\exp{i\tilde{\omega}_j}, \label{eq:zf1} \\
    y_f & = & I_j\ast\exp{i\Omega_j}. \label{eq:yf1}
  \end{eqnarray}
\end{mathletters}
This implies that a body of low mass, such as an asteroid, has as
much impact on a particle's orbit as a body of high mass, such as a
Jupiter mass planet.
The perturbations from a smaller perturber, however, produce longer
secular precession timescales (eq.~[\ref{eq:tsec}]):
\begin{equation}
  t_{sec} = 4[\alpha_j \overline{\alpha}_j b_{3/2}^1(\alpha_j)
             (a_\oplus/a)^{3/2}(M_j/M_\star)\sqrt{M_\star/M_\odot}]^{-1};
          \label{eq:tsec2}
\end{equation}
they would also be similar in magnitude to those of the disk's
self-gravity, which in that case could no longer be ignored.
If there is more than one perturber in the system, $N_{pl} > 1$,
then particles on orbits for which their precession rate equals one of
the system's eigenfrequencies ($A = g_k$, or $-A = f_k$) have infinite
forced elements imposed on their orbits, and so are quickly ejected
from such a "secular resonance" region.

The solution given by equations (\ref{eq:z}) and (\ref{eq:y}) accounts
for the fact that small particles see a less massive star due to the
action of radiation pressure, but not for the P-R drag evolution of their
orbits;
the solution for these particles is discussed in Appendix \ref{app-pr}.
Also, the perturbation theory of \S \ref{sssec-pe} is only valid for particles
with small eccentricities;
i.e., it is not valid for the evolution of a disk's $\beta$ critical particles,
or its $\beta$ meteoroids.
However, if the evolution of a disk's $\beta$ critical particles is
affected by secular perturbations (i.e., if their lifetime is longer than
the secular timescale), then it is probably also affected by
P-R drag (i.e., their lifetime is probably also longer than the P-R drag
timescale), in which case the disk's $\beta$ critical particles do not
contribute much to its observable structure (\S\S \ref{sssec-rf}
and \ref{sssec-pcsig}).
There is no secular evolution to the orbits of $\beta$ meteoroids because
of their short lifetimes.

\subsection{Offset and Warp}
\label{ssec-ow}
The effect of secular perturbations on the structure of a disk can be
understood by considering the effect of the secular evolution of the
constituent particles' orbits on the distribution of their orbital
elements.
The perturbation equations of \S \ref{sssec-pe} show that secular
pertubations affect only the distribution of disk particles' complex
eccentricities, $n(z)$, and complex inclinations, $n(y)$,
while having no effect on their size distribution (and hence the
division of the disk into its particle categories), or on their semimajor
axis distribution (and hence the disk's large-scale radial distribution).

\subsubsection{Offset and Plane of Symmetry of Family Material}
\label{sssec-opsfam}
Consider the family of collisional fragments originating from a
primordial body, the orbital elements of which were described by
$a$, $e_p$, and $I_p$.
Here we consider only fragments that are unaffected by P-R drag
(those that are affected by P-R drag are discussed in Appendix
\ref{app-pr}). 

\begin{figure}
\epsscale{0.45}
  \begin{center}
    \begin{tabular}{cc}
      \epsscale{0.47} \plotone{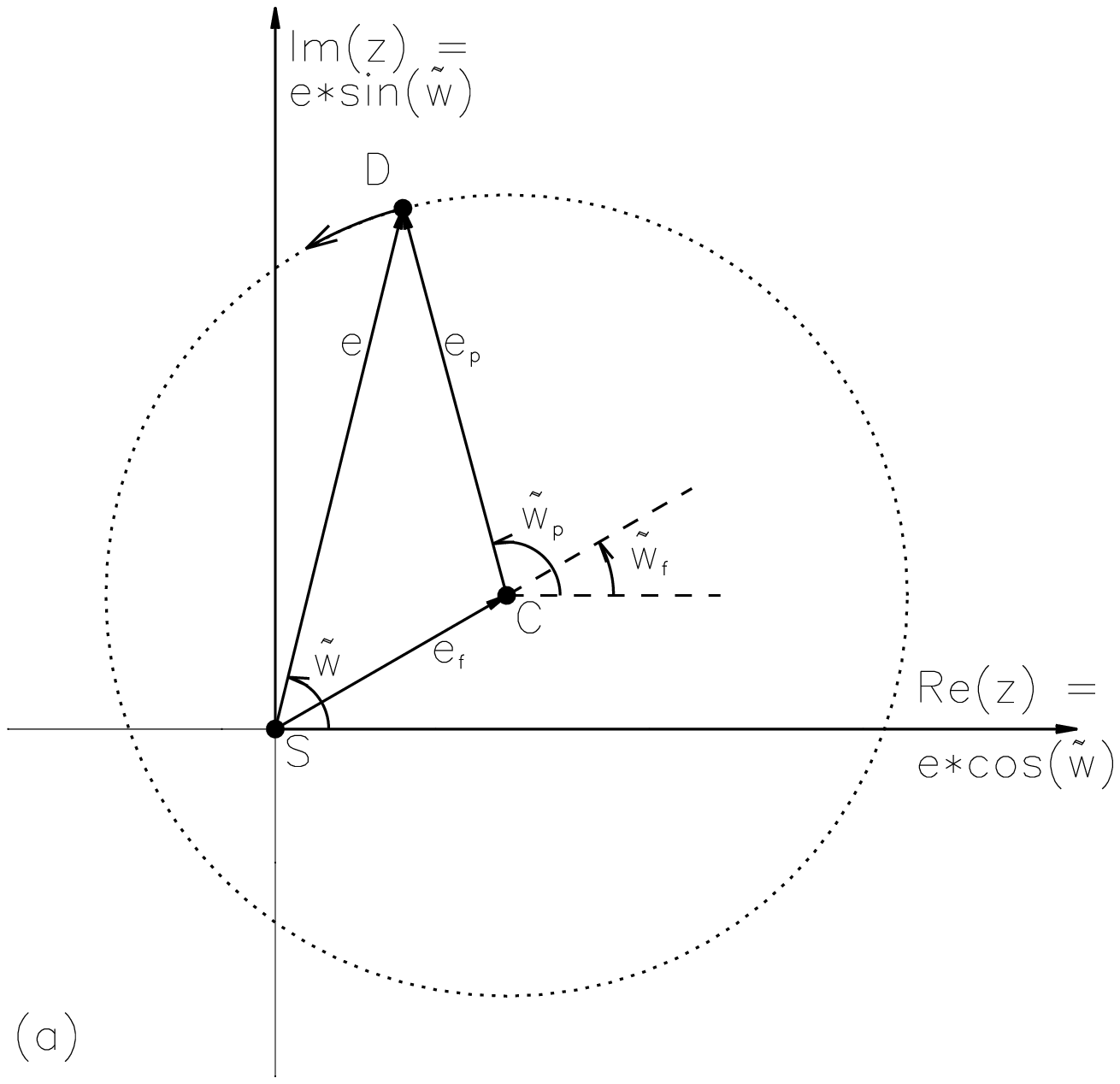} &
      \epsscale{0.43} \hspace{0.1in} \plotone{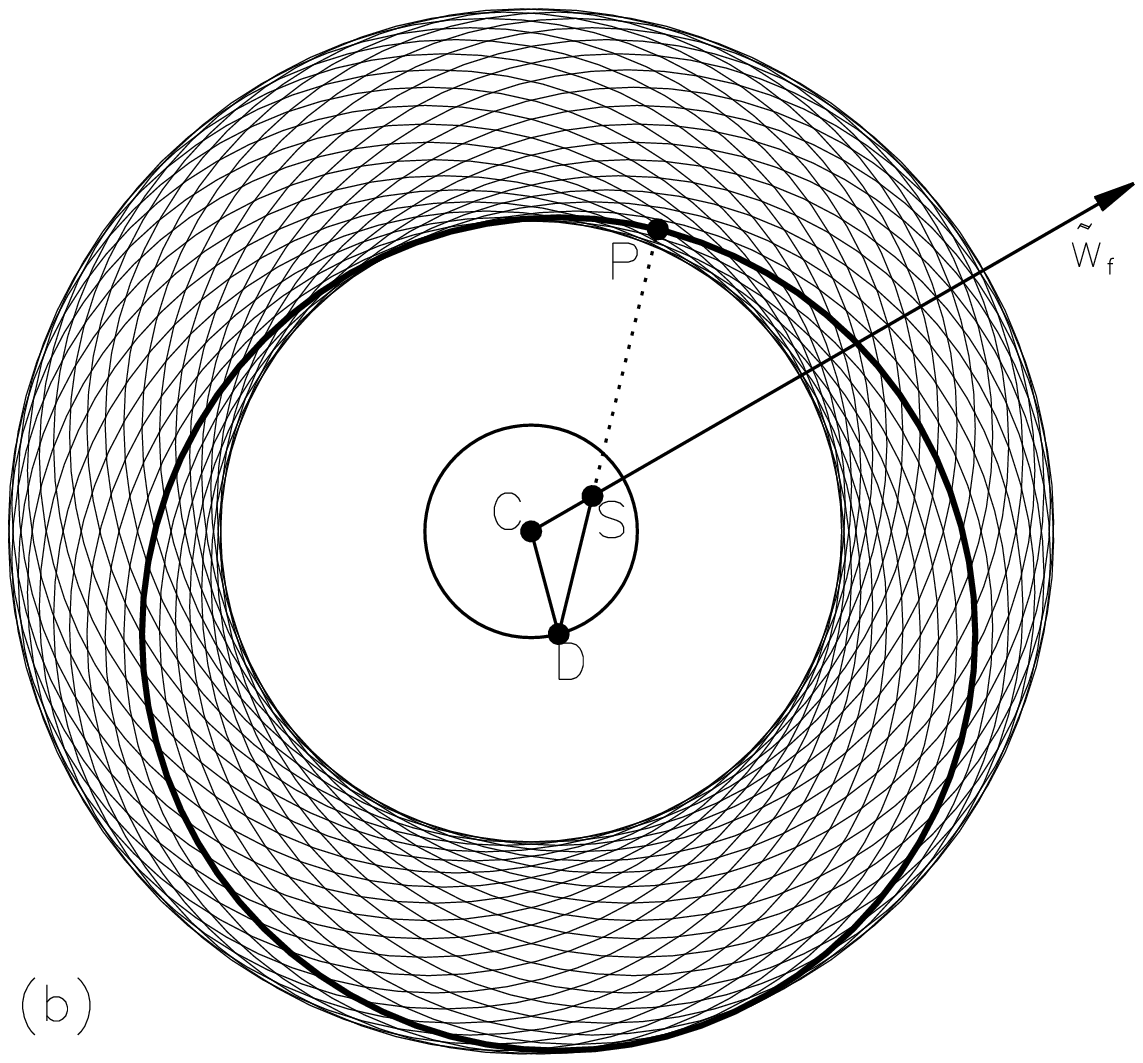}
    \end{tabular}
  \end{center}
  \caption{(\textbf{a}) The osculating (instantaneous) complex eccentricity,
  $z = e \ast \exp{i\tilde{\omega}} = SD$, of the orbit of a particle in a
  system with one or more massive perturbers can be resolved vectorially into
  two components:
  a forced eccentricity, $z_f = e_f \ast \exp{i\tilde{\omega}_f} = SC$, that is
  imposed on the particle's orbit by the perturbers;
  and a proper eccentricity, $z_p = e_p \ast \exp{i\tilde{\omega}_p} = CD$,
  that is the particle's intrinsic eccentricity.
  The secular evolution of its complex eccentricity is to precess
  counterclockwise around the circle in (\textbf{a}), although the forced
  eccentricity may also vary with time.
  Initially, the orbital elements of a family of collisional fragments created
  in the break-up of one large asteroid are the same as those of the original
  asteroid.
  After a few precession timescales, their complex eccentricities are
  evenly distributed around the circle in (\textbf{a});
  this is because each fragment has a slightly different precession
  timescale.
  Thus, these fragments have the same $a$, $e_f$, $\tilde{\omega}_f$,
  $e_p$, but random $\tilde{\omega}_p$, and so their orbits have different
  eccentricities and orientations.
  The spatial distribution of these collisional fragments is shown in
  (\textbf{b}).
  Their elliptical orbits are represented here by circles of radius $a$ with
  centers that are offset from the star, $S$, by $ae$ in a direction opposite
  to the pericenter direction, $\tilde{\omega}$.
  A heavy line is used to highlight the orbit of a fragment with a pericenter
  $P$, and displaced circle center $D$, where $DP = a$ and $SD = ae$;
  the triangle $SCD$, where $SC = ae_f$ and $CD = ae_p$, corresponds
  to a similar one in (\textbf{a}).
  Since the distribution of $\tilde{\omega}_p$ is random, it follows that
  the points $D$ for all the fragments are distributed on a circle of
  radius $ae_p$ and center $C$.
  Thus, the fragments form a uniform torus of inner radius $a(1 - e_p)$, and
  outer radius $a(1 + e_p)$, centered on a point $C$ displaced from the star
  by a distance $ae_f$ in a direction away from the forced pericenter,
  $\tilde{\omega}_f$ (\cite{dnbh85}). }
  \label{fig2}
\end{figure}

\paragraph{Large ($\beta < 0.1$) Fragments}
\label{ssssec-lpfam}
The orbital elements of the largest fragments, those with $\beta<0.1$,
created in the break-up of the primordial body are initially very
close to those of the primordial body;
they do not have identical orbits due to the velocity dispersion imparted
to the fragments in the collision.
The forced elements imposed on the orbits of all of these collisional
fragments are the same as those imposed on the primordial body.
The secular evolution of their osculating complex eccentricities
(eq.~[\ref{eq:z}]) and complex inclinations (eq.~[\ref{eq:y}]), is to precess
about the forced elements (which are also varying with time), but at
slightly different rates (due to their slightly different semimajor axes).
A similar argument applies for all particles created by the collisional
break-up of these fragments.
Thus, after a few precession timescales, the complex eccentricities
and complex inclinations of the collisional fragments of this family
lie evenly distributed around circles that are centered on $z_f(a,t)$
and $y_f(a,t)$, and that have radii of $e_p$ and $I_p$ (e.g., their
complex eccentricities lie on the circle shown in Fig.~\ref{fig2}a),
while their semimajor axes are all still close to $a$.
This is seen to be the case in the asteroid belt:
there are families of asteroids that have similar $a$, $e_p$, and $I_p$,
that are the collisional fragments resulting from the break-up
of a much larger asteroid (\cite{hira18}).

The distribution of the complex eccentricities, $n(z)$, of these particles,
has a distribution of pericenters that is biased towards the orientation
in the disk that is defined by $\tilde{\omega}_f$.
The consequence of this biased orbital element distribution on the
spatial distribution of this family material is best described with the help of
Fig.~\ref{fig2}b.
This shows a face-on view (i.e., perpendicular to the plane of
symmetry), of the family material in orbit around a star $S$.
The resulting disk is made up of particles on orbits that have the same
$a$, $e_f$, $\tilde{\omega}_f$, and $e_p$, but random $\tilde{\omega}_p$.
The contribution of each particle to the spatial distribution of material
in the disk can be described by the elliptical ring of material coincident
with the particle's orbit (see \S \ref{sssec-grav}).
In Fig.~\ref{fig2}b, these elliptical rings have been represented by uniform
circles of radius $a$, with centers that are offset by $ae$ in a direction
opposite to the pericenter direction, $\tilde{\omega}$
(this is a valid approximation to first order in the particles'
eccentricities);
a heavy line is used to highlight the orbital ring with a pericenter located at
$P$, and a displaced circle center located at $D$, where $DP = a$.
The vector $SD$ can be decomposed into its forced and proper
components;
this is shown by the triangle $SCD$, where $SD = ae$, $SC = ae_f$,
and $CD = ae_p$ (there is a similar triangle in Fig.~\ref{fig2}a).
Given that the distribution of $\tilde{\omega}_p$ is random, it follows that
the distribution of the rings' centers, $D$, for the family disk are
distributed on a circle of radius $ae_p$ and center $C$.
Thus, the family forms a uniform torus of inner radius $a(1-e_p)$ and
outer radius $a(1+e_p)$ centered on a point $C$ displaced from
the star $S$ by a distance $ae_f$ in a direction away from the forced
pericenter, $\tilde{\omega}_f$ (\cite{dnbh85}; \cite{dghw98}).

The distribution of the complex inclinations, $n(y)$, of these particles,
is also the distribution of their orbital planes.
Changing the reference plane relative to which the particles' orbital
inclinations are defined to that described by $y_f$, shows that the
secular complex inclination distribution of this family material
leads to a disk that is symmetrical about the $y_f$ plane;
the flaring of this disk is described by $I_p$.

\paragraph{Small ($\beta > 0.1$) Fragments}
\label{ssssec-spfam}
Since the small particles in this family originate from the larger
particles, the orbital elements of the parents of the small particles
have the same $a$, $e_p$ and $I_p$ (they also have the same $z_f$
and $y_f$), but random $\tilde{\omega}_p$ and $\Omega_p$.
Consider the $\beta$ critical particles that are produced at the
same time from the population of family particles that have the
same  $\tilde{\omega}_p$ and $\Omega_p$ at that time;
this parent population is spread out along the orbit defined by the
elements $a$, $e$, and $\tilde{\omega}$, which could be
one of the rings shown in Fig.~\ref{fig2}b.
The average pericenter orientation of these $\beta$ critical particles,
$\langle \tilde{\omega}^{'} \rangle$, is the same as that of the orbit of
the larger particles, $\tilde{\omega}$ (obvious because of the
symmetry of eq.~[\ref{eq:wprime}] with respect to $f$);
their pericenter locations are also the same (eq.~[\ref{eq:rp}]).
Thus, the ring shown in Fig.~\ref{fig2}b defines the inner edge of
the disk of $\beta$ critical particles created in the break-up of
large particles on this ring;
i.e., their disk is offset by an amount $ae$ in the $\tilde{\omega}$
direction.
Consequently, the inner edge of the disk of $\beta$ critical particles
created in the break-up of all large particles in this family
is offset by an amount $ae_f$ in the $\tilde{\omega}_f$ direction.
This disk has the same plane of symmetry as the families' large
particles, since all particle categories from the same family have
the same distribution of orbital planes, $n(y)$.
Similar arguments apply for the families' $\beta$ meteoroids.

\subsubsection{Offset and Warp of Whole Disk}
\label{sssec-owwd}
The disks of material from all of the families that have the same semimajor
axis, or equivalently that are at the same distance from the star, have the
same offset inner edge, and the same plane of symmetry.
This is because their large particles have the same forced elements
imposed on their orbits.
The complex eccentricities and complex inclinations of the large particles
of all of these families lie evenly distributed around circles with the same
centers, $z_f$ and $y_f$, but with a distribution of radii, $n(e_p)$ and
$n(I_p)$, that are the distributions of the proper elements of these
families (defining the width and flaring of the torus consisting of these
families' material).
The whole disk is made up of families with a range of semimajor axes;
the families that are at different semimajor axes can have different
forced elements imposed on their orbits, and in view of the proper
element distributions in the asteroid belt and in the Kuiper belt,
they can also have different distributions of proper elements.

If the forced eccentricity imposed on the disk is non-zero, which it is if
there is at least one perturber in the system that is on a non-circular
orbit (eqs.~[\ref{eq:z}] and [\ref{eq:zf1}]), then the disk's center of
symmetry is offset from the star. 
If the forced inclination imposed on the disk is different for families
at different semimajor axes, which it is if there are two or more
perturbers in the system that are moving on orbits that are not
co-planar (eqs.~[\ref{eq:y}] and [\ref{eq:yf1}]), then the disk's plane
of symmetry varies with distance from the star;
i.e., the disk is warped.

\subsubsection{Physical Understanding of Offset and Warp}
\label{sssec-puow}
There is also a physical explanation for the secular perturbation
asymmetries.
The secular perturbations of a massive body are equivalent to the
gravitational perturbations of the elliptical ring that contains the
mass of the perturber spread out along its orbit, the line density of
which varies inversely with the speed of the perturber in its orbit
(Gauss' averaging method, \cite{bc61}; \cite{md99}).
The ring's elliptical shape, as well as its higher line density at the
perturber's apocenter, mean that the center of mass of the "star-ring"
system is shifted from the star towards the perturber's apocenter.
The focus of the orbits of particles in such a system is offset
from the star;
i.e., the center of symmetry of a disk in this system is offset from
the star.
The gravitational perturbations of the ring also point to the plane
coincident with the perturber's orbital plane.
In systems with two or more perturbers, the system's plane of
symmetry (that in which the perturbing forces out of this plane
cancel) varies with distance from the star;
i.e., a disk in such a system is warped.

\subsection{Observational Evidence of Offset and Warp in the Zodiacal Cloud}
\label{ssec-owzc}

\begin{figure}
\epsscale{0.45}
  \begin{center}
    \begin{tabular}{rlrl}
      (\textbf{a}) & \hspace{-0.4in} \plotone{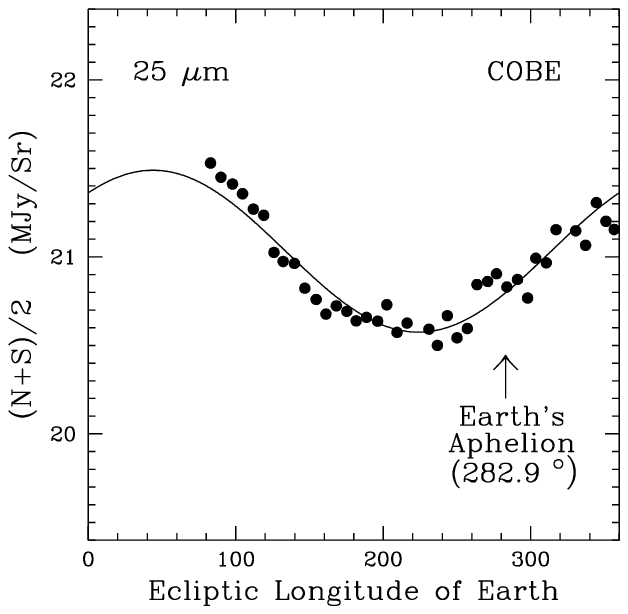} &
      \hspace{0.3in} (\textbf{b}) & \hspace{-0.45in} \plotone{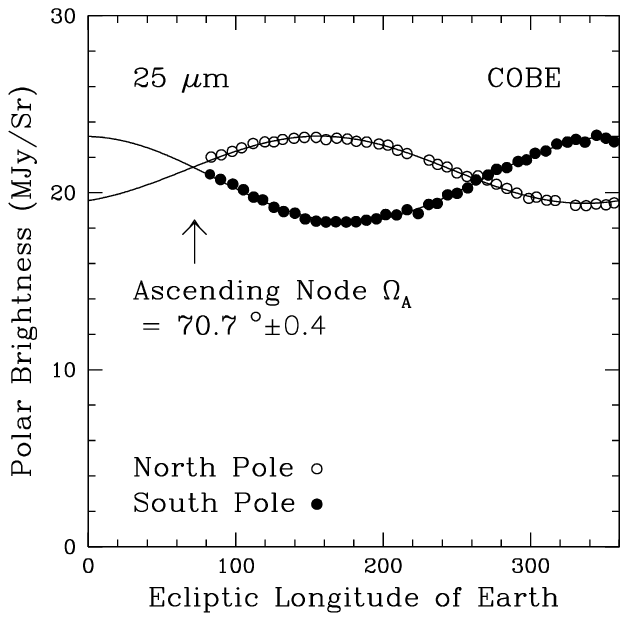}
    \end{tabular}
  \end{center}
  \caption{COBE observations in the 25 $\mu$m waveband of the variation of
  brightnesses at the North, $N$, and South, $S$, ecliptic poles with
  ecliptic longitude of the Earth, $\lambda_\oplus$ (\cite{dghk99}).
  (\textbf{a}) $N + S$ is at a minimum at $\lambda_\oplus
  = 224 \pm 3^{\circ}$.
  The displacement of this minimum from the Earth's aphelion at
  $\lambda_\oplus = 282.9^{\circ}$ implies that the center of symmetry of
  the zodiacal cloud is displaced from the Sun.
  (\textbf{b}) $N$ and $S$ are equal when the Earth is at either
  the ascending or the descending node of the plane of symmetry of the
  cloud at 1 AU, giving an ascending node of $70.7 \pm 0.4^\circ$.
  This plane of symmetry is different from the plane of symmetry of the cloud
  at distances $>1$ AU from the Sun, implying that the cloud is
  warped. }
  \label{fig3}
\end{figure}

\placefigure{fig3}

Mid-IR geocentric satellite observations (such as the IRAS, COBE, and
ISO observations) are dominated by the thermal emission of the
zodiacal cloud's P-R drag affected particles in all directions except
that of the galactic plane (\cite{lg90}).
Such observations contain detailed information about the spatial
structure of the zodiacal cloud, especially since their observing
geometry changes throughout the year as the Earth moves around
its orbit.
Since there are 9 massive perturbers in the solar system,
the resulting secular perturbation asymmetries should be apparent
in the IRAS, COBE, and ISO data-sets (see Appendix \ref{app-pr}
for a discussion of why the distribution of a disk's P-R drag affected
particles should also contain the signatures of the system's secular
perturbations).

Fig.~\ref{fig3}a shows COBE observations of the sum of the brightnesses
in the 25 $\mu$m waveband at the north and south ecliptic poles, $(N+S)/2$
(\cite{dghk99}), where there is no contamination from the galactic plane.
If the zodiacal cloud was rotationally symmetric with the Sun at the
center, then the cross-sectional area density of particles in the near Earth
region would vary according to $\sigma(r,\theta,\phi) \propto r^{-\nu}f(\phi)$,
where $\nu$ is a constant.
Because the Earth's orbit is eccentric, geocentric observations
sample the zodiacal cloud at different radial distances from the Sun.
Thus, the minimum of the $(N+S)/2$ observation is expected to occur
either at the Earth's aphelion, $\lambda_\oplus = 282.9^{\circ}$,
or perihelion, $\lambda_\oplus = 102.9^{\circ}$, depending on whether
$\nu > 1$ or $\nu < 1$, which is determined by the collisional evolution
of particles in the near-Earth region (e.g., \cite{lg90} discuss the
observational evidence and conclude that $\nu \approx 1.3$ as found
by the Helios zodiacal light experiment).
However, the minimum in the 25 $\mu$m waveband observations occurs at
$\lambda_\oplus = 224^{\circ}$, and a similar result is found in the
12 $\mu$m waveband.
This is expected only if the Sun is not at the center of symmetry of the
zodiacal cloud.
Parametric models of the zodiacal cloud have also shown the need for an
offset to explain the observations (e.g., \cite{kwfr98}).

Fig.~\ref{fig3}b shows the variation of the brightnesses of the ecliptic
poles with ecliptic longitude of the Earth (\cite{dghk99}).
The north and south polar brightnesses are equal when the Earth is at
either the ascending or descending node of the local (at 1 AU)
plane of symmetry of the cloud, giving an ascending node of
$\Omega_{asc} = 70.7 \pm 0.4^\circ$.
However, COBE observations of the latitudes of the peak brightnesses
of the zodiacal cloud measured in the directions leading and trailing
the Earth's orbital motion give $\Omega_{asc} = 58.4^\circ$
(\cite{djxg96}).
Since such observations sample the cloud external to 1 AU,
this implies that the plane of symmetry of the zodiacal cloud varies with
heliocentric distance, i.e., that the zodiacal cloud is warped.

To observe the zodiacal cloud's offset and warp asymmetries, an
observer outside the solar system, would, at the very least, need an
observational resolution greater than the magnitude of these
asymmetries;
e.g., to observe the offset asymmetry, the observer needs a
resolution of $>(ae_f/a_\oplus)/R_\odot$ arcseconds, where the distance
from the observer to the Sun, $R_\odot$, is measured in pc.
An offset would be more observable in a disk such as HR 4796 due to its
central cavity, since this causes a brightness asymmetry in the
emission from the inner edge of the disk.

\section{The Secularly Perturbed HR 4796 Disk Model}
\label{sec-hr4796}
This section describes our model of the HR 4796 disk that accounts for the
brightness distribution seen in the IHW18 waveband observation of T99
(Fig.~\ref{fig8}a).
There are three components of the observation that had to be included
in the model (\S \ref{sec-ocd}):
the disk's structure, $\sigma(r,\theta,\phi)$, defined in \S \ref{ssec-dds};
the combination of the optical properties and the size distribution of the
disk's particles given by $P(\lambda,r)$, defined in \S \ref{ssec-los};
and the disk's orientation.
While the modeling techniques used here are new to the study of
circumstellar disks, they have already been widely used to study the
observed structure of the zodiacal cloud (see, e.g.,
\cite{djxg94}; \cite{gdjx97}), and so can be used with a certain degree
of confidence.

\subsection{Model of Offset Disk Structure, $\sigma(r,\theta,\phi)$}
\label{ssec-mds}
The disk's structure was modeled as that of a secularly perturbed dynamic
disk.
A combination of the theory of \S\S \ref{sec-ddc} and \ref{sec-spd}, and
inferences from the observation (Fig.~\ref{fig8}a), was used to parameterize
the distribution of the orbital elements of the disk's large particles
(\S \ref{sssec-doe}).
This was used to create parameterized models of the spatial distribution
of these large particles (\S \ref{sssec-simul}), that could then be
compared with the observed spatial distribution.
Since the observed spatial distribution is that of the "emitting"
particles, the underlying assumption is that these emitting particles
either are the disk's large particles, or that their spatial distribution is
the same as that of the disk's large particles (whether they are large or
not).
The implications of this assumption are discussed in the interpretation
of the model (\S \ref{sec-disc}).

\subsubsection{Distribution of Orbital Elements,
$\sigma(a,e,I,\Omega,\tilde{\omega})$}
\label{sssec-doe}
The quintessentially secular part of the distribution of the orbital
elements of the large particles in a secularly perturbed disk is the
distributions of their complex eccentricities, $n(z)$, and complex
inclinations, $n(y)$.
For a particle with an orbit of a given semimajor axis, $a$, its complex
eccentricity, $z$, and complex inclination, $y$, are the addition of
forced elements, $z_f(a)$ and $y_f(a)$, to proper elements
that have $\tilde{\omega}_p$ and $\Omega_p$ chosen at random,
while $e_p$ and $I_p$ are chosen from the distributions $n(e_p)$
and $n(I_p)$.

Since there is insufficient information available to determine the
variation of the forced and proper elements with semimajor axis
in this disk, they were assumed to be constant across the disk.
The forced elements were left as model variables:
the forced eccentricity, $e_f$, defines the magnitude of the offset
asymmetry in the disk model;
the forced pericenter orientation, $\tilde{\omega}_f$, defines the
orientation of this asymmetry;
and the forced inclination, $y_f$, defines the plane of symmetry of
the disk model.
When creating a disk model, both $\tilde{\omega}_f$ and $y_f$ were
set to zero;
these were incorporated later into the description of the disk's
orientation to our line of sight (see \S \ref{ssec-do}).
The distributions of the proper eccentricities, $n(e_p)$, and proper
inclinations, $n(I_p)$, of particles in the disk model were taken to be
like those of the main-belt asteroids with absolute magnitudes
$H < 11$ (\cite{bowe96}).
These large asteroids constitute a bias-free set (\cite{bowe96}) and have
mean proper eccentricities and proper inclinations of
$\langle e_p \rangle = 0.130$ and $\langle I_p \rangle = 10.2^\circ$.
Not enough Kuiper belt objects have been discovered yet to infer a
bias-free distribution for their orbital elements.

The distribution of the semimajor axes, $n(a)$,  of particles in the
disk defines its radial distribution.
There is no way of guessing this distribution from theoretical
considerations, since it depends on the outcome of the system's
planetary formation process, which varies from system to system
(compare the distribution of the solar system's planets,
and its disk material, with those found in exosolar systems, e.g.,
\cite{bp93}; \cite{mb98}).
Thus, it had to be deduced purely observationally.
Fig.~\ref{fig8}a shows that the disk has an inner edge, inside of
which there is a negligible amount of dust;
this was modeled as a sharp cut-off in the distribution of
semimajor axes at $a_{min}$, a model variable.
The observation also shows that the disk has an outer edge at
$\sim 130$ AU;
this was modeled as a sharp cut-off in the distribution of
semimajor axes at $a_{max} = 130$ AU (this is a non-critical
parameter, since particles near the outer edge of the disk contribute
little to the observation, see \S \ref{sssec-pg}).
The distribution between $a_{min}$ and 130 AU was taken as
$n(a) \propto a^\gamma$, where $n(a)da$ is the number of particles
on orbits with semimajor axes in the range $a \pm da/2$, and
$\gamma$ is a model variable.
To get an idea of the radial distribution resulting from this semimajor
axis distribution, consider that if the particles had zero eccentricity,
this distribution would result in a volume density (number of particles
per unit volume) distribution that is $\propto r^{\gamma-2}$
(since the number of particles in a spherical shell of width $dr$, the
volume of which is $\propto r^2dr$, would contain a number of particles
that is $\propto r^\gamma dr$).

\subsubsection{Conversion to Spatial Distribution, $\sigma(r,\theta,\phi)$}
\label{sssec-simul}
Disk models were created from the orbital element distribution of
\S \ref{sssec-doe} using the "SIMUL" program;
SIMUL was developed by the solar system dynamics group at the
University of Florida (\cite{dgdg92}).
A disk model is a large three-dimensional array,
$\sigma(r,\theta,\phi)$, that describes the spatial distribution of the
cross-sectional area of material in the disk model per unit volume binned
in: $r$, the radial distance from the star;
$\theta$, the longitude relative to an arbitrary direction
(set here as the forced pericenter direction, $\tilde{\omega}_f$);
and $\phi$, the latitude relative to an arbitrary plane
(set here as the forced inclination, $y_f$, or symmetry, plane).
SIMUL creates a disk model by taking the total cross-sectional
area of material in the disk (specified by the model variable
$\sigma_{tot}$), and dividing it equally among a large number of
orbits (5 million in this case), the elements of each of which are
chosen randomly from the specified distribution (\S \ref{sssec-doe}).
The disk model is populated by considering the contribution of each
orbit to the cross-sectional area density in each of the cells it crosses.

\placefigure{fig4}

The spatial distribution of material in one of our models of the HR 4796
disk can be described by the three variables $a_{min}$, $\gamma$,
and $e_f$;
$\sigma_{tot}$ simply scales the amount of material in the model,
and $\tilde{\omega}_f$ and $y_f$ describe the orientation of the
disk to our line of sight.
Fig.~\ref{fig4} is a plot of the surface density of material in a disk
model with $a_{min} = 62$ AU, $e_f = 0.02$, and $\gamma = -2$
(this is our final model of \S \ref{ssec-mpr}).
This illustrates how the specified distribution of orbital elements
affects the spatial distribution of material in the disk model:
the sharp cut-off in semimajor axes at $a_{min}$ determines the radial
location of the inner hole, which has a sloping cut-off in $r$ due
to the particles' eccentricities;
as predicted in \S \ref{sssec-opsfam}, particles at the inner edge of
the disk in the forced pericenter direction are closer to the star than
those in the forced apocenter direction by $\sim 2a_{min}e_f$;
the distribution of semimajor axes has produced a surface density
distribution that is $\propto r^{\gamma-1}$, but only exterior to 70 AU.

\subsection{Model of $P(\lambda,r)$}
\label{ssec-mpp}

\subsubsection{Optical Properties of Disk Particles}
\label{sssec-pop}
The optical properties of the disk particles were found assuming the
particles to be made of astronomical silicate (\cite{dl84}; \cite{ld93}),
a common component of interplanetary dust found in both the zodiacal cloud
(\cite{lg90}) and exosolar systems (e.g., \cite{tk91}; \cite{ftk93};
\cite{sglr99});
such an assumption can be tested at a later date using spectroscopy to look
for silicate features in the HR 4796 disk emission.
Furthermore, the particles were assumed to be solid, spherical, and have a
density of $\rho = 2500\textrm{kg}/\textrm{m}^3$.
Their optical properties were calculated using Mie theory, assuming that
HR 4796A has a luminosity and temperature of $L_\star = 21L_\odot$,
$T_\star = 9500^{\circ}$K (\cite{jmwt98}), and using for its spectrum,
that of the A0V star Vega (\cite{cohe99}).
In all calculations, the mass of HR 4796A was assumed to be
$M_\star = 2.5M_\odot$ (\cite{jfht98}).

\placefigure{fig5}

The properties of particles of different sizes, and at different distances
from HR 4796A, are shown in Fig.~\ref{fig5}.
The temperatures of the particles are plotted in Figs.~\ref{fig5}a
and \ref{fig5}b.
The form of Fig.~\ref{fig5}a can be understood by consideration of
equation (\ref{eq:tdr}), and the wavelengths at which the star and a
particle, if it was a black body, emit most of their energy:
$\lambda_\star \approx 2898/T_\star$ $\mu$m, and
$\lambda_{bb} \approx 2898/T_{bb} =
10\sqrt{r/a_\oplus}(L_\odot/L_\star)^{0.25}$ $\mu$m.
As a crude approximation, a particle with diameter $D$ has
$Q_{abs} \approx 1$ for $\lambda \ll \pi D$ and $Q_{abs} \rightarrow 0$ for
$\lambda \gg \pi D$.
Thus, in terms of their thermal properties, disk particles can be divided
into four categories:
the largest particles, $D \gg \lambda_{bb}/\pi$, are efficient absorbers and
emitters at all relevant wavelengths and so achieve nearly black body
temperatures, $T_{bb}$;
particles with $D \ll \lambda_{bb}/\pi$, are inefficient emitters at
their black body temperature, and so need temperatures higher
than $T_{bb}$ to re-radiate all of the incident energy;
the smallest particles, $D \ll \lambda_\star/\pi$, are also
inefficient absorbers at the stellar temperature, and so do not need as high
temperatures as slightly larger particles to re-radiate the absorbed energy;
and particles with $D \approx 20$ $\mu$m have temperatures below that of
a black body --- this is because these particles are super-efficient
emitters at their black body temperatures (due to silicate resonances,
$Q_{abs}$ can go up as high as 2), and so need lower temperatures to
re-radiate the incident energy.
The form of Fig.~\ref{fig5}b can be understood in the same way.
The fall-off of a large (e.g., $D = 1000$ $\mu$m) particle's temperature with
distance from HR 4796A is like that of a black body, i.e.,
$T \propto 1/\sqrt{r}$, while the fall-off for smaller particles is not that
steep because these particles emit less efficiently the further they
are from the star (due to their lower temperatures, and consequently
higher $\lambda_{bb}$);
e.g., the fall-off for $D = 2.5$ $\mu$m particles is $T \propto 1/r^{0.34}$,
which is close to the $1/r^{1/3}$ fall-off expected for particles with an
emission efficiency that decreases $\propto 1/\lambda^2$ (e.g., \cite{bp93}).

More important observationally is the variation of the particles'
$Q_{abs}(\lambda,D)B_\nu[\lambda,T(D,r)]$, since this determines
the contribution of a particle's thermal emission to the flux density
received at the Earth (eqs.~[\ref{eq:flux}],[\ref{eq:fnu}] and [\ref{eq:p}]).
This is plotted for $\lambda = 18.2$ $\mu$m in Figs.~\ref{fig5}c and
\ref{fig5}d, and for $\lambda = 10.8$ $\mu$m in Figs.~\ref{fig5}e
and \ref{fig5}f.
The form of Figs.~\ref{fig5}c and \ref{fig5}e can be explained in the
same way that Fig.~\ref{fig5}a was explained:
all three figures have similar forms, which is to be expected since
the particles' temperature also appears in $B_\nu$;
Figs.~\ref{fig5}c and \ref{fig5}e are, however, attenuated for
$D \ll \lambda/\pi$, since these particles are inefficient emitters at that
wavelength.
The fall-off with distance of the different particles shown in
Figs.~\ref{fig5}d and \ref{fig5}f is due solely to their different temperature
fall-offs (Fig.~\ref{fig5}b);
e.g., for $\lambda = 18.2$ $\mu$m, the approximate fall-off for
$D = 1000$ $\mu$m particles is $\propto 1/r^{5.4}$, while that for
$D = 2.5$ $\mu$m particles is $\propto 1/r^{2.6}$.

\subsubsection{Cross-sectional Area Distribution}
\label{sssec-dtyp}
The definition of $P(\lambda,r)$ (eqs.~[\ref{eq:p}] and [\ref{eq:p2}])
shows that it is the convolution of $Q_{abs}B_\nu$
(Fig.~\ref{fig5}c - \ref{fig5}f) with the cross-sectional area distribution,
$\bar{\sigma}(D,r)$.
Since theoretical arguments cannot supply an accurate size distribution
(\S \ref{sssec-pcsig}), we use the assumption of equation (\ref{eq:pdtyp}),
which is that $P(\lambda,r)$ is equal to the $Q_{abs}B_\nu$ of particles
in the disk with characteristic size, $D_{typ}$.
We further assume that this $D_{typ}$ is constant across the disk
in the IHW18 waveband;
i.e., the brightness of a disk model observation in the IHW18 waveband
is calculated using $P(\lambda,r)$ from the line on Fig.~\ref{fig5}c
corresponding to $D_{typ}$, where $D_{typ}$ is a model variable.
A qualitative understanding of $D_{typ}$ comes from Fig.~\ref{fig5}c:
since $Q_{abs}B_\nu$ is fairly flat for particles larger than $\sim 8$
$\mu$m, an IHW18 waveband observation is dominated by those
particles with the most cross-sectional area, unless there are a
significant amount of particles smaller than 8 $\mu$m;
Fig.~\ref{fig5}c also shows that the observation is unlikely to be dominated
by particles smaller than $O$(0.01 $\mu$m), unless their contribution to
the disk's cross-sectional area is much higher than that of
larger particles.

In addition to the IHW18 waveband, T99 observed HR 4796 in the
N ($\lambda = 10.2$ $\mu$m) waveband.
The N band observations also show the double-lobed feature,
but the inaccuracy of the subtraction of the image of
HR 4796A from the observations means that they cannot be used to
constrain the disk's structure.
The N band observations can, however, be used to constrain the
disk's brightness in this waveband;
the brightness of a disk in different wavebands differs only in
the factor $P(\lambda,r)$ (eq.~[\ref{eq:fnu}]), and so the N band
observation can be used to obtain information about the disk's
size distribution.
Since Figs.~\ref{fig5}c and \ref{fig5}e have similar forms, the two
observations should be dominated by the emission of similarly sized
particles;
i.e., they should have the same $D_{typ}$, and show similar
structures.
The same characteristic particle size, $D_{typ}$, was used to
calculate the brightness of a disk model in both the IHW18 and N
wavebands, using $P(\lambda,r)$ from the appropriate
lines on the plots of Figs.~\ref{fig5}d and \ref{fig5}f.
There is just one $D_{typ}$ that simultaneously matches the disk's
observed brightnesses in both wavebands.

\placefigure{fig6}

We made an initial estimate for $D_{typ}$ based on the flux densities
observed in the two wavebands:
the flux densities of the disk in the IHW18 and N wavebands are
857 and 40 mJy, respectively (T99);
i.e., the observed flux density ratio (N/IHW18) is $O(0.05)$.
The expected flux density ratio of the two wavebands (the ratio of
$P(\lambda,r)$ for different $D_{typ}$) is plotted in Fig.~\ref{fig6}a.
Assuming the emission to arise mostly from particles near the inner
edge of the disk, $r =$ 60-80 AU, Fig.~\ref{fig6}a shows that the
observed emission can be fairly well-constrained to come from
particles with $D_{typ} = 2-3$ $\mu$m.

\subsubsection{Pericenter Glow}
\label{sssec-pg}

\placefigure{fig7}

Fig.~\ref{fig7}a shows a contour plot of an unsmoothed IHW18 waveband
observation of the disk model of Fig.~\ref{fig4} viewed face-on
(i.e., perpendicular to the disk's plane of symmetry, $y_f$).
This shows the observational consequence of the offset center of symmetry
of the disk model.
Because the particles at the inner edge of the disk (those that contribute most
to the disk's brightness) are closer to the star in the forced pericenter
direction, $\tilde{\omega}_f$, than those in the forced apocenter direction
(Fig.~\ref{fig4}), they are hotter and so contribute more to the disk's thermal
emission than those at the forced apocenter.
This is the "pericenter glow" phenomenon, which leads to the
horseshoe-shaped highest contour line (the filled-in 1.02 mJy/pixel line),
which is pointed in the $\tilde{\omega}_f$ direction.
This asymmetry is a consequence of $a_{min}$ and $e_f$ only,
and its magnitude is determined by $e_f$ only.
In particular, if there is a gradient of $e_f$ across the disk, then it is
$e_f$ at the inner edge of the disk that controls the magnitude of the
asymmetry.
The outermost contour plotted on Fig.~\ref{fig7}a, which is an
offset circle with a radius of 95 AU, is that corresponding to 0.17 mJy/pixel.
Thus, there is little emission from the outer edge of the disk,
justifying the arbitrary use of $a_{max} = 130$ AU in the modeling.

\subsection{Disk Model Orientation}
\label{ssec-do}
The two variables that define the orientation of the HR 4796 disk to our
line of sight are $\tilde{\omega}_f$ and $I_{obs}$;
how they define this orientation is best explained using Fig.~\ref{fig7}a.
Imagine that the disk starts face-on with the forced pericenter
direction pointing to the left.
It is then rotated clockwise by $\tilde{\omega}_f$ (this is shown in
Fig.~\ref{fig7}a, where $\tilde{\omega}_f = 26^{\circ}$),
and then tilted by $90^{\circ} - I_{obs}$ about the dotted line on
Fig.~\ref{fig7}a.
The direction of this tilt, whether the top or bottom of the disk ends
up closer to the observer, is not constrained in the modeling,
since no account was made for either the extinction of the disk's
emission by the disk itself, or for the disk's scattered light (e.g., an
observer would see forward-scattered starlight from the closest
part of the disk and back-scattered starlight from the farthest part,
a phenomenon that could produce an apparent asymmetry in a
symmetric disk, \cite{kj95}).
If the resulting inclination of the disk's symmetry plane to our
line of sight, $I_{obs}$, is small, then the resulting nearly edge-on
observation shows two lobes, one either side of the star.
Since the hotter, brighter, pericenter glow material is predominantly
in one of the lobes (unless $\tilde{\omega}_f = 90^\circ$),
the lobes have asymmetric brightnesses.

\subsection{Modeling Process and Results}
\label{ssec-mpr}

\placefigure{fig8}

Pseudo-observations of disk models in the IHW18 and N wavebands were
produced that mimicked the real OSCIR (the University of Florida mid-IR
imager) observations in both pixel size,
\hbox{1 pixel $= 0\farcs0616 = 4.133$ AU} at 67 pc (\cite{jmwt98}),
and smoothing, using the observed PSFs (which are asymmetric and slightly
fatter than diffraction limited, T99), and including the post-observational
gaussian smoothing of FWHM = 3 pixels.
The model variables: $a_{min}$, $\gamma$, $I_{obs}$, $e_f$,
$\tilde{\omega}_f$, and $\sigma_{tot}$, were optimized so that the
modeled IHW18 observation correctly predicts the observed IHW18 brightness
distribution;
at the same time, the variable, $D_{typ}$, was optimized so that the modeled
N band observation correctly predicts the observed N band brightness.

The model observations were compared with the real observations using the
following diagnostics:
the lobe brightnesses, $F_{ne,sw}$, and their projected radial offsets from
HR 4796A, $R_{ne,sw}$, that were found by fitting a quintic polynomial surface
to a $10 \times 10$ pixel region around each of the lobes with 0.1 pixel
resolution (the location of HR 4796A in the real observations was also found
in this way);
and line-cuts through the disk both parallel (e.g., Figs.~\ref{fig8}d and
\ref{fig9}), and perpendicular (e.g., Figs.~\ref{fig8}c and \ref{fig8}e)
to the line joining the two lobes in the IHW18 observations.
An understanding of how the different model variables affect the
different diagnostics allowed the modeling process to be decoupled
into solving for:
the disk's symmetrical structure, defined by $I_{obs}$, $a_{min}$, and
$\gamma$;
the particle size $D_{typ}$;
and the disk's asymmetrical structure, defined by $e_f$ and
$\tilde{\omega}_f$.
Throughout the modeling, the amount of material in a model,
$\sigma_{tot}$, was scaled so that the model observation predicted
the correct observed mean brightness of the lobes in the IHW18
waveband, $F_{mean} = (F_{ne}+F_{sw})/2 = 1.40 \pm 0.02$ mJy/pixel;
its final value, $\sigma_{tot} = 2.03\times 10^{24}$ m$^2$, was calculated
once the other variables had been constrained.

\subsubsection{Symmetrical Disk Structure}
\label{sssec-dss}
Since $e_f$ and $\tilde{\omega}_f$ only pertain to the disk's asymmetrical
structure, then for a given $D_{typ}$, the variables pertaining to the
disk's symmetrical structure, $I_{obs}$, $a_{min}$, and $\gamma$, could
be solved using a model with $e_f = 0$;
such a model is axisymmetric, and so the variable $\tilde{\omega}_f$ is
redundant.
The inclination of the disk's plane of symmetry to the line of sight,
$I_{obs} = 13 \pm 1^{\circ}$, was constrained to give the best fit to the
line-cuts perpendicular to the lobes (Figs.~\ref{fig8}c and \ref{fig8}e);
this is in agreement with that found by previous models of HR 4796
disk observations (\cite{krwb98}; \cite{ssbk99}).
Since our model is that of a "fat" disk, a different proper inclination
distribution would lead to a different $I_{obs}$;
e.g., if the disk is actually thinner than modeled here,
$\langle I_p \rangle < 10.2^\circ$, then the inferred $I_{obs}$ is an
underestimate, and vice versa.
The inner edge of the disk, \hbox{$a_{min} = 62 \pm 2$ AU},
was constrained such that the model observation reproduces the
observed mean radial offset of the lobes from HR 4796A,
\hbox{$R_{mean} = (R_{ne}+R_{sw})/2 = 58.1 \pm 1.3$ AU}.
The semimajor axis distribution, $\gamma = -2 \pm 1$, was constrained
to give the best fit to the cut along the line joining the two lobes
(Fig.~\ref{fig8}d).

\subsubsection{Particle Size}
\label{sssec-tps}

\placefigure{fig9}

Since $D_{typ}$ was already estimated to be about 2-3 $\mu$m
(\S \ref{sssec-dtyp}), the modeling of the disk's symmetrical structure
was repeated for $D_{typ} =$ 2, 2.5, and 3 $\mu$m.
Adjusting $D_{typ}$ by such a small amount did not affect the inferred
symmetrical structure parameters.
This was expected, since the $P$(18.2 $\mu$m$,r)$ for each of
these $D_{typ}$ are very similar (Fig.~\ref{fig5}d).
Remembering that the model is always normalized to predict the
observed mean IHW18 lobe brightnesses, the predicted N band lobe
brightnesses are compared for the three values of $D_{typ}$ in
Fig.~\ref{fig9}.
This shows that the particle size can be constrained to be
$D_{typ} = 2.5 \pm 0.5$ $\mu$m;
i.e., the crude method of calculating the particle size of
\S \ref{sssec-dtyp} gives a very good estimate of this size.
This particle size means that the total mass of emitting particles in
the disk model is $\sim 1.4 \times 10^{-3} M_\oplus$, where
$M_\oplus = 3 \times 10^{-6}M_\odot$ is the mass of the Earth.
However, this is not a useful constraint on the disk's mass, since
the disk's mass is expected to be concentrated in its largest particles. 
The best estimate of the mass of the HR 4796 disk, from
submillimeter observations, is that it is between $0.1M_\oplus$ and
$1.0M_\oplus$ (\cite{jgwm95}; \cite{jmwt98}),
which, as expected, is well above the mass of our disk model.

\subsubsection{Asymmetrical Disk Structure}
\label{sssec-das}
The disk's observed asymmetries are defined by:
the lobe brightness asymmetry,
$(F_{ne} - F_{sw})/F_{mean} = 5.1 \pm 3.2\%$,
and the radial offset asymmetry,
$(R_{sw} - R_{ne})/R_{mean} = 6.4 \pm 4.6\%$.
These asymmetries are also apparent in the disk model, and their
magnitudes are determined by both $e_f$ and $\tilde{\omega}_f$.
The lobe brightness asymmetry was used to constrain $e_f$ and
$\tilde{\omega}_f$, and it was found that the $e_f$ necessary to cause
the $5.1 \pm 3.2\%$ asymmetry depends on the geometry of the
observation according to the relation shown in Fig.~\ref{fig7}b.
Thus, for the majority of the geometries, a forced eccentricity of between
0.02 and 0.03 is sufficient to cause the observed brightness asymmetry.
In the context of this modeling, the observed brightness asymmetry
implies a radial offset asymmetry of $\sim 5\%$, which is within
the limits of the observation.
Our final model shown in Figs.~\ref{fig4} and \ref{fig7}-\ref{fig9}
assumes a modest value of $e_f = 0.02$, which corresponds to an
$\tilde{\omega}_f$ orientation of $26^\circ$ (Fig.~\ref{fig7}b).
Whether this rotation puts the pericenter glow material above or below
the horizontal, $\tilde{\omega}_f = 26^\circ$ or $-26^\circ$, is not
constrained here, since it has a minimal effect on the observation:
in the model observation of Fig.~\ref{fig8}, the top of the disk is
brighter than the bottom of the disk by a fraction that would be
undetectable in the observation due to noise and the disk's unknown
residual structure.
The line-cuts of Fig.~\ref{fig8} show how well the model fits all aspects
of the observation --- the vertical structure, the horizontal structure,
and the lobe location and asymmetry.

\subsubsection{Statistical Significance}
\label{sssec-ss}
The standard deviations of the OSCIR lobe observations quoted in this
section were found using model observations that mimicked the noise
present in the OSCIR observations.
The background sky noise in the IHW18 observation was found to be
approximately gaussian with zero mean and a $1\sigma$ noise per pixel
of $0.15$ mJy (T99);
this was included after the PSF smoothing, but before the
post-observational smoothing.
Observations of the model were repeated for 50,000 different noise
fields to obtain the quoted standard deviations.
Since the observed PSF was asymmetric (T99), this introduces
an apparent lobe asymmetry of $-0.8\%$ in an observation of a
symmetric disk (one with $e_f=0$), and so the observed lobe
asymmetry is $5.9 \pm 3.2\%$ from the mean, and its statistical
significance is $1.8\sigma$.
While this is small, it does show that the pericenter glow phenomenon is
observable with current technology:
HR 4796 was observed with the infrared imager OSCIR for only one hour
on Keck II (T99);
in the background-limited regime, the significance level of any asymmetry
increases at a rate $\propto \sqrt{t}$;
thus, one good night on a 10 meter telescope should be enough to get a
definitive observation of the HR 4796 lobe asymmetry.
The real significance of the HR 4796 asymmetry may also be higher
than quoted above, since it also seems to be apparent in other
observations of this disk (\cite{krwb98}; \cite{ssbk99}).
Even if subsequent observations happen to disprove the existence of
the asymmetry, this is still significant, since, as we show in
\S \ref{ssec-ias}, an asymmetry is to be expected if the companion
star, HR 4796B, is on an eccentric orbit.

\section{Interpretation of the HR 4796 Disk Model}
\label{sec-disc}
The interpretation of our HR 4796 disk model is broken down into
sections that cover discussions of:
the dynamics of the disk particles, \S \ref{ssec-ddch};
the lobe asymmetry, \S \ref{ssec-ias};
the emitting particle category, \S \ref{ssec-deps};
the origin of the inner hole, \S \ref{ssec-lsd};
and the residual structure of the observation once the model has been
subtracted, \S \ref{ssec-rs}.

\subsection{The Dynamic HR 4796 Disk}
\label{ssec-ddch}
Our interpretation of the observed structure of the HR 4796 disk starts
with a discussion of the dynamics of the particles in the disk,
and where the emitting particles fit into our understanding of the
dynamic disk (\S \ref{sec-ddc}).

\subsubsection{Radiation Forces, $\beta$}
\label{sssec-rfh}
The radiation forces, defined by $\beta$ (eq.~[\ref{eq:beta}]), acting on
particles in the HR 4796 disk can be found from their optical properties.
Fig.~\ref{fig6}b shows the $\beta$ of particles with the optical properties
assumed in our model (\S \ref{sssec-pop}).
Thus, particles in the disk with $D<8$ $\mu$m are $\beta$ meteoroids
(even the submicron particles have $\beta>0.5$), and a good approximation for
the non-$\beta$ meteoroids is:
\begin{equation}
  \beta \approx 4/D, \label{eq:betah}
\end{equation}
where $D$ is measured in $\mu$m (in agreement with eq.~[\ref{eq:beta2}]).
Fig.~\ref{fig9} shows that the emitting particles have
$D_{typ} \approx 2.5$ $\mu$m; these particles have $\beta = 1.7$.
So, the modeling implies that the emitting particles are $\beta$ meteoroids,
i.e., that they are blown out of the system on hyperbolic orbits.
Since the lifetime of $\beta$ meteoroids is $O$(370 years) for those
created at $\sim 70$ AU, which is much shorter than the age of the
HR 4796 system, any $\beta$ meteoroids that are currently in the disk
cannot be primordial particles, rather they must be continuously
created from a reservoir of larger particles;
i.e., the existence of $\beta$ meteoroids implies the existence of a
dynamically stable population of larger particles.

If the emitting particles are on hyperbolic orbits, we can use the mass of
the disk's emitting particles, $\sim 1.4 \times 10^{-3}M_\oplus$, and the
emitting lifetime of these particles, $\sim 370$ years, to estimate the mass
loss rate of the disk to be $\sim 4 \times 10^{-6}M_\oplus/$year.
If this mass loss rate has been sustained over the age of the system,
the original disk must have been $\sim 40M_\oplus$ more massive than
it is today.
In fact, a more massive disk would have had a higher mass loss rate, since
this rate increases proportionally with the total cross-sectional
area in the disk (i.e., $\propto m_{tot}^{2/3}$).
This means that if the current disk has a mass $1.0M_\oplus$, the original
disk must have had a mass $\sim 7 \times 10^4M_\oplus$;
i.e., the HR 4796 disk may provide evidence for the type of collisional
mass loss that may have happened in the early Kuiper belt (\cite{sc97}). 

\subsubsection{Collisional Processes}
\label{sssec-collh}
The collisional lifetime of the disk's emitting particles can be calculated
directly from the disk model using equations (\ref{eq:tcol3}) and
(\ref{eq:fccdtyp}):
$t_{coll}(D_{typ}) < 10^{4}$ years across most of the disk (55-85 AU),
with a minimum at $\sim 70$ AU of $\sim 4500$ years.
This collisional lifetime is much less than the age of the HR 4796
system.
Thus, the emitting particles cannot be primordial particles (irrespective
of whether they are $\beta$ meteoroids).
The collisional lifetime of the emitting particles can be corroborated
using equation (\ref{eq:tcol6}).
The model's effective optical depth (eq.~[\ref{eq:taueffdefn}]) is its
surface density, plotted in Fig.~\ref{fig4}, multiplied by the cross-sectional
area of a particle in the model, $\sigma = \pi D_{typ}^2/4$.
This peaks at 70 AU, where $\tau_{eff}$(70 AU) = $5 \times 10^{-3}$, and
$t_{per} = O$(370 years), giving a collisional lifetime that has a minimum
of $t_{coll}(D_{typ}) \approx 6000$ years.
The disk's effective optical depth at 70 AU, $\tau_{eff}$(70 AU), can
also be corroborated directly from the observation using equation
(\ref{eq:taueff}).
The observed edge-on, smoothed lobe brightness is $\sim 1.40$ mJy/pixel;
this can be scaled to the unsmoothed face-on brightness by the factor
1.10/1.40 (see Figs.~\ref{fig7}a and \ref{fig8}b), which, since each pixel
subtends $(0.0616*\pi/648000)^2$ sr, gives a brightness of
$F_\nu$(18.2 $\mu$m,70 AU)$/\Omega_{obs} = 12 \times 10^9$ Jy/sr.
For 2.5 $\mu$m particles, Figs.~\ref{fig5}c and \ref{fig5}d give
$P$(18.2 $\mu$m,70 AU) $\approx 2.2 \times 10^{12}$ Jy/sr, thus
confirming that $\tau_{eff}$(70 AU) $\approx 5 \times 10^{-3}$ (this is
also in agreement with $\tau \approx 5 \times 10^{-3}$ found by
\cite{jgwm95}).
In fact, assuming that the disk's total IHW18 flux density, 857 mJy (T99),
comes from particles between $70 \pm 15$ AU, the unsmoothed face-on
brightness of the disk can also be corroborated;
equation (\ref{eq:fnuobs}) with $R_\star = 67$ pc gives
$F_\nu$(18.2 $\mu$m, 70 AU)$/\Omega_{obs} = O(10^{10}$ Jy/sr).

Since particles are only broken up by collisions with particles that
have diameters more than a tenth of their own (eq.~[\ref{eq:dimp}]),
the collisional lifetime of the disk's large particles must be longer
than that of the smaller emitting particles.
Assuming the cross-sectional area distribution to follow equation
(\ref{eq:nd}) with $q = 11/6$ down to particles of size $D_{typ}$, the
collisional lifetime of particles with $D \gg D_{typ}$ can be estimated
as (eq.~[\ref{eq:tcol7}]):
\begin{equation}
  t_{coll} \approx 4 \times 10^7 \sqrt{D},
\end{equation}
where $t_{coll}$ is measured in years, and D in km.
Since particles for which $t_{coll} < t_{sys}$ cannot be primordial,
this implies that particles currently in the HR 4796 disk that are smaller
than 60 m cannot be original, rather they must have been created in the
break-up of larger particles;
i.e., these disk particles form a collisional cascade, through
which the spatial distributions of the smaller particles are related
to that of the larger particles.
The disk's particles that are larger than 60 m are its primordial
particles.

\subsubsection{P-R Drag}
\label{sssec-prh}
The disk's high effective optical depth means that $\beta_{pr} = 132$
at 70 AU (eq.~[\ref{eq:btr}]), which in turn means that there is not
expected to be significant P-R drag evolution for any of the disk's
particles.
Analysis of the P-R drag evolution of disk particles (eq.~[\ref{eq:tpr}])
shows that even the most affected particles, those with $\beta=0.5$, have
a P-R drag lifetime of $t_{pr} = O$(1.6 Myr) at 70 AU, and that these
particles would only make it to $\sim 69.8$ AU before they are broken up by
collisions and blown out of the system by radiation pressure.
Appendix \ref{app-pr}, which gives a discussion of lobe asymmetries
in disks for which P-R drag is important, gives an analysis that shows
that the cumulative effect of P-R drag on the orbits of all of a
disk particle's ancestors is also insignificant.

\subsubsection{The Dynamic HR 4796 Disk}
\label{sssec-ddch}
Since P-R drag is not an important process in the HR 4796 disk's evolution,
there are just three categories of particles in the disk:
large particles, $\beta$ critical particles, and $\beta$ meteoroids.
If we are to believe the modeled emitting particle size (discussed further
in \S \ref{ssec-deps}), the particles that are seen in both the IHW18 and
the N band observations are the disk's $\beta$ meteoroids.
Since we modeled the disk as if the emitting particles are large particles,
this inconsistency needs to be borne in mind in our interpretation of
the model.

\begin{figure}
\epsscale{0.9}
  \begin{center}
    \begin{tabular}{c}
      \plotone{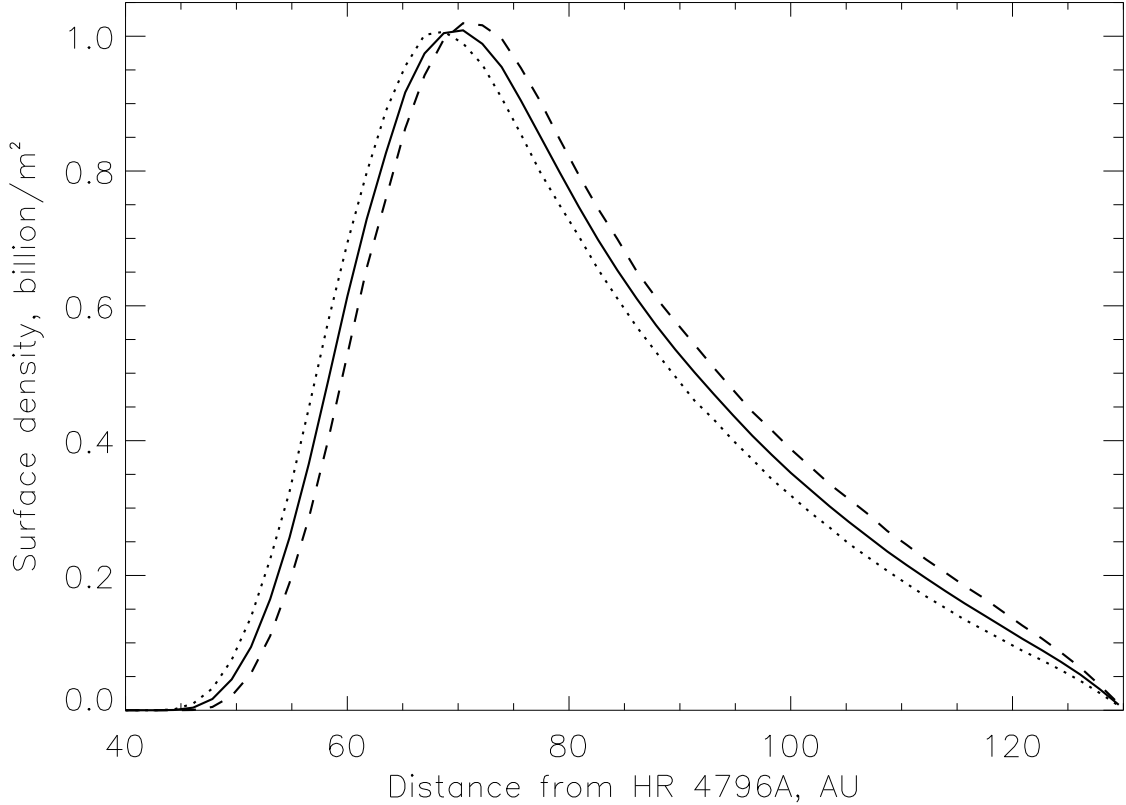}
    \end{tabular}
  \end{center}
  \caption{ The surface number density of the 2.5 $\mu$m dust grains
  in the HR 4796 disk derived from the 18.2 $\mu$m brightness
  distribution (T99).
  The solid curve is the azimuthal average of the surface number density,
  and the dotted and dashed lines indicate the density through the
  disk towards and away from the forced pericenter direction of the model,
  respectively.
  The offset is a result of the forced eccentricity imposed on the disk
  model;
  the inner edge of each side of the disk is offset by
  $\sim a_{min}e_f \approx 1$ AU.
  The disk's surface density peaks at $\sim 1.02 \times 10^9$ m$^{-2}$
  at $\sim 70$ AU.
  Interior to this, the surface density falls to zero by 45 AU;
  the sloping cut-off is due to the eccentricities of the disk model
  particles' orbits.
  Exterior to 70 AU, the surface density falls off $\propto r^{-3}$;
  this is due to the distribution of the disk model particles' semimajor
  axes, $n(a) \propto a^{-2}$.}
  \label{fig4}
\end{figure}

\begin{figure}
  \begin{center}
    \begin{tabular}{cccc}
      & \epsscale{0.52} \plotone{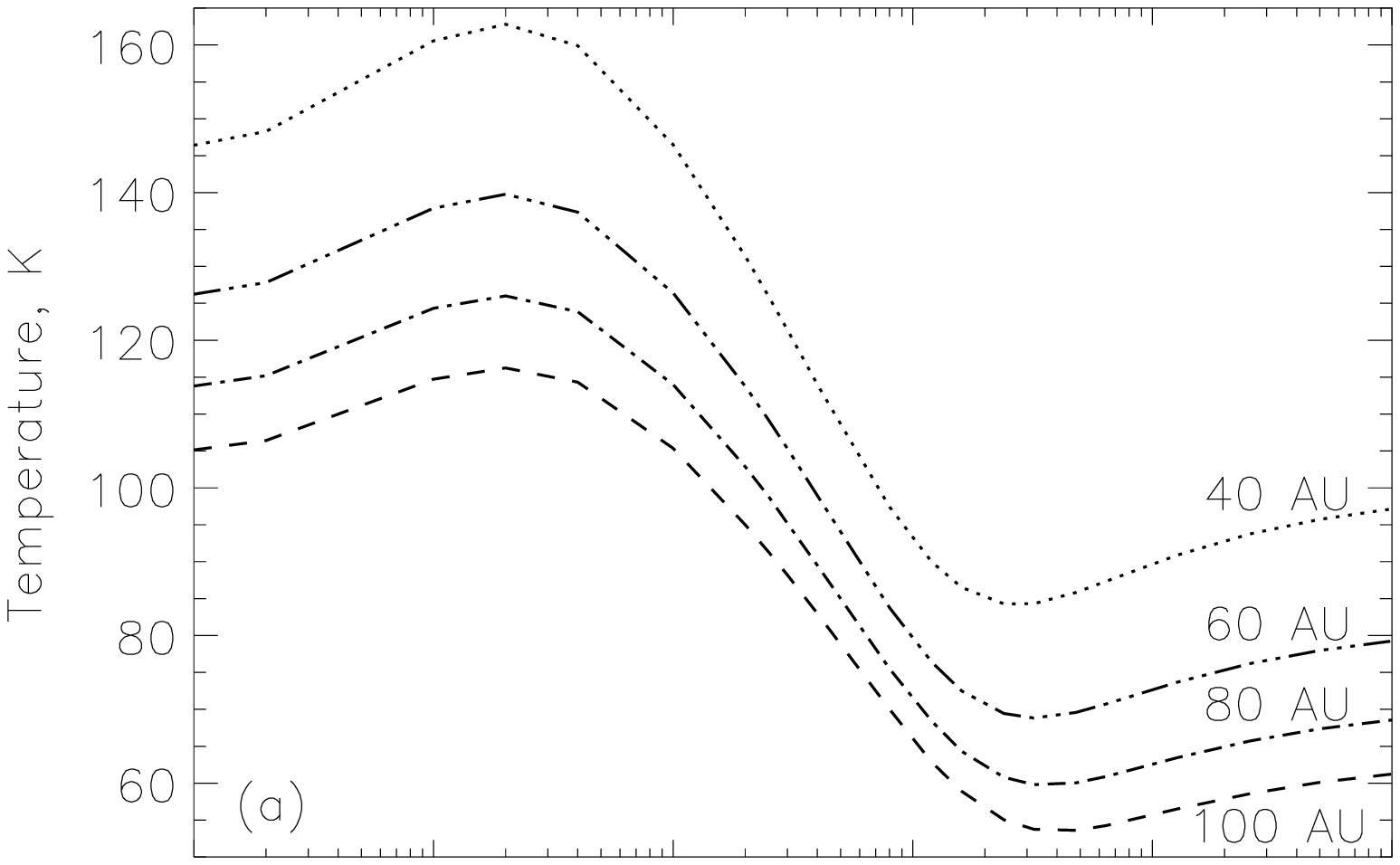} & \hspace{-0.2in}
        \epsscale{0.45} \plotone{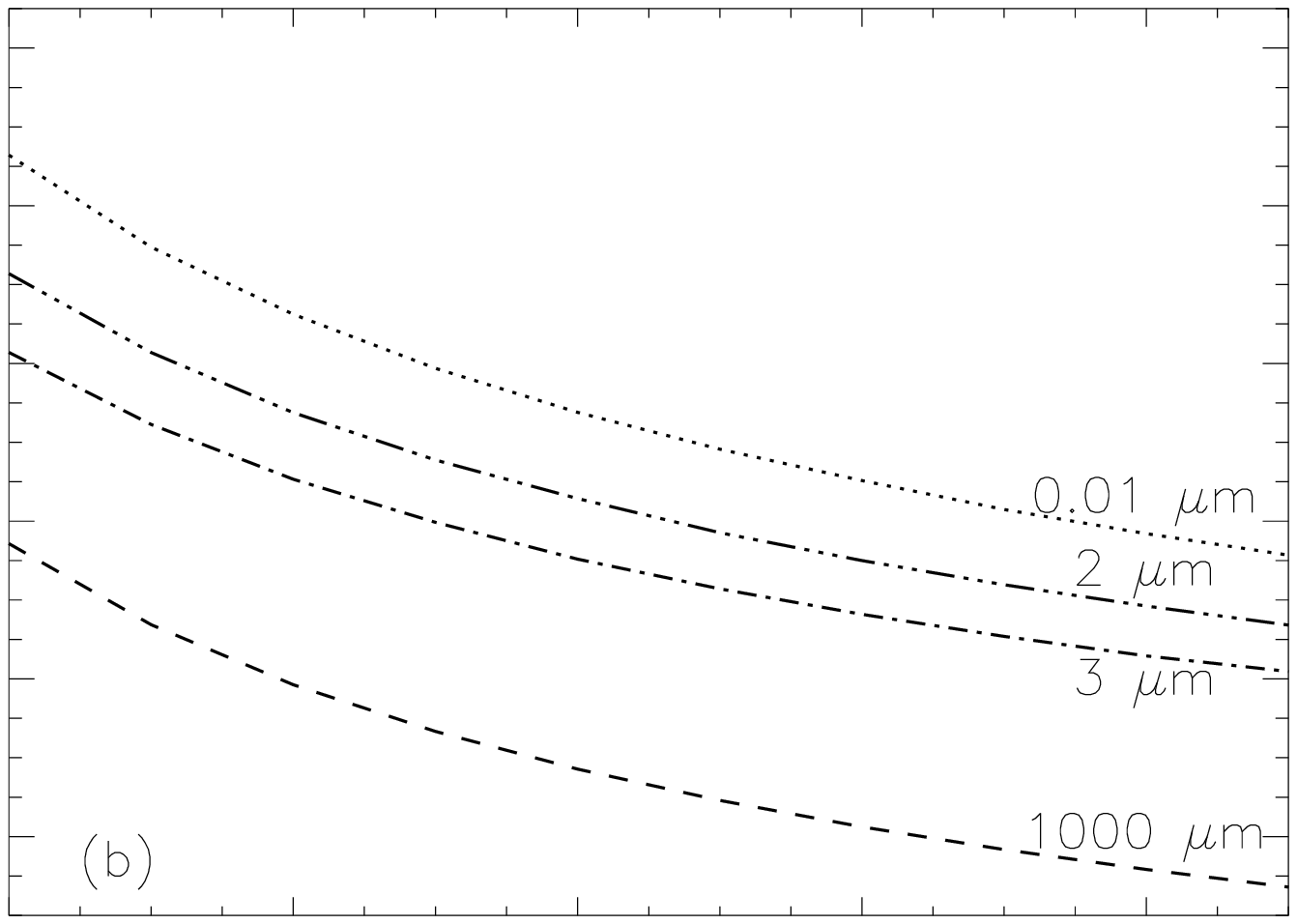} & \vspace{-0.01in} \\
      & \hspace{-0.15in} \epsscale{0.534} \plotone{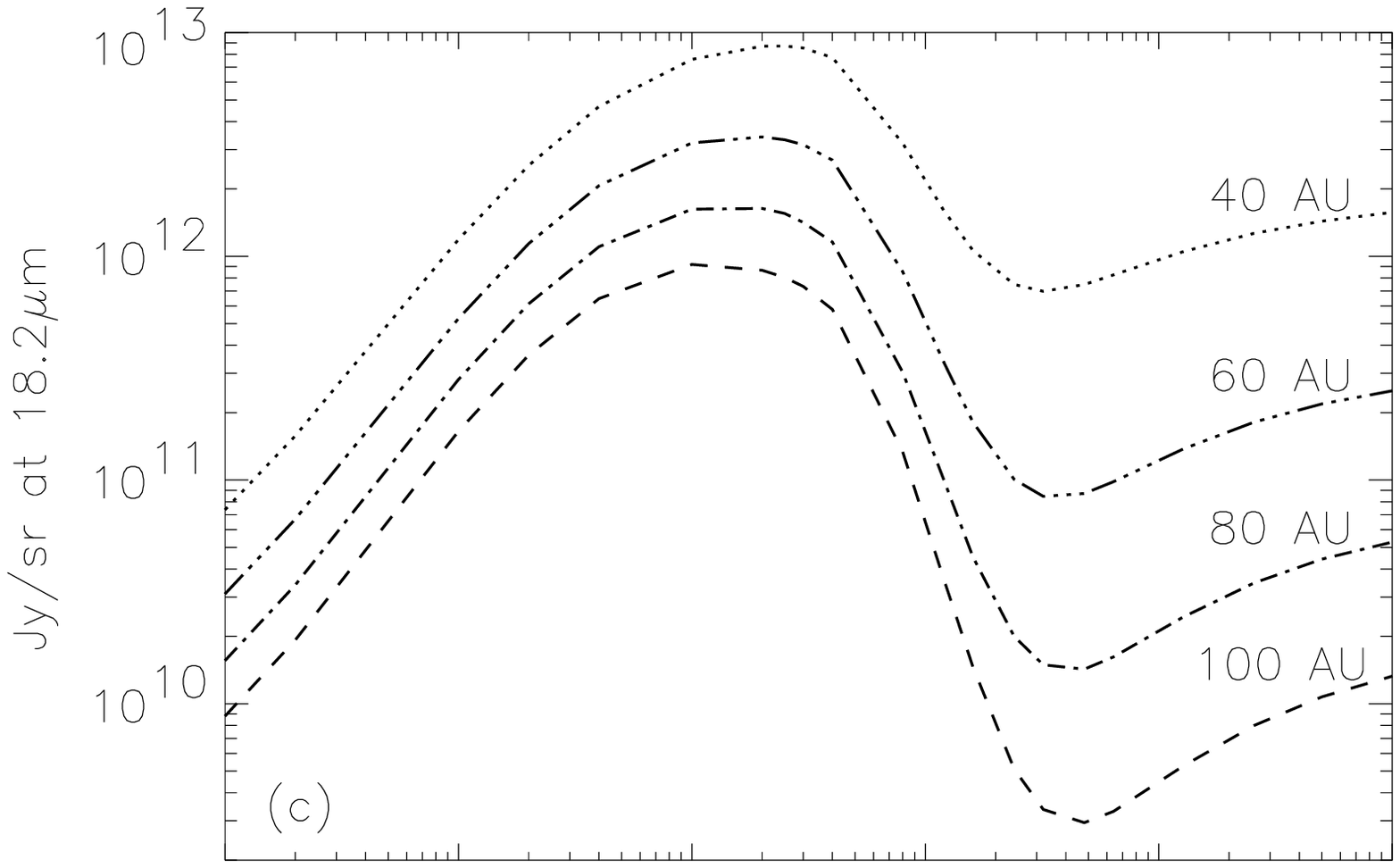} & \hspace{-0.2in}
        \epsscale{0.45} \plotone{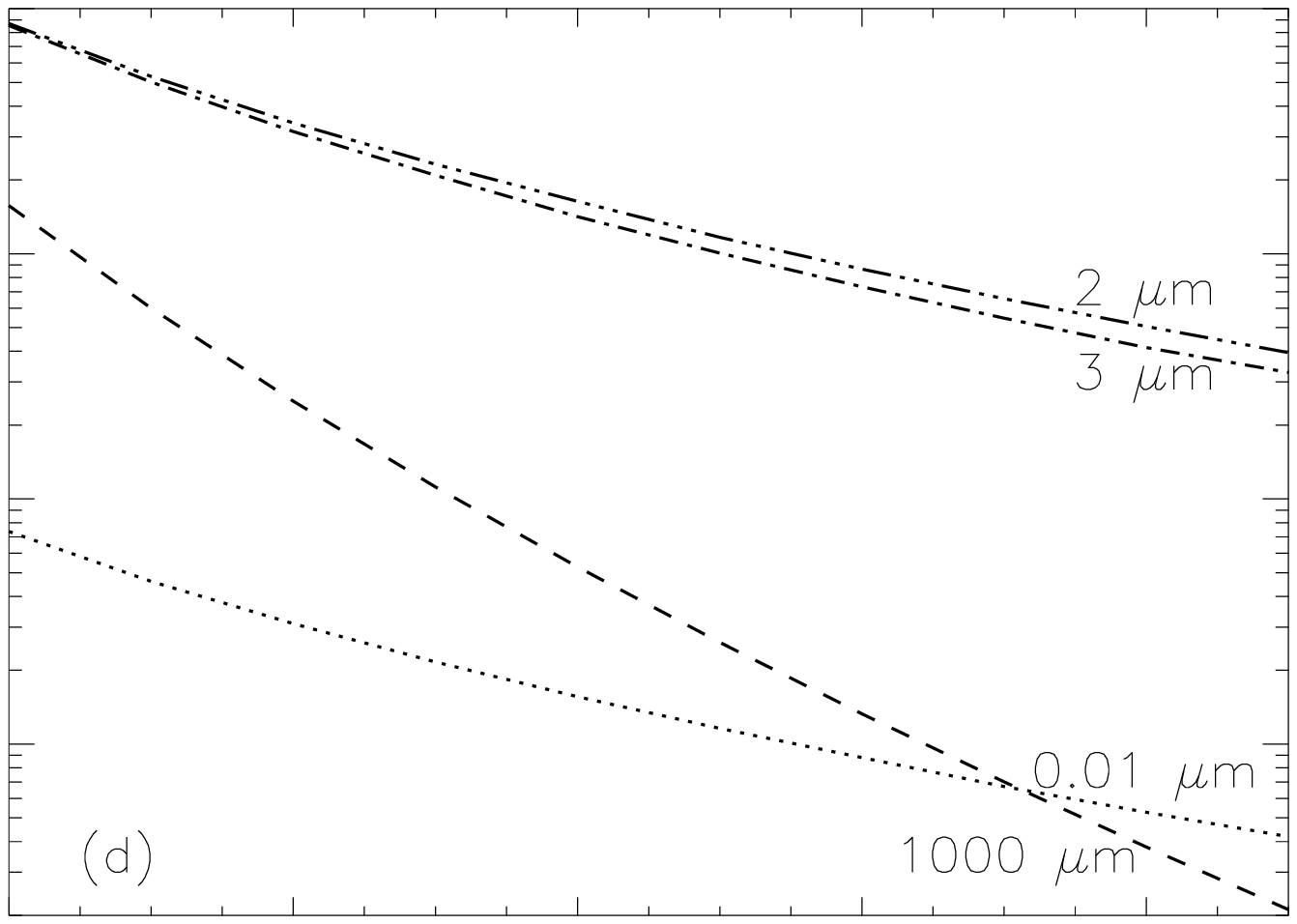} & \vspace{-0.03in} \\
      & \hspace{-0.15in} \epsscale{0.534} \plotone{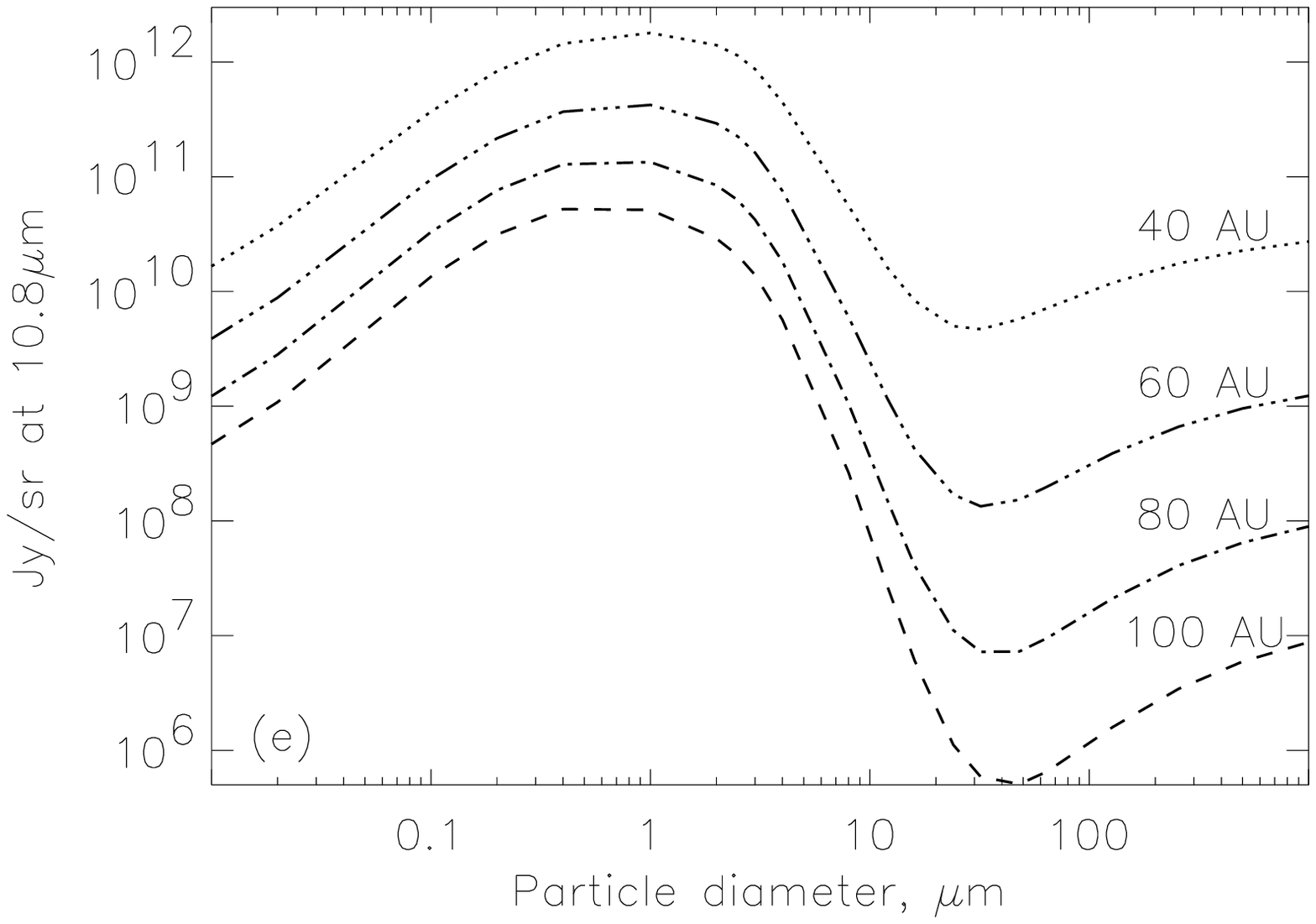} & \hspace{-0.2in}
        \epsscale{0.45} \plotone{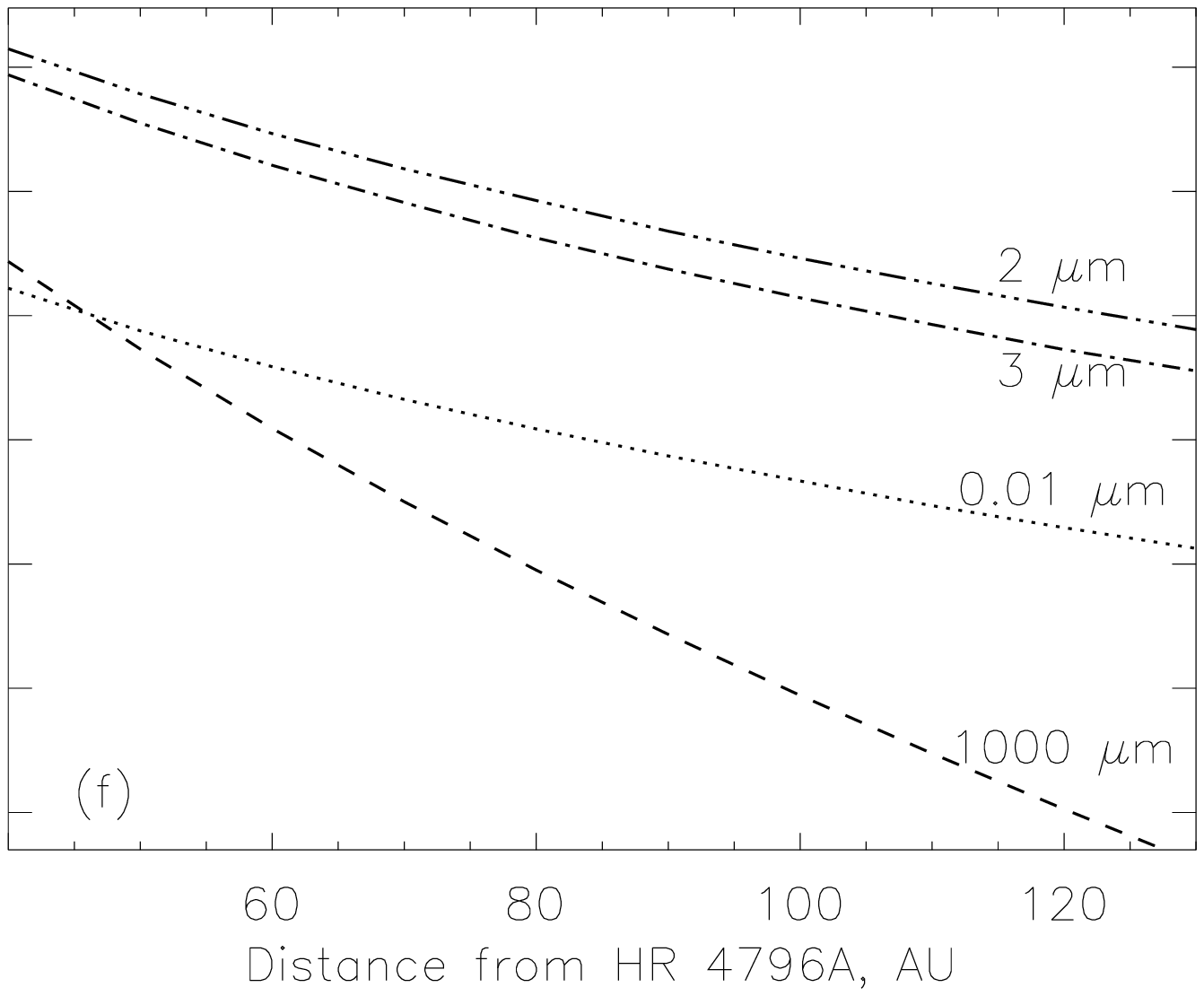} & 
    \end{tabular}
  \end{center}
  \caption{ The thermal properties of astronomical silicate Mie spheres
  in the HR 4796 disk, plotted for
  particles of different sizes at 40, 60, 80, and 100 AU from
  HR 4797A (\textbf{a}), (\textbf{c}), and (\textbf{e}),
  and for 0.01, 2, 3, and 1000 $\mu$m diameter particles at different
  distances from HR 4796A (\textbf{b}), (\textbf{d}), and (\textbf{f}).
  The temperatures that these particles attain is plotted in (\textbf{a})
  and (\textbf{b}).
  The contribution of a particle's thermal emission to the flux density
  received at the Earth per solid angle that its cross-sectional area
  subtends there, $Q_{abs}(D,\lambda)B_\nu[T(D,r),\lambda]$
  (eq.~[\ref{eq:flux}]), is plotted for observations in the IHW18,
  18.2 $\mu$m, (\textbf{c}) and (\textbf{d}), and N, 10.8 $\mu$m,
  (\textbf{e}) and (\textbf{f}) wavebands.
  The brightnesses of disk models in these two wavebands were
  calculated by taking $P(\lambda,r)$ from the lines on (\textbf{d})
  and (\textbf{f}) corresponding to particles of diameter $D_{typ}$. }
  \label{fig5}
\end{figure}

\begin{figure}
\epsscale{0.5}
  \begin{center}
    \begin{tabular}{ccc}
      & \vspace{-0.04in} \epsscale{0.64}  \plotone{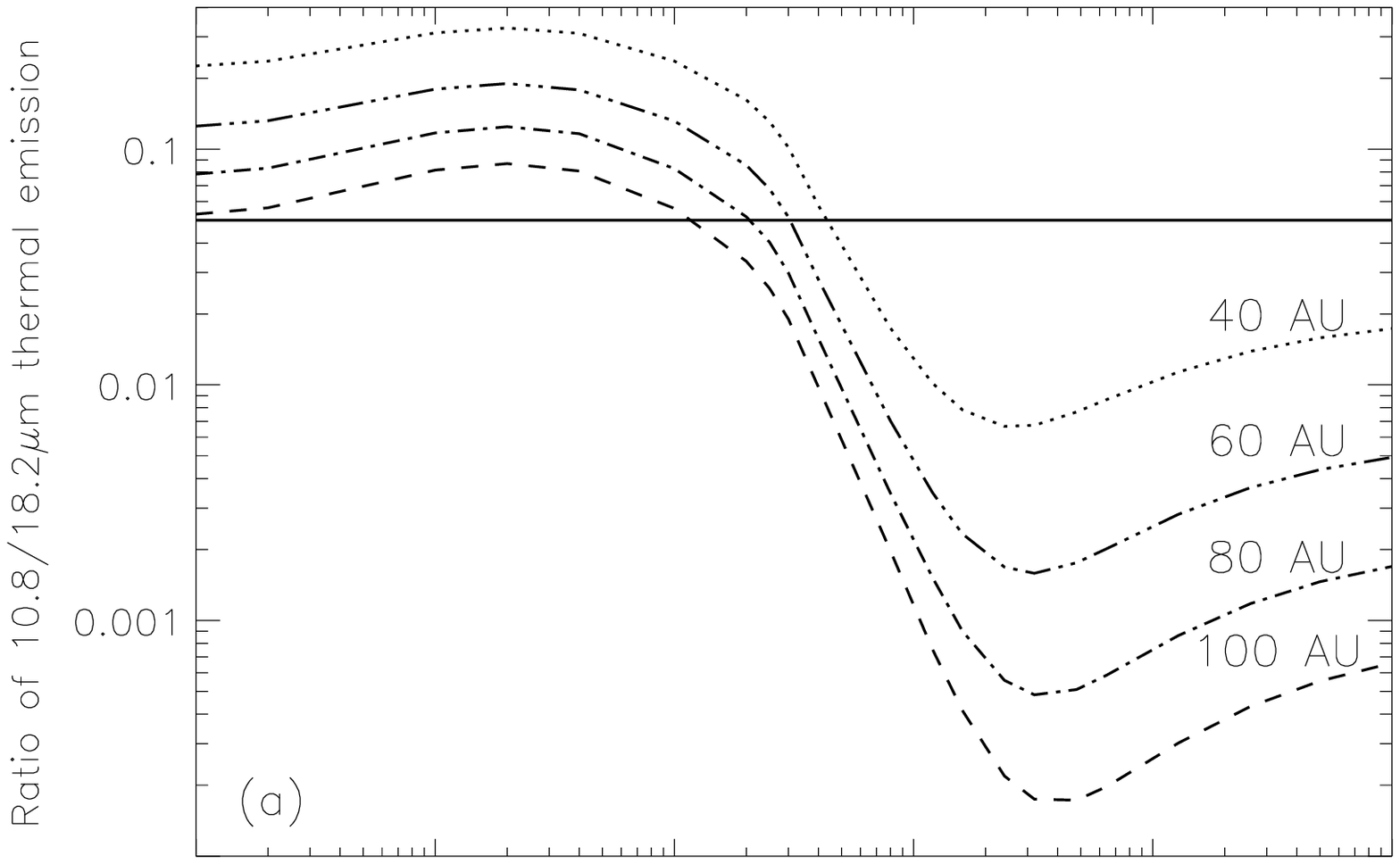} & \\
      & \hspace{0.04in}  \epsscale{0.63} \plotone{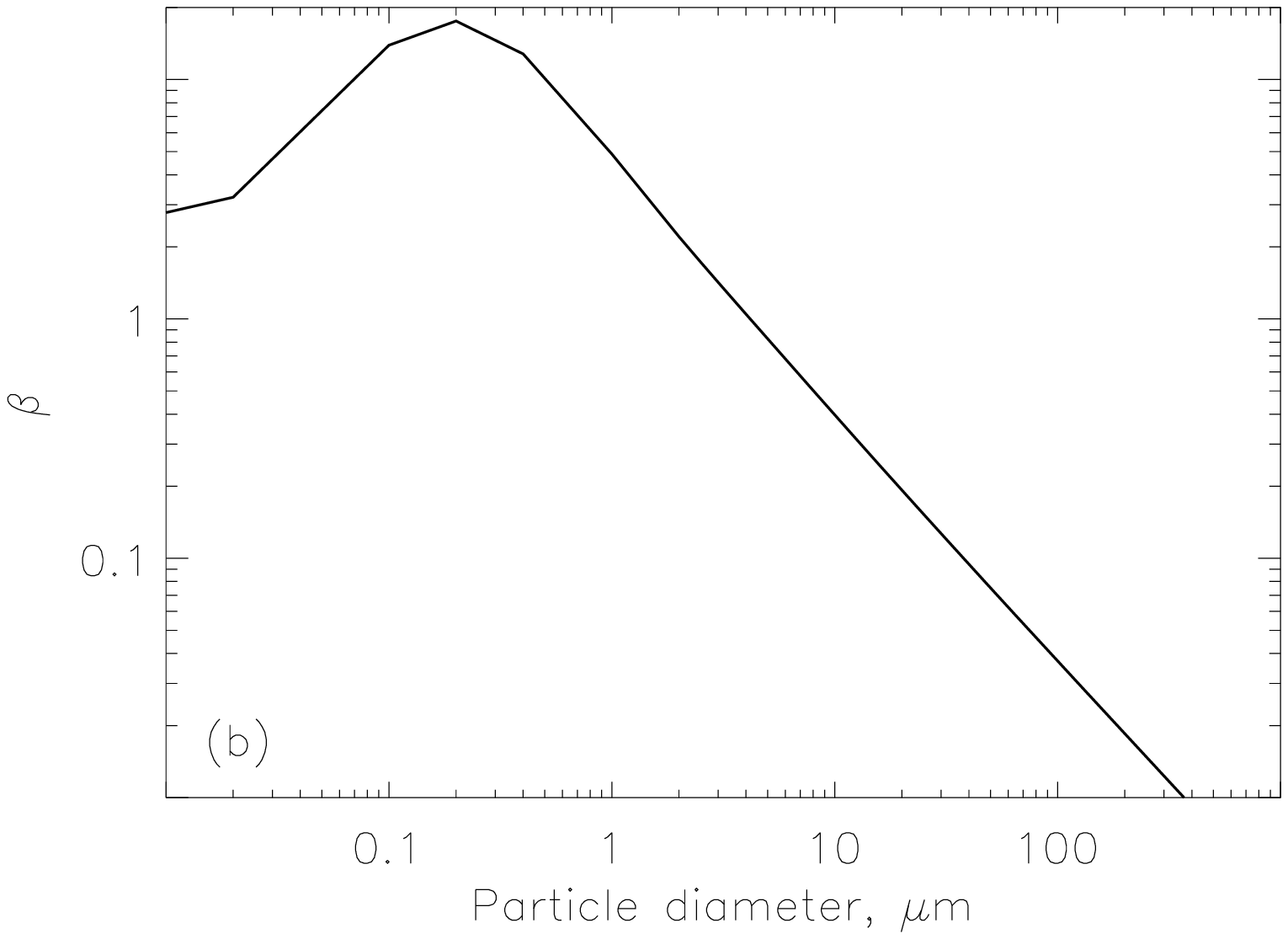} &
    \end{tabular}
  \end{center}
  \caption{ (\textbf{a}) The ratio of the thermal emission in the N, 10.8
  $\mu$m, and IHW18, 18.2 $\mu$m, wavebands, of astronomical
  silicate Mie spheres of different sizes at 40, 60, 80, and 100 AU
  from HR 4796A (i.e., Fig.~\ref{fig5}e divided by Fig.~\ref{fig5}c).
  Assuming the disk's flux densities in the two wavebands to be
  dominated by the emission of particles at 60-80 AU, the observed
  ratio of flux densities, $O(0.05)$ (T99; \S \ref{sssec-dtyp}), can
  be used to estimate that the disk's emitting particles have
  $D_{typ} = 2-3$ $\mu$m.
  (\textbf{b}) The ratio, $\beta$, of the radiation pressure force to
  the gravitational force acting on astronomical silicate Mie spheres
  of different sizes in the HR 4796 disk. }
  \label{fig6}
\end{figure}

\begin{figure}
  \begin{center}
    \begin{tabular}{cc}
      \epsscale{0.45} \plotone{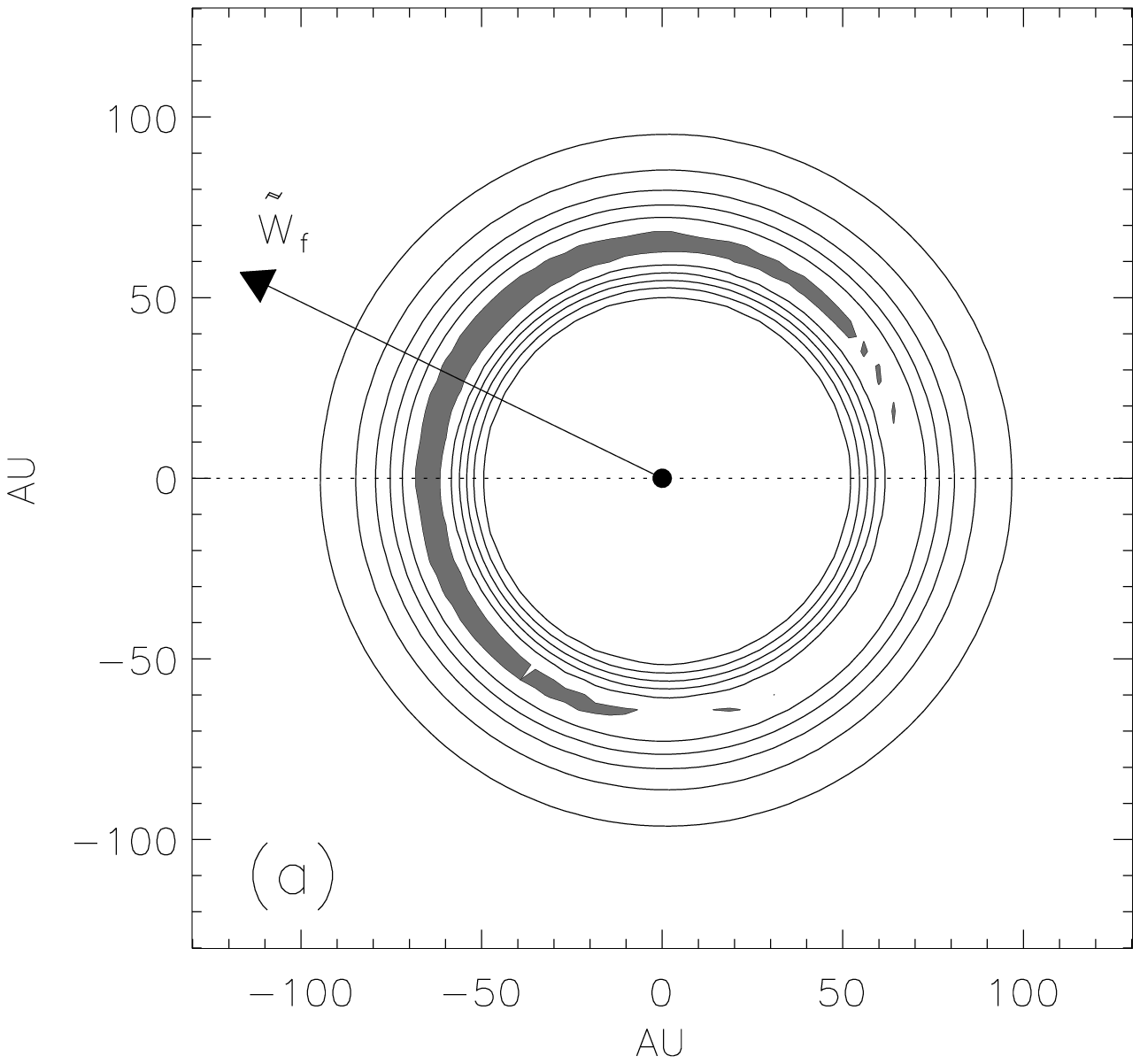} &
      \epsscale{0.44} \plotone{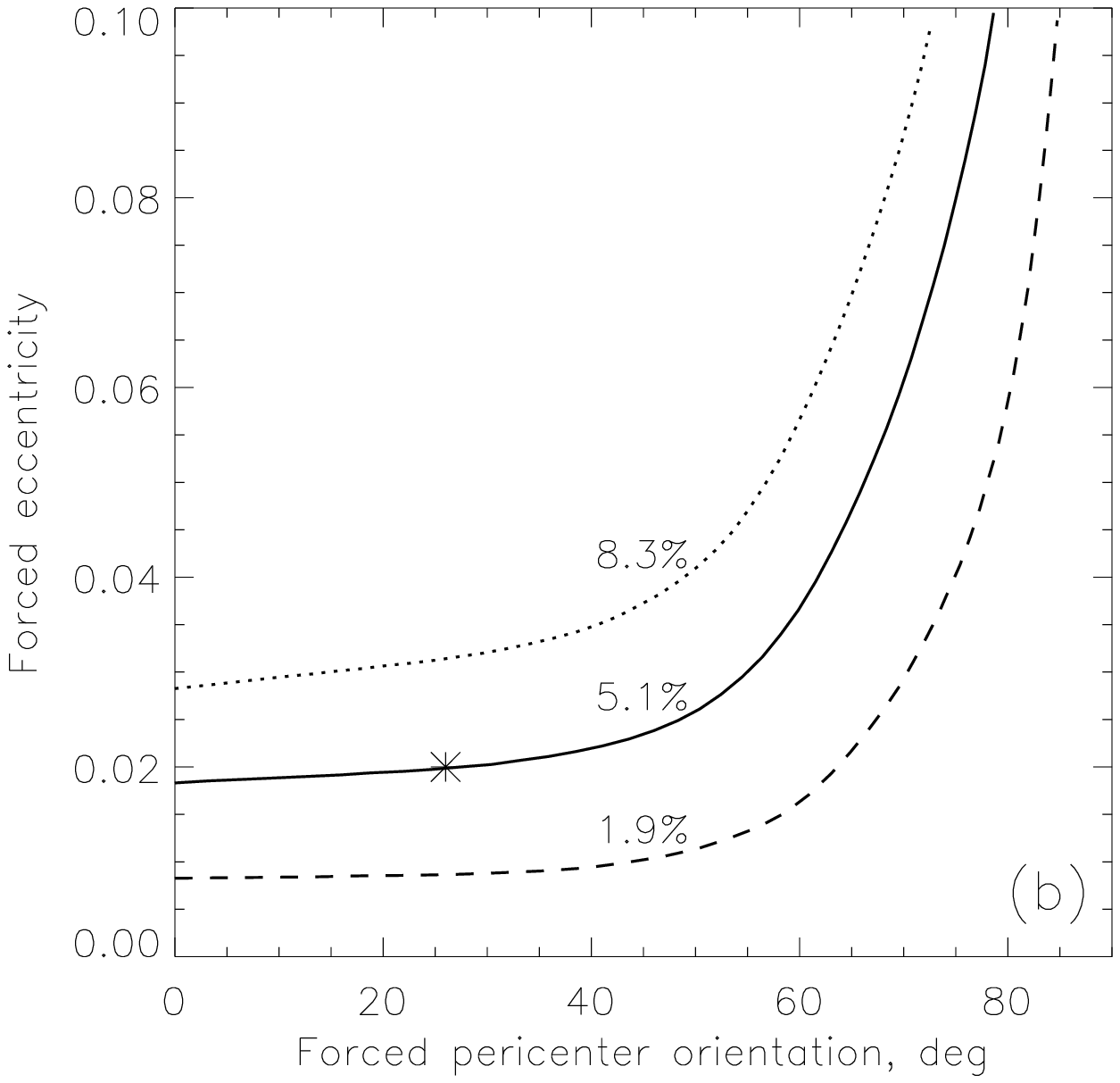}
    \end{tabular}
  \end{center}
  \caption{(\textbf{a}) Contour plot of an unsmoothed face-on view of the
  HR 4796 disk model (shown also in Fig.~\ref{fig4}) seen in the IHW18
  (18.2 $\mu$m) waveband.
  The contours are spaced linearly at 0.17, 0.34, 0.51, 0.68, 0.85, and
  1.02 mJy/pixel.
  The disk's offset causes particles in the forced pericenter direction,  
  located at a position angle of $90^\circ - \tilde{\omega}_f$, where
  $\tilde{\omega}_f = 26^\circ$, as measured from North in a
  counterclockwise direction, to be hotter, and hence brighter, than
  those in the forced apocenter direction;
  this is the "pericenter glow" phenomenon, evident in this figure
  by the shape of the brightest (filled-in) contour.
  The geometry of the observation is defined such that the
  disk as it is shown here is rotated by a further $90^\circ - I_{obs}$
  about the dotted line, where $I_{obs} = 13^\circ$.
  (\textbf{b}) Relation of forced eccentricity, $e_f$, to the orientation
  of the longitude of forced pericenter, $\tilde{\omega}_f$, in the model
  to achieve the observed lobe brightness asymmetry of $5.1 \pm 3.2\%$
  (a similar relationship is necessary to achieve the observed radial
  offset).
  A forced eccentricity as small as 0.02 would suffice to achieve
  the observed asymmetry, but a higher forced eccentricity could be
  necessary if the forced pericenter is aligned in an unfavorable
  direction.
  Our final model, shown in Figs.~\ref{fig4}, \ref{fig7}-\ref{fig9}, has
  $e_f = 0.02$ and $\tilde{\omega}_f = 26^\circ$, and this point is
  shown with an asterisk on this plot. }
  \label{fig7}
\end{figure}

\begin{figure}
  \begin{center}
    \begin{tabular}{rlll}
      (\textbf{a}) & \hspace{-0.5in} \epsscale{0.53} \plotone{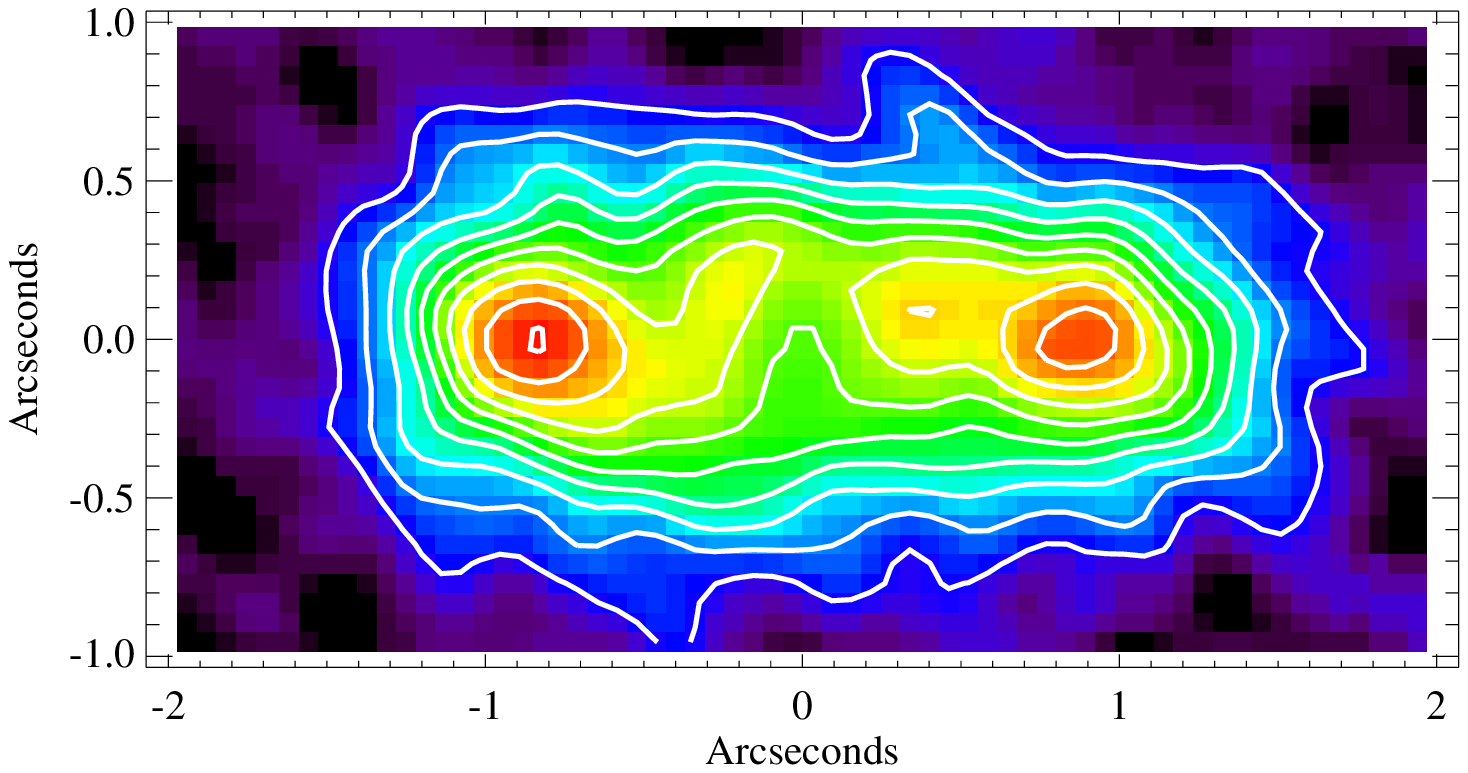} &
      \hspace{-0.22in} \epsscale{0.482} \plotone{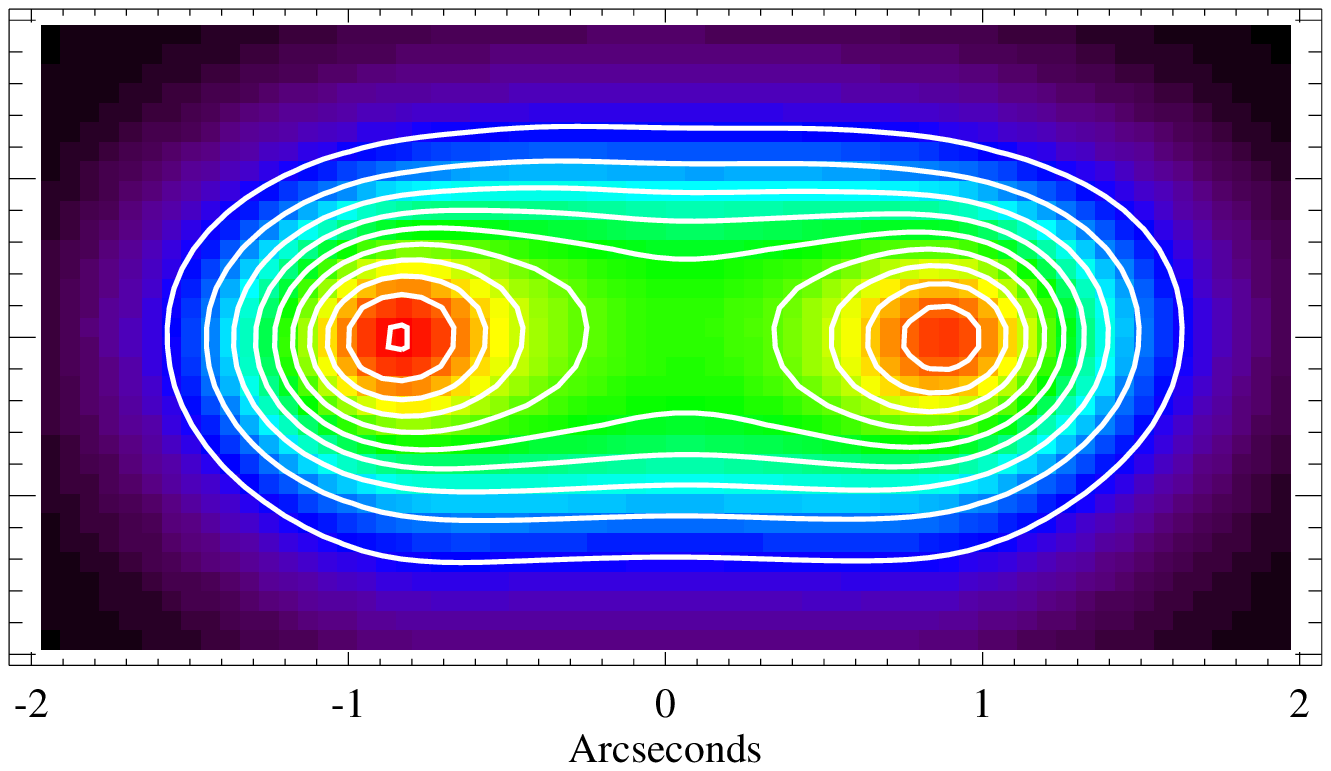} &
      \hspace{-0.2in} (\textbf{b}) \vspace{0.1in}
    \end{tabular}
    \begin{tabular}{ccc}
      \epsscale{0.306} \plotone{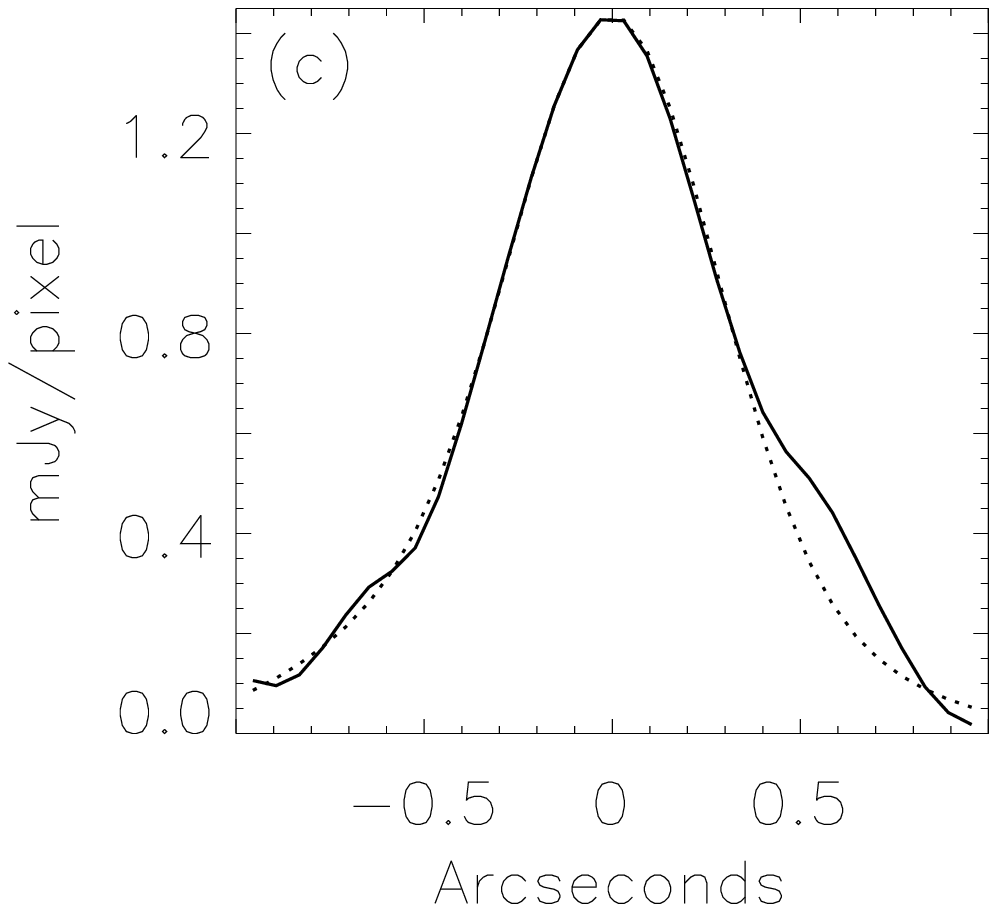} &
      \epsscale{0.464} \hspace{-0.22in} \plotone{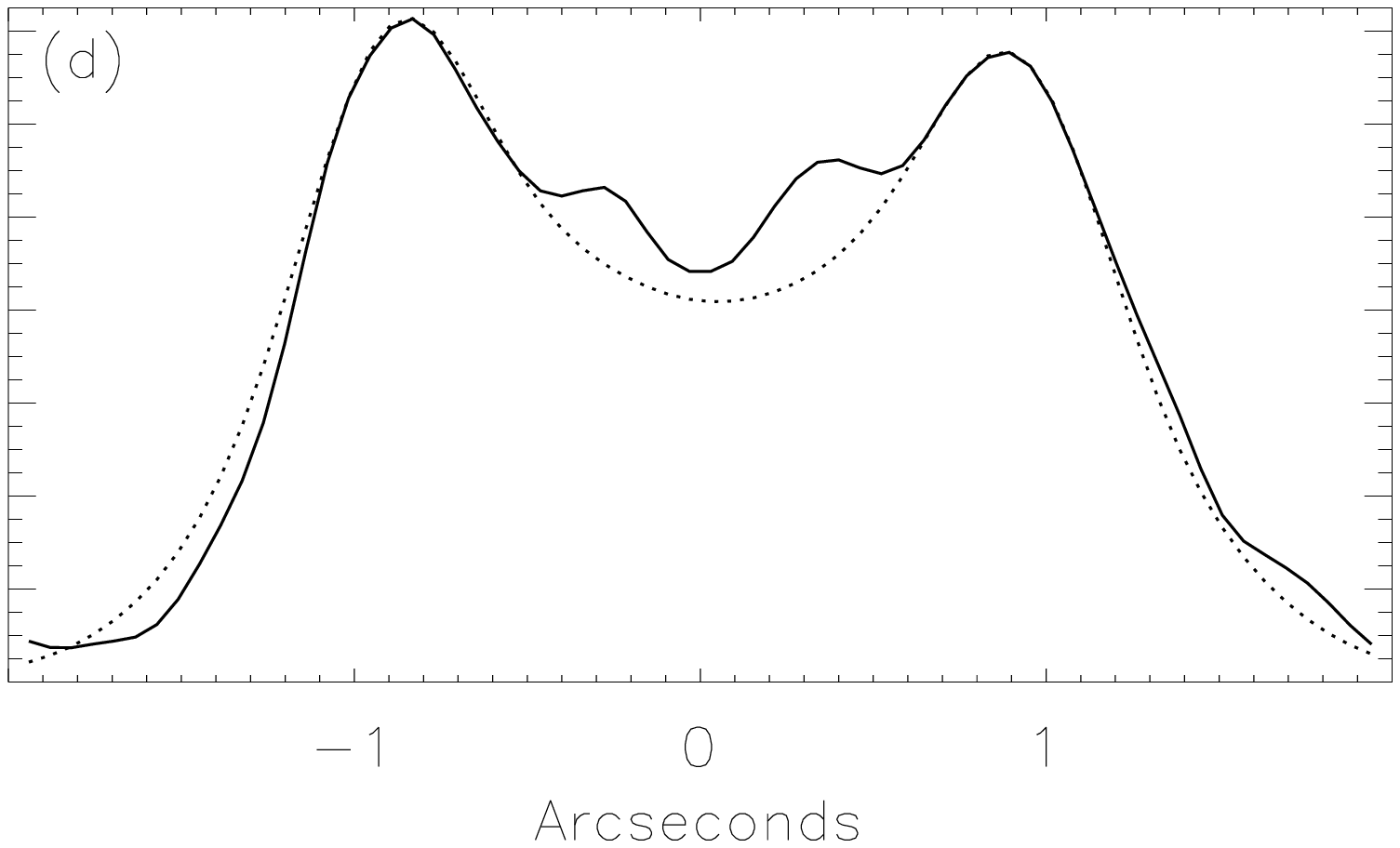} &
      \epsscale{0.236} \hspace{-0.22in} \plotone{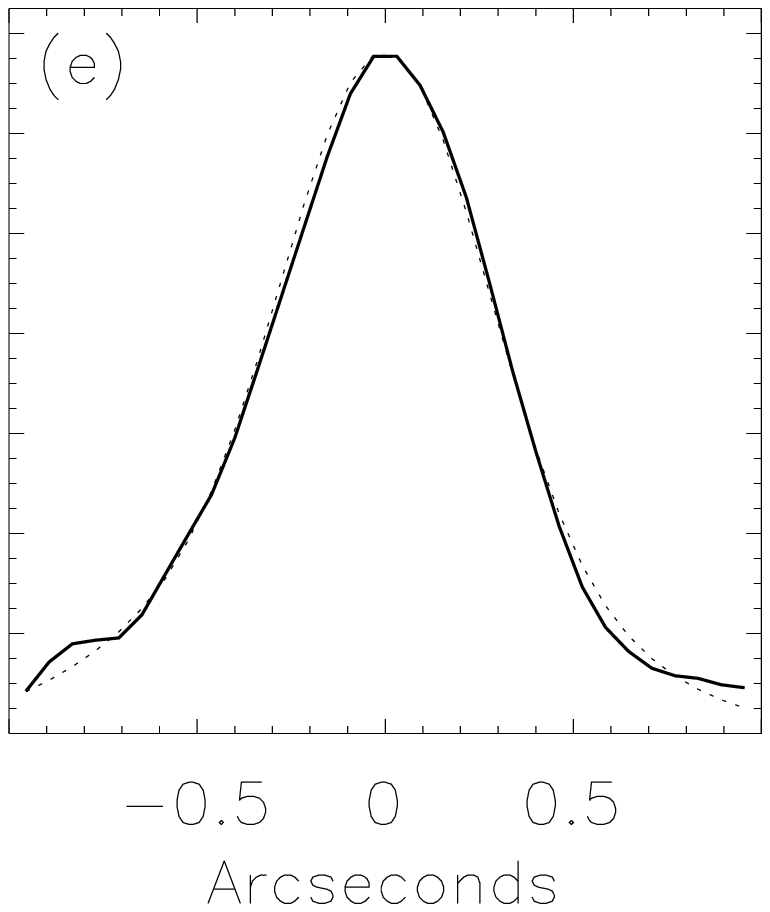}
    \end{tabular}
  \end{center}
  \newpage
  \caption{\textbf{Top} --- False color images of HR 4796 in the IHW18
  (18.2 $\mu$m) waveband.
  Both the observation (\textbf{a}), on the left, and the model (\textbf{b}),
  on the right, have been rotated to horizontal with the NE lobe on the left.
  The contours are spaced linearly at 0.22, 0.35, 0.49, 0.62, 0.75, 0.89,
  1.02, 1.15, 1.29, and 1.42 mJy/pixel.
  The observation has had the photospheric emission of HR 4796A
  subtracted, and a 3 pixel FWHM gaussian smoothing applied.
  The image of the model mimics the observation both in pixel
  size (1 pixel $= 0\farcs0616 = 4.133$ AU) and
  smoothing (using an observed PSF and including the 3 pixel
  post-observational smoothing).
  \textbf{Bottom} --- Line-cuts in the vertical direction through the NE
  (\textbf{c}) and SW (\textbf{e}) lobes, and in the horizontal
  direction through the center of both lobes (\textbf{d}).
  The observations are shown with a solid line and the
  model with a dotted line. }
  \label{fig8}
\end{figure}

\begin{figure}
  \begin{center}
    \epsscale{0.54} \plotone{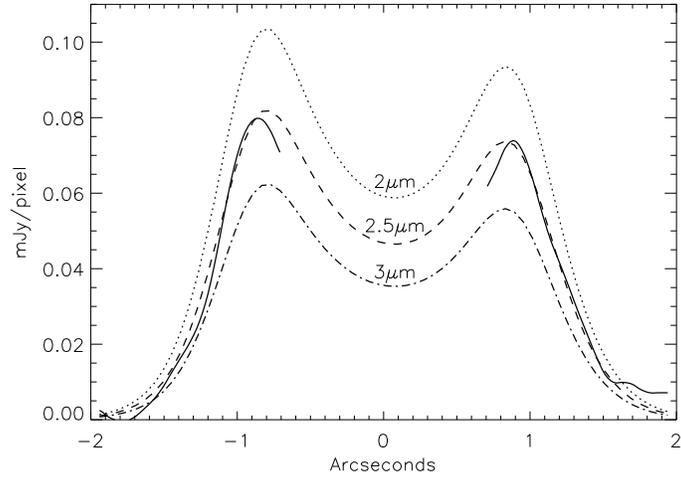}
  \end{center}
  \caption{Horizontal line-cuts along the plane of the lobes in the
  N (10.8 $\mu$m) band.
  The observation is shown with a solid line and models with particle
  diameters of $D_{typ}$ = 2, 2.5, and 3 $\mu$m are shown with dotted,
  dashed, and dash-dot lines.
  The total amount of cross-sectional area in the models, $\sigma_{tot}$,
  has been scaled to fit the observed mean brightness of the lobes in the
  IHW18 (18.2 $\mu$m) waveband;
  the model with $D_{typ} = 2.5$ $\mu$m gives the best fit to the observed
  lobe brightnesses in the N band.
  The observed N band flux density is not well constrained within
  0\farcs8 of HR 4796A due to imperfect subtraction of the stellar
  photosphere from the image (T99), and so it is not shown here. }
  \label{fig9}
\end{figure}

While the modeling used a distribution of orbital elements that is
only appropriate for the disk's large particles, it was the spatial
distribution of the disk's emitting material, $\sigma(r,\theta,\phi)$,
that was constrained by the modeling, not the distribution of orbital
elements, $\sigma(a,e,I,\Omega,\tilde{\omega})$.
Therefore, the inferred distribution, $\sigma(r,\theta,\phi)$, which is
that shown in Fig.~\ref{fig4}, is an accurate description of the spatial
distribution of the disk's emitting particles, whatever their size.
Indeed, it is in excellent agreement with that inferred from other
observations of the HR 4796 disk
(\cite{jfht98}; \cite{krwb98}; \cite{ssbk99}).
If the emitting particles are the disk's large particles, then the
inferred model variables have physical interpretations
for the distribution of the orbits of the disk's large particles.
If, as appears to be the case, the emitting particles are $\beta$
meteoroids, then further modeling of $\sigma(r,\theta,\phi)$ needs
to be done to infer the distribution of the orbits of these particles.

However, since the disk's $\beta$ critical and $\beta$ meteoroid particles
are created from its large particles, the spatial distributions of all of these
particles share a great deal in common (see \S\S \ref{ssec-ddpc} and
\ref{ssec-ow}):
they all have the same plane of symmetry, the same flaring,
and the same offset and warp asymmetries, but the radial distributions
of the smaller particles are more extended than that of the large particles.
This means that if we are seeing the disk's $\beta$ meteoroids, then
the spatial distribution of its large particles has a plane of symmetry that
is defined by $I_{obs}$, and an inner edge that is at the same radial
location, and that is offset by the same amount and in
the same direction, as that shown in Fig.~\ref{fig4};
their radial distribution, however, would not be as extended as that of
Fig.~\ref{fig4}.
Thus, the model parameters $a_{min}$, $e_f$, and $\tilde{\omega}_f$
have physical interpretations for the distribution of the orbits of the
disk's large particles, irrespective of the size of the emitting particles.

\subsection{Interpretation of Lobe Asymmetry: HR 4796's Secular Perturbations}
\label{ssec-ias}
The lobe asymmetry in the model of \S \ref{sec-hr4796} is due solely to the
offset inner edge of the disk.
The model shows that secular perturbations amounting to a comparatively
small forced eccentricity, $e_f = 0.02$, imposed on the orbits of large
particles at the inner edge of the disk, $a = 62$ AU, would cause the disk's
inner edge to be offset by a sufficient amount to cause the observed
5\% lobe asymmetry.
This section considers what kind of a perturber system would impose such
a forced eccentricity on the inner edge of the disk, and whether such a
system is physically realistic.
A discussion of the system's secular perturbations also allows
interpretation of the disk's inferred orientation, defined by the
parameters $\tilde{\omega}_f$ and $I_{obs}$.

\begin{figure}
  \begin{center}
    \begin{tabular}{cccc}
      & \epsscale{0.452} \plotone{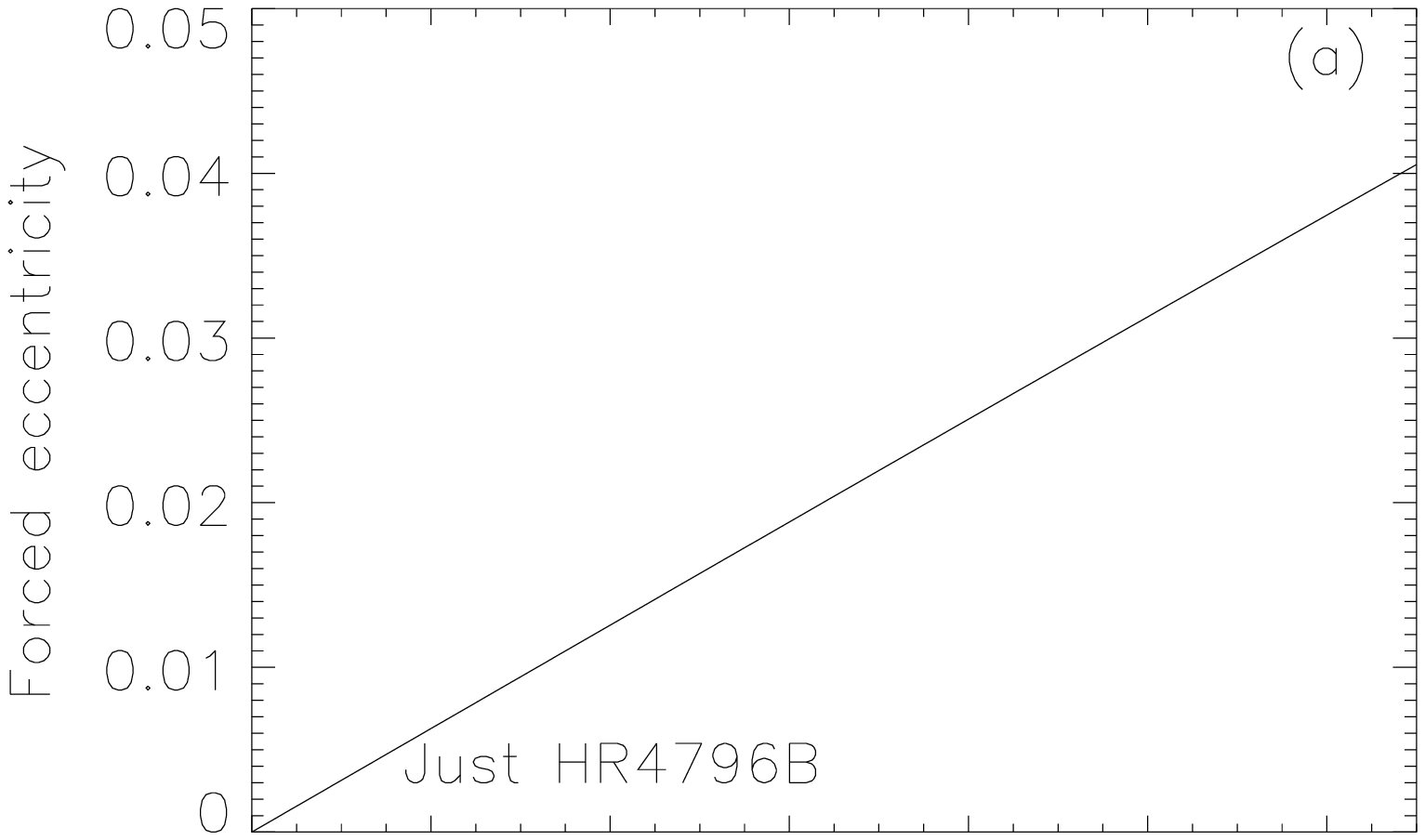} &
        \epsscale{0.452} \plotone{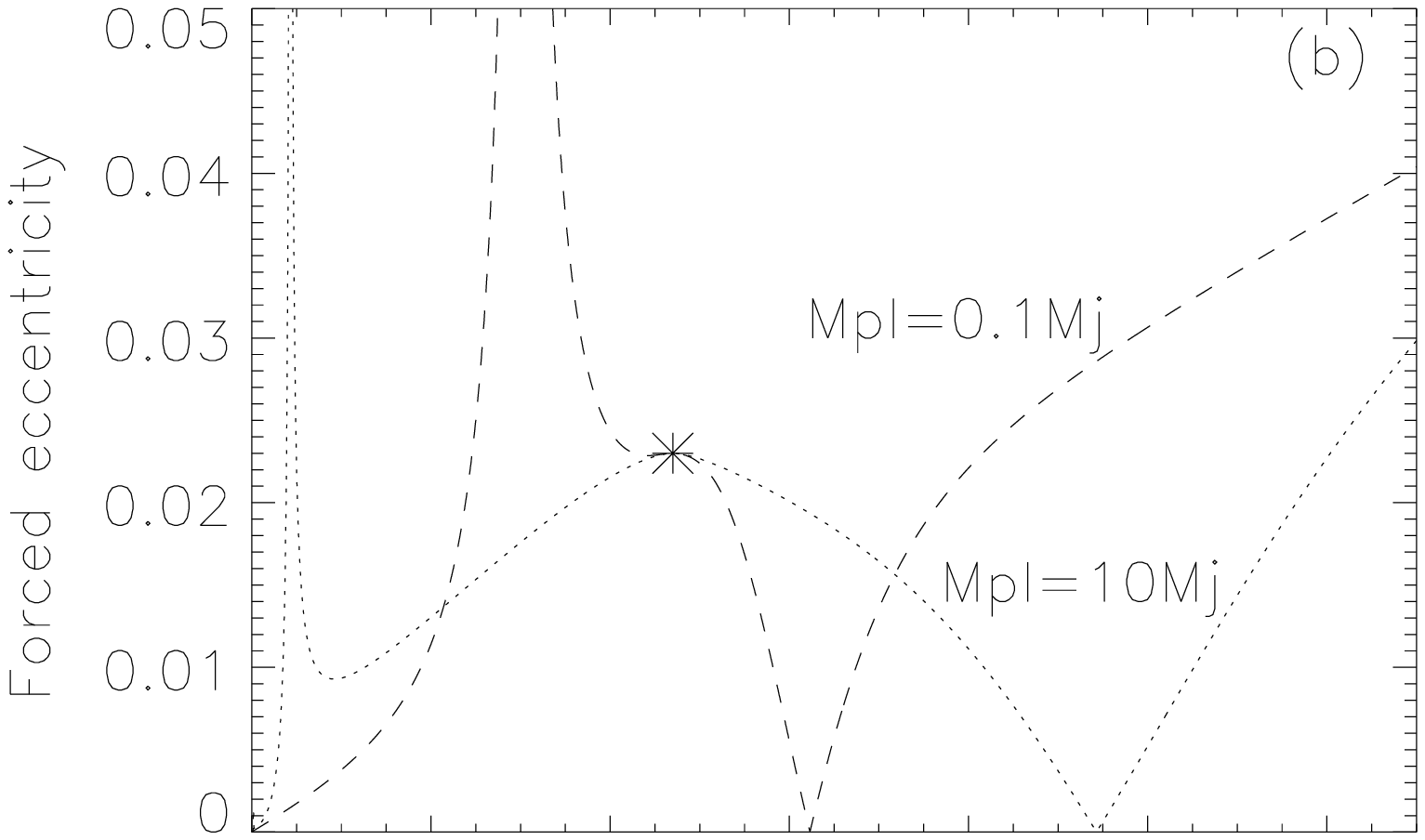} & \\ \vspace{-0.02in}
      & \epsscale{0.452}  \plotone{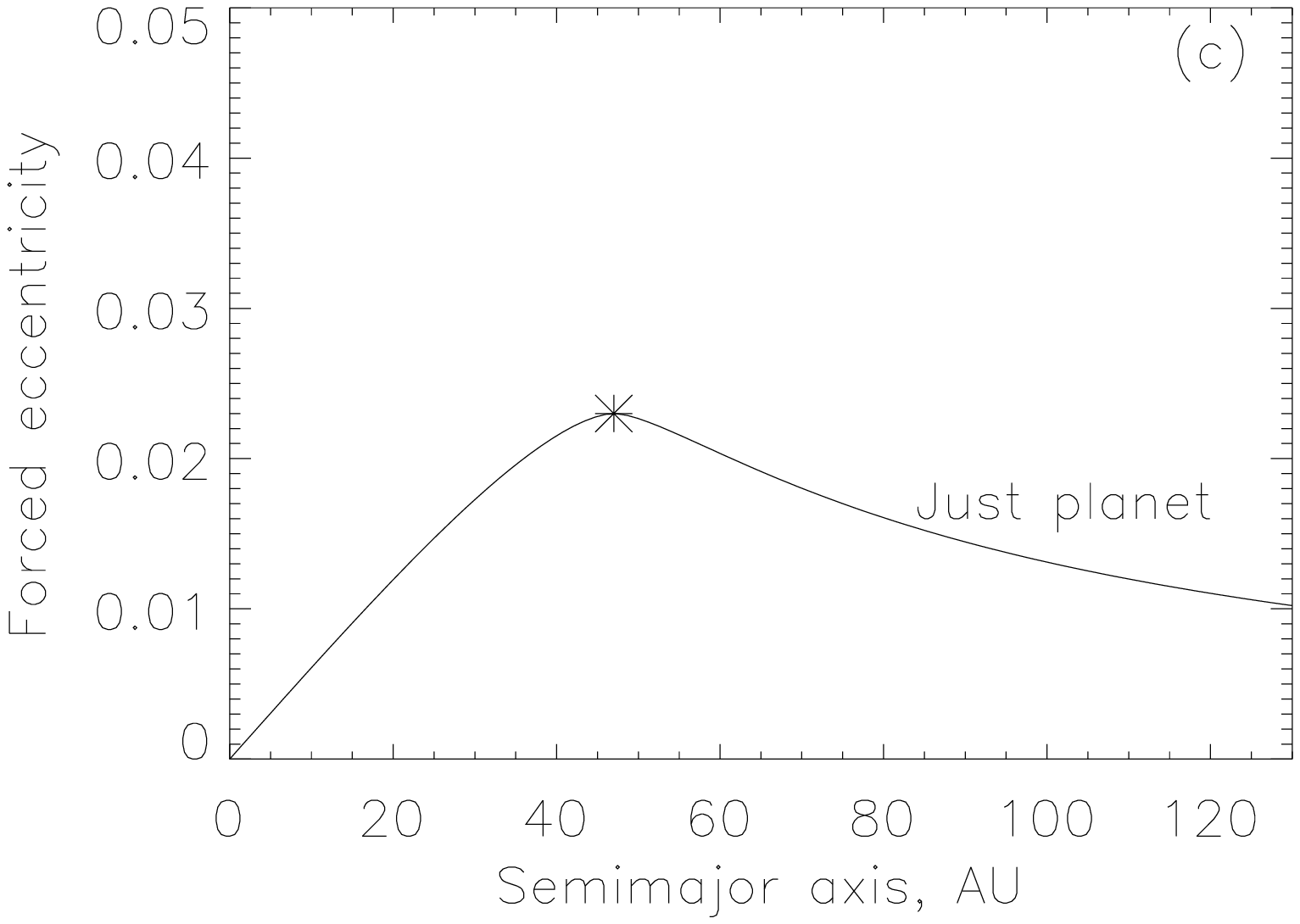} &
        \epsscale{0.432} \hspace{0.04in} \plotone{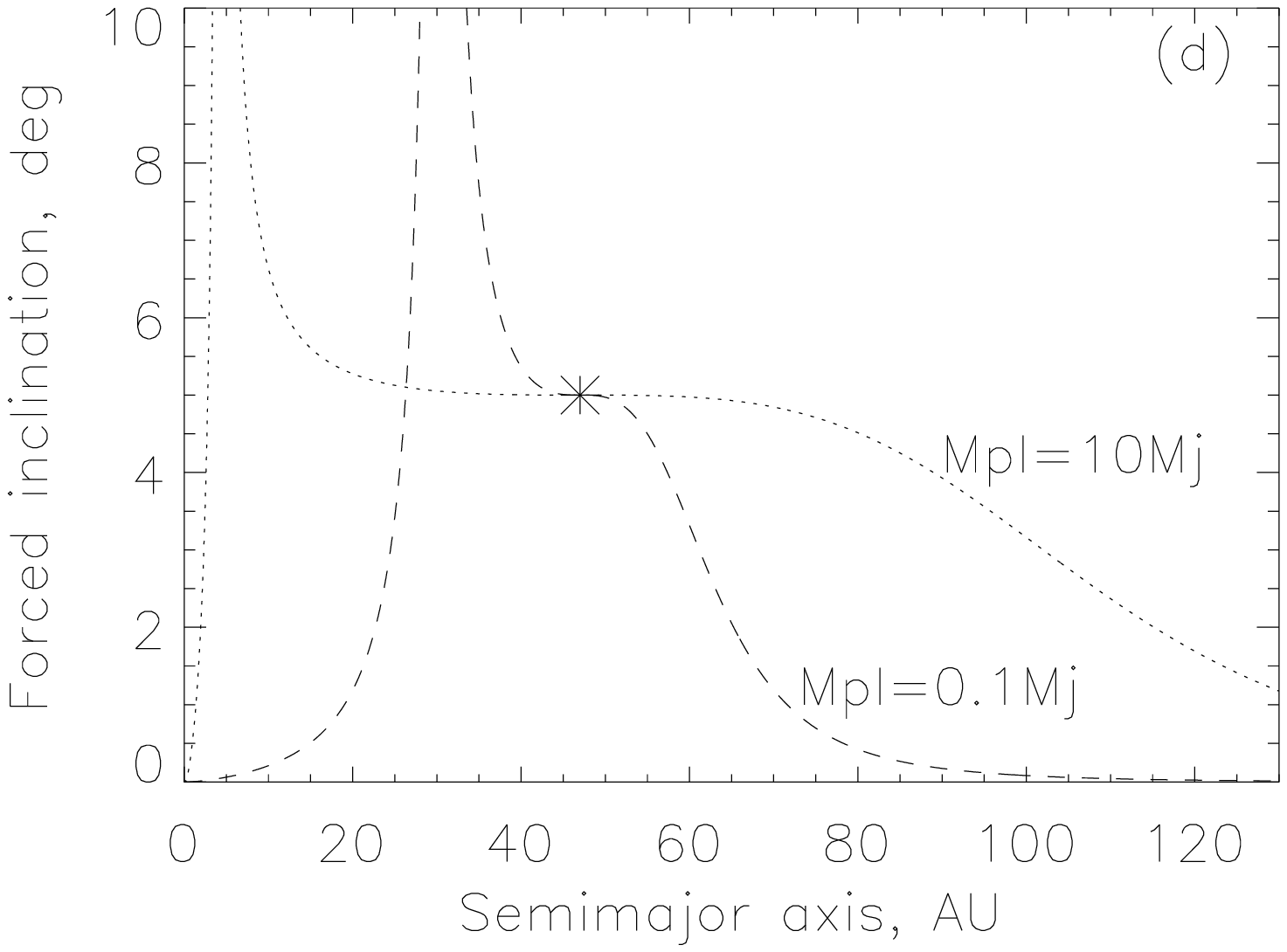} & \\
    \end{tabular}
  \end{center}
  \caption{Plots of the forced eccentricities imposed on the orbits of
  particles in the HR 4796 system as a function of their semimajor axes,
  assuming different combinations of perturbers in the system:
  (\textbf{a}) HR 4796B only;
  (\textbf{b}) HR 4796B and a planet at the inner edge of the disk;
  (\textbf{c}) a planet only.
  The forced inclinations imposed by the two perturber system are
  shown in (\textbf{d}).
  The inner edge of the disk is at $a = 62$ AU.
  HR 4796B and the planet are assumed to have:
  $M_B = 0.38M_\odot$, $a_B = 517$ AU, $e_B = 0.13$, and $I_B = 0$;
  $M_{pl} = 0.1$ and 10 $M_J$, where $M_J$ is the mass of Jupiter,
  $a_{pl} = 47$ AU, $e_{pl} = 0.023$, and $I_{pl} = 5^\circ$.
  The orbital elements of the planet are marked by an asterisk on
  the forced elements plots.
  In a one perturber system the forced eccentricity is independent
  of the mass of the perturber, and the forced inclination is the plane of
  the perturber's orbit.
  In a two perturber system, the shapes of the forced element plots depend
  on the mass of the planet, and the forced eccentricity also
  depends on the orientations of the perturbers' orbits.
  The forced eccentricity is plotted in (\textbf{b}) assuming that
  $\tilde{\omega}_{pl} = \tilde{\omega}_{B} + 180^\circ$.
  This means that $\tilde{\omega}_f = \tilde{\omega}_{pl}$ for $a < a_{crit}$,
  and $\tilde{\omega}_f = \tilde{\omega}_{B}$ for $a > a_{crit}$, where
  $a_{crit}$ is the semimajor axis of a particle's orbit for which $e_f = 0$;
  a similar alignment with the perturbers' orbital planes seen in
  the plot of the particles' forced inclinations.
  Thus, the lobes have both their asymmetries and their plane of symmetry
  aligned with the orbit of the planet if $M_{pl} > 0.1M_J$, and with
  the orbit of HR 4796B if $M_{pl} < 0.1M_J$. }
  \label{fig10}
\end{figure}

If HR 4796A's binary companion, HR 4796B, is on an eccentric orbit,
it would have imposed a forced eccentricity on the disk particles.
However, a forced eccentricity could also have been imposed on the disk
by an unseen planet close to the inner edge of the disk, a planet which
could be responsible for clearing the inner region
(e.g., \cite{rsss94}).
The secular perturbations imposed on the HR 4796 disk by a perturber
system that includes HR 4796B and a putative planet located at the inner
edge of the disk are shown in Fig.~\ref{fig10} for the four cases:
$M_{pl} = 0$, $M_{pl} = 0.1M_J$, $M_{pl} = 10M_J$, and just planet.
The parameters of the two perturbers are assumed to be:
\begin{description}
\item{\textbf{HR 4796B}}
  $M_B = 0.38M_\odot$(\cite{jfht98});
  the orbit of HR 4796B is unknown at present (\cite{jzbs93}), so the
  semimajor axis of its orbit is arbitrarily taken as its projected distance,
  $a_B = 517$ AU (\cite{jmwt98}; note that we are not assuming that
  this is the semimajor axis of the ellipse that the star's orbit traces on
  the sky) --- this gives an orbital period of $\sim 7000$ years
  (eq.~[\ref{eq:tper}]), and a timescale for secular perturbations from
  HR 4796B to have built up at 62 AU of $O$(1 Myr) (eq.~[\ref{eq:tsec2}]);
  $e_B = 0.13$, the eccentricity necessary to cause $e_f = 0.02$ at
  $a = 62$ AU if there were no unseen perturbers (eq.~[\ref{eq:zf1}]);
  $I_B = 0^{\circ}$, defining the reference plane for the analysis.
\item{\textbf{Planet}}
  $M_{pl}$ is a variable measured in Jupiter masses,
  where $M_J = 10^{-3}M_\odot$ (current observations have limited
  the size of a planet in the system to $M_{pl}<20M_J$, \cite{jmwt98});
  $a_{pl} = 47$ AU (see \S \ref{ssec-lsd});
  $e_{pl} = 0.023$, the eccentricity necessary to cause $e_f = 0.02$ at
  $a = 62$ AU if the planet was the only perturber (eq.~[\ref{eq:zf1}]);
  $I_{pl} = 5^{\circ}$, an arbitrary choice that represents the fact that
  the orbital plane of the planet is not necessarily be aligned with
  that of HR 4796B.
\end{description}

\placefigure{fig10}

\subsubsection{Just HR 4796B}
\label{sssec-jb}
For the cases when there is just one perturber in the system,
the forced elements in the system can be found from equations (\ref{eq:zf1})
and (\ref{eq:yf1}):
the forced eccentricity, $e_f $, is determined by the ratio of the
semimajor axes of the perturber and the particle, and by the eccentricity
of the perturber's orbit, but is independent of the perturber's mass;
the forced pericenter, $\tilde{\omega}_f$, is aligned with the pericenter of
the perturber;
and the plane of symmetry of the disk, $y_f$, is constant across the disk,
and is the orbital plane of the perturber.

So, if the only perturber is HR 4796B, then to impose $e_f = 0.02$
at $a = 62$ AU, the eccentricity of its orbit would have to be $e_B = 0.13$;
the consequent forced eccentricity imposed on the disk is plotted in
Fig.~\ref{fig10}a.
This also means that if $e_B > 0.1$, then a brightness asymmetry in this
disk of $>5\%$ would be expected unless adverse geometrical conditions
prevented it.
The position angle from north of HR 4796B relative to HR 4796A is
$225^\circ$ (\cite{jfht98}), while that of the SW lobe (i.e., the least
bright lobe) is $206^\circ$ (T99).
For the lobe asymmetry to be the consequence of perturbations from
HR 4796B only, HR 4796B must currently be close to its apastron, and
its orbital plane must be the plane of symmetry of the disk,
i.e., inclined at $13^{\circ}$ to the line of sight.
All of these conclusions are consistent with our initial estimate that
the semimajor axis of the star's orbit is equal to its observed projected
distance, 517 AU.

\subsubsection{HR 4796B and a Planet}
\label{sssec-bp}
Consider the effect of adding a planet at the inner edge of the disk into
the HR 4796 system\footnote{The orbital elements of a low-mass planet
in the system would, just like the disk particles, have forced and
proper components;
a high-mass planet would perturb the orbit of HR 4796B.}.
If there are two perturbers in the system, then the forced element
variation with semimajor axis depends both on the masses of the
perturbers and on the orientations of their orbits.
In the plot of Fig.~\ref{fig10}b, $\tilde{\omega}_{pl} = \tilde{\omega}_B +
180^{\circ}$ was chosen so that the forced eccentricity (and hence the
lobe asymmetry) is aligned with the planet's pericenter for $a < a_{crit}$,
and aligned with HR 4796B's pericenter for $a > a_{crit}$, where
$a_{crit}$ is the semimajor axis for which $e_f = 0$.
Since the two perturbers were also chosen to have different orbital planes,
a similar change in the disk's alignment is seen in the plot of $I_f$
(Fig.~\ref{fig10}d):
the disk's plane of symmetry is aligned with the planet's orbital plane at
its inner edge, and with the orbital plane of HR 4796B at its outer edge;
this could cause an image of the disk to appear warped.
Such a warp could be modeled using the same modeling techniques
that are described in this paper, and would provide further constraints
on the perturbers in the system (even if no warp was observed).

So, it is possible that the brightness asymmetry, and the symmetry plane of
the lobes, are determined by a planet close to the edge of the disk
that has $M_{pl} > 0.1M_J$, rather than by HR 4796B.
Using such an analysis, the pericenter glow phenomenon could be used to
test for the existence of a planet in the HR 4796 system, but only after
the orbit of HR 4796B has been determined;
e.g., if $\tilde{\omega}_B$, $e_B$ or the plane of HR 4796B's orbit
contradicted the observed asymmetry orientation, brightness asymmetry
magnitude, or the plane of symmetry of the lobes, then the existence of a
planet at the inner edge of the disk with $M_{pl} > 0.1 M_J$ could be
inferred.

\subsubsection{Just Planet}
\label{sssec-jp}
A double-lobed disk structure could also be observed in a system with no
observable companion.
The only possible perturber in such a system is an unseen planet,
the secular perturbations of which warrant the same kind of discussion
as for the case when HR 4796B was the only perturber (\S \ref{sssec-jb}).
Depending on the planet's mass, radial location and eccentricity,
it too could give rise to a detectable pericenter glow.
The only constraint on the planet's mass is that the disk must be old enough
for its secular perturbations to have affected the distribution of orbital
elements of the disk particles over the age of the system.
Since it takes of the order of one precession timescale to distribute the
complex eccentricities of collisional fragments around the circle centered
on the forced eccentricity, the constraint on the planet's mass can be
approximated as that for which is that the age of the system, $t_{sys}$, is
greater than the secular timescale, $t_{sec} \propto 1/M_{pl}$
(eq.~[\ref{eq:tsec2}]);
for the HR 4796 system this limit is $M_{pl} > 10M_\oplus$.
The constraint on the planet's eccentricity is even less stringent than for
the binary companion because the planet is closer to the edge of the disk:
a planet in the HR 4796 system would only need an eccentricity of
$e_{pl} > 0.02$ to produce the observed $5\%$ lobe asymmetry,
and the forced eccentricity imposed on the disk by a planet with
$e_{pl} = 0.023$ is plotted in Fig.~\ref{fig10}c.
So, the signature of even a low-mass planet would not escape detection
and symmetrical double-lobed features are unlikely to be observed in
systems that contain planets.

\subsubsection{Other Considerations}
\label{sssec-oc}
If the disk itself is massive enough to cause significant gravitational
perturbations to the orbits of the disk particles, then the mass of the
disk should be incorporated into the analysis of the secular perturbations
in the system.
A massive disk could dampen the eccentricity of a planet at the inner
edge of the disk (\cite{wh98}), thus reducing the offset asymmetry.

\subsubsection{Fomalhaut}
\label{sssec-fom}
The Fomalhaut disk lobes may have asymmetric brightnesses
(\cite{hgzw98}), but the statistical significance of this asymmetry is low.
Fomalhaut is a wide visual binary system.
Gliese 879 is Fomalhaut's common proper motion companion
(\cite{bshb97});
the two stars are separated by $\sim 2^\circ$, which corresponds to a
projected separation of $O$(55,000 AU) at $7.7$ pc.
At such a distance, the forced eccentricity imposed on the disk
by the binary star is insignificant (eq.~[\ref{eq:zf1}]).
A secular perturbation offset asymmetry in this disk would be expected
only if there is a planet in the disk that has a non-circular orbit.

\subsection{Discussion of Emitting Particle Category}
\label{ssec-deps}
The emitting lifetime of the disk's $\beta$ meteoroids, $O$(370 years),
is less than the emitting lifetime of their parents, $O(10^4$ years);
equivalently, $\tau_{eff} < 0.1$.
Thus, our understanding of the dynamic disk
implies that the disk's cross-sectional area distribution should not
contain a significant amount of $\beta$ meteoroids.
Rather, since there are no P-R drag affected particles in the disk, the
disk's emission is expected to come from its $\beta$ critical
particles, and its smallest large particles (\S \ref{sssec-pcsig}).
Not all disk particles have the same composition and morphology;
even if this were a close approximation, there is, as yet, no evidence
to suggest whether the particle properties chosen in our modeling
(\S \ref{sssec-pop}) are correct.
Are we indeed seeing the disk's $\beta$ meteoroid particles, or did
the assumptions of the modeling lead us to this conclusion?

Consider the initial crude estimate of the particle size (\S \ref{sssec-dtyp});
this proved to be an accurate method for estimating the particle size
(\S \ref{sssec-tps}).
If different assumptions had been made about the particles'
properties (e.g., if the particles had been assumed to be made of ice)
or morphologies (e.g., if the particles had been assumed to be like the
"bird's nest" structures of \cite{gust94}), both Figs.~\ref{fig6}a and
\ref{fig6}b would be different, and different conclusions might have been
drawn about the $\beta$ of the emitting particles.
If a size distribution had been included in the modeling, this would also
have affected our conclusions.
These are considerations that should be modeled before any firm
conclusions about the dynamics of the emitting particles can be reached.
However, since irrespective of their assumed properties large particles
have black body temperatures and brightness ratios similar to those
of the $D > 100$ $\mu$m particles in Fig.~\ref{fig6}b, a flexible interpretation
of the observed brightness ratio is that the emitting particles must have
temperatures that are hotter than black body.
Thus, the emitting particles are either small (e.g., the simple analysis of
\S \ref{sssec-pop} implies that $D < 10$ $\mu$m), in which case
they are likely to be $\beta$ meteoroids (e.g., 10 $\mu$m particles have
$\beta > 0.5$ unless they have densities $> 3000$ kg/m$^3$,
eq.~[\ref{eq:beta2}]), or they are large particles that are made up of
smaller hotter particles.

The assumptions about the particle properties in the model could also
have affected our conclusions about the collisional lifetime of the
emitting particles.
Consider the estimate of $\tau_{eff}$(70 AU) derived from the IHW18
lobe brightness (\S \ref{sssec-collh}).
Changing the properties of particles in the model would change the
estimate of $\tau_{eff}$ because of the resultant changes in
$P(18.2$ $\mu$m$,r)$;
e.g., if we had modeled the disk using 30-50 $\mu$m particles, we would
have had to put more cross-sectional area in the model for it to give the
observed lobe brightness.
Fig.~\ref{fig5}c shows that for astronomical silicate Mie spheres that are
larger than 0.01 $\mu$m in diameter, $P(18.2$ $\mu$m,70 AU) must lie in the
range $0.034-2.3 \times 10^{12}$ Jy/sr, depending on whether the disk's
cross-sectional area is concentrated in its 30-50 $\mu$m particles that
emit at cool temperatures, or in its 2-3 $\mu$m particles that are
small and hot, but large enough to emit efficiently at 18.2 $\mu$m.
For particles with different optical properties, $P(18.2$ $\mu$m,70 AU) could
be below $0.034 \times 10^{12}$ Jy/sr if the particles have temperatures well
below black body (or if they have low emission efficiencies);
equally, it could be higher than $2.3 \times 10^{12}$ Jy/sr if the particles
are hotter than the 2-3 $\mu$m astronomical silicate Mie spheres.
Taking $0.034 \times 10^{12}$ Jy/sr as a lower limit for $P(18.2$ $\mu$m,70 AU)
implies that $\tau_{eff}$(70 AU) $< 0.35$, giving a collisional
lifetime for the emitting particles that could be as low as 85 years.
However, given the temperature of the emitting particles inferred from
the observed brightness ratio, $O(10^4$ years) remains the best
estimate for their collisional lifetime.
A further note of caution is necessary about the inferred collisional
lifetime:
Appendix \ref{app-tcoll} assumes collisions between disk particles
to be either catastrophic or irrelevant.
While this may be appropriate for the disk's larger particles, since
these are likely to be solid bodies, collisions between its smaller
particles, which may have fluffy "bird's nest" structures (\cite{gust94}),
could be more erosive than destructive, and could lead to significant
grain growth.

In conclusion, neither observational, nor theoretical considerations can
provide a definitive answer as to the dynamics of the emitting particles.
However, the confirmation of the emitting particles' collisional lifetimes
means that we can be sure that these particles are not primordial,
and that there are no P-R drag affected particles in the disk.
$\beta$ meteoroids remain the most likely candidate for the
emitting particles.
Mid-IR emitting particles that are on hyperbolic orbits have been inferred
from observations of the disks around both $\beta$ Pictoris
(\cite{tdbw88}) and HD 141569 (\cite{ftpk99}).

\subsection{Origin of the Inner Hole}
\label{ssec-lsd}
Whatever the size of the emitting particles, analysis of the optical depth
of the disk's inner region (T99) shows that it is a few hundred times less
than that of the outer disk, and so there are very few emitting
particles in this region.
Because this central hole is necessary for the secular perturbation offset
asymmetry to be observed (without the hole only the radial offset
could be observed), its physical origin requires attention.
Since the existence of small emitting particles in the disk requires the
existence of large particles, the question to answer is why there
are so many large particles in the outer disk, but so few in the inner
disk?
Either the physical conditions were such that they were able to form in the
outer region, but not in the inner region, or they formed across the whole disk,
but those formed in the inner region have since been removed.
Rather than discussing the planetary formation process and the stage of the
system's evolution (although these are of utmost importance in determining
the physics of the disk), this section offers possible dynamical
explanations for the removal of the large particles.

If a planet (or planets) exists interior to the inner edge of the disk, then
resonance overlap removes all material from a region of radial width
$\sim 1.3a_{pl}(M_{pl}/M_\star)^{2/7}$ about the planet's orbit within about
1000 orbital periods (\cite{wisd80}; \cite{dqt89}).
Material is also removed from the secular resonance regions (\cite{lf97})
--- the origin of these regions, which cover the range
of semimajor axes on the plots of Figs.~\ref{fig10}b and \ref{fig10}d where
$e_f, I_f \rightarrow \infty$, was discussed in \S \ref{sssec-lp}.
The radial distribution of material would also be affected by planetary
radial migration (\cite{malh95}; \cite{tgbl98}).
However, such mechanisms can only explain the total lack of large particles
in the inner region by invoking a system with either many planets, or just one
planet that is either very large, or on a very eccentric orbit.
Since all of these mechanisms take longer than a particle's orbital period
to take effect, they would cause an inner cut-off in the disk particles'
semimajor axes (as opposed to a cut-off in radial distance from the
star).

To estimate the orbit of a putative planet at the inner edge of the HR 4796
disk that is causing the cut-off, consider the inner edge of the Kuiper belt.
There is almost no Kuiper belt material on orbits with semimajor axes
interior to that of Pluto (\cite{jewi99}), which is in \textbf{2:3} resonance
with Neptune.
This is supposed to be the result of resonance sweeping that occurred as
Neptune's orbit expanded early in the history of the solar system due
to the clearing of planetesimal debris from the inner solar system and the
formation of the Oort cloud (\cite{malh95}).
By analogy, assuming that the inner cut-off of the disk's large particles
occurs at the planet's \textbf{2:3} resonance location, and that this cut-off
can be described by $a_{min} = 62$ AU, we can estimate the orbit of
the planet to have a semimajor axis of
$a_{pl} = a_{min}[\textbf{2/3}]^{2/3} = 47$ AU, giving an orbital period
of $\sim 200$ years (eq.~[\ref{eq:tper}]).

\subsection{Interpretation of the Residual Structure}
\label{ssec-rs}
So far, no explanation has been offered for the structure of the
residuals (what is left after subtracting the model from the observation,
see T99).
Analogy with the zodiacal cloud implies that there could be a population
of warmer dust in the inner region that may be unrelated to the dust in
the outer disk.
Depending on the perturbers in the inner region, such dust could
contain considerable structure.
Analysis of emission from such regions would reveal a great deal about the
system's perturbers, however, this would not be easy, since the resolution
required to map such small-scale structure is at the limit of current
technological capabilities.
In addition, such emission is masked by that of both the stellar
photosphere and the outer disk, the accurate subtractions of which are
difficult.

There may also be residual structure associated with the outer regions of
the disk.
If there is a planet orbiting HR 4796A close to the inner edge of the
disk, then the distribution of large particles in the outer disk would contain
structure associated with the planet's gravitational perturbations in
addition to the secular perturbations already discussed
(\cite{djxg94}; \cite{malh96}; \cite{dghw98}; \cite{wh98}).
Some of the emitting particles might be trapped in its resonant ring.
Such a ring could be responsible for some of the observed lobe asymmetry.
The existence of such a ring would give the inner edge of the disk structure
that co-orbits with the planet;
i.e., observations of this structure would vary on timescales of
$\sim 200$ years, offering a method of distinguishing between this
structure and the large-scale background structure, which would vary on
secular timescales of $O$(1 Myr).
The asymmetric structure of the Earth's resonant ring includes a
$0.2^3$ AU$^3$ cloud of dust located permanently in the Earth's wake
with a number density $\sim 10\%$ above the background (\cite{djxg94}).
Observations of such an asymmetric structure in an exosolar disk could
be modeled using the same techniques that were used to model the Earth's
resonant ring (\cite{djxg94}; \cite{jd99}), possibly allowing us
to determine the presence, location, and even the mass of the
perturbing planet (\cite{dghw98}).
However, the evidence suggests that such observations may not be possible
with currently available technology.
In fact, calculations show that when viewed from a distant point in space
normal to the ecliptic plane, the Earth's "wake" would only have
an IR signal $O(0.1)$ times that of the Earth (\cite{back98}).
So, regardless of the resolution requirements, if one were to observe the solar
system from outside, it would be easier to detect the Earth
directly than to infer its existence from the structure of the zodiacal
cloud.

The Earth's resonant ring is a result of the trapping of particles
that are evolving into the inner solar system due to P-R drag.
Another method of forming a resonant ring is for a planet to undergo
radial migration of its orbit, trapping all particles exterior to its orbit
into its strongest resonances;
such an interaction is supposed to have happened between Neptune
and the Kuiper belt (\cite{malh95}).
The amount of disk material that is trapped in a Kuiper belt ring depends
on how much radial migration has taken place to the planets orbit.
A ring with all of its particles trapped in the 1:2 and 2:3 resonances
with the planet (these are the strongest resonances) would have three
lobes, with the planet residing in an "empty" fourth lobe.
Could this be the cause of the tri-lobed structure observed in
$\epsilon$ Eridani (\cite{ghmj98})?

\section{Conclusions}
\label{sec-con}
The primary intent of this paper was to show how the long-term
effect of the gravitational perturbations, i.e., the secular perturbations,
of a massive perturber could be the cause of the $\sim 5$\%
brightness asymmetry of the double-lobed feature seen in observations
of the HR 4796 disk (T99):

\begin{enumerate}
\item We showed how the secular perturbations of a massive perturber
in a disk impose a forced eccentricity on the orbits of particles in the
disk, thus causing the disk's center of symmetry to be offset from
the star towards the perturber's apastron.
We also showed how the same perturbations impose a forced inclination
on the particles' orbits, which, if there is more than one perturber in the
disk, could cause the disk to be warped.

\item We produced a model of the HR 4796 disk that accurately maps
the 18.2 $\mu$m brightness distribution observed by T99;
this model is based on a consideration of the dynamics of the
particles in the disk.
The model shows how the brightness of a disk that has an inner clear region,
that also has an offset center of symmetry caused by a forced eccentricity
imposed on the disk particles' orbits, would be asymmetric,
since the inner edge of one side of the disk is closer to the star, and so
is hotter and brighter, than the other side.
We showed that a forced eccentricity as small as 0.02 is all that is
necessary to cause a 5\% lobe brightness asymmetry in the
HR 4796 disk.

\item If the eccentricity of orbit of the companion star, HR 4796B, is
larger than 0.13, then a forced eccentricity of 0.02 is to be expected.
However, if there is a planet of mass $> 0.1M_J$ located close
to the inner edge of the disk, then the forced eccentricity, and
hence the asymmetry, imposed on material in the disk's lobes
is controlled by the planet rather than the binary companion;
this could also cause the disk to be warped.
If a forced eccentricity is indeed the cause of the observed lobe
asymmetry then observations that constrain the orbit of HR 4796B
would help to clarify whether such a planet exists.
If the HR 4796 system had no binary companion, a forced eccentricity
of 0.02 could have been imposed on the disk by a lone planet with a
mass of $>10M_\oplus$, and an eccentricity of $> 0.02$.

\item The statistical significance of the HR 4796 disk's lobe asymmetry in the
observations of T99 is only at the $1.8 \sigma$ level, however, it is also
apparent in the observations of other authors (\cite{krwb98}; \cite{ssbk99}).
It would take one good night on a 10 meter telescope to get a clear
observation of the HR 4796 asymmetry.
Thus, the indirect detection of planets, even small planets, hiding in
circumstellar disks is clearly within reach using these dynamic modeling
techniques.
This is particularly important, since the direct detection of planets
around even nearby stars is well beyond current capabilities
(\cite{back98}), and indirect detection techniques such as radial velocity
(\cite{mb98}) or astrometric (\cite{gate96}) techniques, permit detection
only of very massive planets that are close to the star.

\item If there is a planet close to the inner edge of the disk, many of the
disk's particles could be trapped in resonance with that planet, thus
forming a resonant ring.
Such a ring would give the inner edge of the disk structure that
rotates on the timescale of the orbital period of the planet, $\sim 200$
years.
This structure could be contributing to the observed lobe asymmetry,
and may also be present in the residuals of the observation.
This possibility could be explored with further observations.
Resonant rings may be the predominantly observable structures of
some exosolar systems, such as that recently observed around
$\epsilon$ Eridani (\cite{ghmj98}).
\end{enumerate}

The HR 4796 disk modeling also revealed important information about
the large-scale symmetrical structure of the disk, as well as about
the dynamic properties of its emitting particles:

\begin{enumerate}
\setcounter{enumi}{6}
\item The spatial distribution of material in the disk inferred from our
modeling matches that inferred by other authors (\cite{jfht98};
\cite{krwb98}; \cite{ssbk99}).
The surface density of cross-sectional area in the disk peaks at $\sim 70$ AU
from HR 4796A, falling off $\propto r^{-3}$ outside this radius,
dropping by a factor of $\sim 2$ between 70 and 60 AU, and falling to zero
by 45 AU.
This soft inner edge to the disk is to be expected if the disk particles'
orbits are eccentric.

\item Assuming the particles to be astronomical silicate Mie spheres, the
diameter of the emitting particles was estimated to be $D_{typ} = 2-3$ $\mu$m.
Particles this small have radiation forces that are characterized by
$\beta \approx 2$, and so are blown out of the system on hyperbolic orbits on
timescales of $\sim 370$ years.
The HR 4796 disk is very dense;
the collisional lifetime of its emitting particles is $\sim 10^4$ years.
Thus, the emitting particles cannot be primordial particles, rather they must
be continuously created from a reservoir of larger particles.
The collisional lifetimes of all of the disk's particles are shorter than
their P-R drag lifetimes;
i.e., none of the disk's particles are affected by P-R drag.
Further investigation of the particles' properties needs to be done
before any firm conclusions can be reached about whether the disk
is dense enough to support a population of particles on hyperbolic
orbits that is large enough to dominate the disk's emission.
If the emitting particles are on hyperbolic orbits, the modeling implies
a mass loss rate for the disk $\sim 4 \times 10^{-6}M_\oplus/$year,
and further modeling would have to be done to ascertain the spatial
distribution of the dynamically stable population of large particles
from which these emitting particles originated.
\end{enumerate}


\appendix

\section{Collisional Lifetime of Disk Particles}
\label{app-tcoll}
Consider a collision between two disk particles, the larger of which
is denoted by the subscript 1, and the smaller by the subscript 2.
For this collision to be "catastrophic", that is, for it to
result in the break-up of the larger particle, the impact energy of the
collision must be large enough both to overcome the tensile strength
of the larger particle, and to impart enough energy to the collisional
fragments to overcome its gravitational binding energy.
In the asteroid belt this limit means that a collision is only catastrophic
if $m_2/m_1 \geq 10^{-4}$ (\cite{dohn69}).
Since the impact energy of a collision is $\propto m_2 v_{rel}^2$,
assuming that exosolar disk particles have similar tensile strengths to the
solar system's asteroids, this limit can be scaled to exosolar disks
by the square of the ratio of the mean relative velocity of collisions in
the asteroid belt (at $\sim 3$ AU), $v_{rel} \approx 5$ km/s
(\cite{vedd98}), to that of collisions in the exosolar disk.
The mean relative velocity of collisions in exosolar disks can
be described by: 
\begin{equation}
  v_{rel}(r) = f(e,I)v(r), \label{eq:vrel}
\end{equation}
where $f(e,I)$ is some function of the disk particles' eccentricities and
inclinations, and $v(r) = 30\sqrt{(M_\star/M_\odot)(a_\oplus/r)}$ km/s
is the average velocity of particles at $r$ (eq.~[\ref{eq:vgrav}] with
$a$ replaced by $r$).
Thus, assuming $f(e,I) \approx 0.3$ as for collisions in the asteroid belt,
an exosolar disk particle of diameter $D \propto m^{1/3}$, would only
suffer a catastrophic collision if the other particle in the collision had a
diameter $\geq D_{cc}(D)$, where
\begin{equation}
  D_{cc}(D) = 0.03[(M_\odot/M_\star)(r/a_\oplus)]^{1/3}D.
  \label{eq:dimp}
\end{equation}

The "collisional lifetime", i.e., the mean time between catastrophic
collisions, of a particle of diameter $D$, at a location in a disk
denoted by $r,\theta,$ and $\phi$, is the inverse of its catastrophic
collision rate (\cite{kess81}):
\begin{equation}
  t_{coll}(D,r,\theta,\phi) = [R_{coll}(D,r,\theta,\phi)]^{-1},
  \label{eq:tcol1}
\end{equation}
where
\begin{equation}
  R_{coll}(D,r,\theta,\phi) = \sigma_{cc}(D,r,\theta,\phi)v_{rel}(r),
  \label{eq:rcoll}
\end{equation}
$\sigma_{cc}(D,r,\theta,\phi)$ is the catastrophic collision
cross-section seen by the particle, and $v_{rel}(r)$ is the mean
encounter velocity of disk particles at $r$ (eq.~[\ref{eq:vrel}]).
Using the definition of a disk's structure given by equation
(\ref{eq:sig2}), this catastrophic collision cross-section is:
\begin{equation}
  \sigma_{cc}(D,r,\theta,\phi) = f_{cc}(D,r)\sigma(r,\theta,\phi),
\end{equation}
where
\begin{equation}
  f_{cc}(D,r) = \int_{D_{cc}(D)}^{D_{max}} (1+D/D^{'})^2
          \bar{\sigma}(D^{'},r) dD^{'}, \label{eq:fcc}
\end{equation}
and $D_{cc}(D)$ is the smallest particle with which a catastrophic
collision could occur (eq.~[\ref{eq:dimp}]).
  
However, unless $t_{coll} \ll t_{per}$, the particle's orbit takes
it through a range of $\theta$ and $\phi$ before a collision occurs
(there is also a variation of $r$ along the particle's orbit due to the
eccentricity of its orbit).
Thus, it is more appropriate to calculate the particle's collisional
lifetime using the mean catastrophic collision rate of the particles
in the size range $D \pm dD/2$ that are in the spherical shell of radius,
$r$, and width $dr$.
Consider an element of this shell that has a volume,
$dV = r^2drd\theta\cos{\phi}d\phi$.
The number of particles in the diameter range $D \pm dD/2$ in this
element is given by $n(D,r,\theta,\phi)dDdV$, and each of these particles
has a catastrophic collision rate given by equation (\ref{eq:rcoll}).
Integrating over the whole shell gives:
\begin{equation}
  t_{coll}(D,r) = \frac{ \int_{-I_{max}}^{+I_{max}}\int_0^{2\pi}
      \sigma(r,\theta,\phi) d\theta \cos{\phi} d\phi }
    { \int_{-I_{max}}^{+I_{max}}\int_0^{2\pi}
      [\sigma(r,\theta,\phi)]^2 d\theta \cos{\phi} d\phi
      \ast f_{cc}(D,r) v_{rel}(r)},
\label{eq:tcol3}
\end{equation}
where $I_{max}$ is the maximum inclination of the disk particles' orbits
to the reference plane.

Equation (\ref{eq:tcol3}) can be simplified by considering a cylindrical
shell, defined by $r,\theta,$ and $z$, rather than a spherical one.
An element of the cylindrical shell has a volume $dV = r dr d\theta dz$,
and the corresponding collisional lifetime of a particle in the shell is
given by equation (\ref{eq:tcol3}), but with $\phi$, $\cos{\phi}d\phi$ and
$\pm I_{max}$, replaced by $z$, $dz$, and $\pm h$, where
$h = r\sin{I_{max}}$.
Here we introduce the parameter $\tau_{eff}$, the disk's face-on effective
optical depth:
\begin{equation}
  \tau_{eff}(r) = \int_{-h}^{+h} \sigma(r,\theta,z) dz,
  \label{eq:taueffdefn}
\end{equation}
where the dependence on $\theta$ has been dropped since orbits sample the
full range of $\theta$.
This is not a true optical depth, since that would include a consideration
of the particles' extinction coefficients ($Q_{ext} = Q_{abs} + Q_{sca}$);
rather, it is the disk's face-on surface density of cross-sectional area,
which is equal to its true optical depth if its particles had $Q_{ext} = 1$.
Assuming that $\sigma(r,\theta,z)$ is independent of $z$, so that
$\int_{-h}^{+h} [\sigma(r,\theta,z)]^2 dz = 0.5\tau_{eff}(r)^2/h$,
and that the encounter speed is determined by the vertical motion of
particles in the disk, so that $f(e,I) \approx \sin{I_{max}}$,
equation (\ref{eq:tcol3}) can be simplified to:
\begin{equation}
  t_{coll}(D,r) = \frac{t_{per}(r)}{\pi f_{cc}(D,r) \tau_{eff}(r)},
\label{eq:tcol5}
\end{equation}
where $t_{per}(r)$ is the average orbital period of particles at $r$
(eq.~[\ref{eq:tper}] with $a$ replaced by $r$).

A disk's effective optical depth, $\tau_{eff}$, can be estimated
observationally from equation (\ref{eq:brightness}):
\begin{equation}
  \tau_{eff}(r) \approx (F_\nu(\lambda,r)/\Omega_{obs})/P(\lambda,r),
  \label{eq:taueff}
\end{equation}
where $F_\nu/\Omega_{obs}$ is the disk's face-on unsmoothed
brightness.
The disk's face-on unsmoothed brightness can be calculated either
from the observed brightness, making corrections to account for
both the disk's orientation, as well as for the PSF smoothing, or
from the disk's total flux density, $F_\nu(\lambda)$, and assuming
the disk material to be evenly distributed between $r \pm dr/2$:
\begin{equation}
  F_\nu(\lambda,r)/\Omega_{obs} = F_\nu(\lambda)C_fR_\star^2/rdr,
  \label{eq:fnuobs}
\end{equation}
where $C_f = 6.8 \times 10^9$ AU$^2$/pc$^2$/sr, and
$R_\star$ is the distance of the star from the observer.

\subsection{Collisional Lifetime of Particles with the Most
Cross-sectional Area}
\label{app-tcoll1}
Consider the particles in a disk that make up most of the disk's
cross-sectional area, i.e., those with diameters close to $D_{typ}$.
By definition, these particles are most likely to collide with each other
(a collision that would definitely be catastrophic), and so their collisional
lifetime can be found from equation (\ref{eq:tcol3}) using the
approximation:
\begin{equation}
  f_{cc}(D_{typ},r) \approx 4. \label{eq:fccdtyp}
\end{equation}
Applying this approximation to equation (\ref{eq:tcol5}) gives
equation (\ref{eq:tcol6}), which is also the collisional lifetime
determined by Artymowicz (1997) for particles in $\beta$ Pictoris.

\subsection{Collisional Lifetime of a Disk's Large Particles}
\label{app-tcoll2}
The collisional lifetime of particles of different sizes in a disk differ
only by the factor $f_{cc}(D,r)$.
This factor can be ascertained by making assumptions about the
disk particles' size distribution. 
Assuming that the size distribution of equation (\ref{eq:nd}) holds for
disk particles between $D_{min}$ and $D_{max}$, the
normalized cross-sectional area distribution is given by:
\begin{equation}
  \bar{\sigma}(D) = (3q-5)D^{4-3q}/D_{min}^{5-3q}.
  \label{eq:sigd}
\end{equation}
Substituting into equation (\ref{eq:fcc}) gives:
\begin{equation}
  f_{cc}(D) = \left( \frac{XD}{D_{min}} \right)^{5-3q}
    \left[ 1+\frac{6q-10}{(3q-4)X} + \frac{3q-5}{(3q-3)X^2} \right],
  \label{eq:fcc2}
\end{equation}
where $X = D_{cc}(D)/D$ for $D_{cc}(D) > D_{min}$, and $X = D_{min}/D$
for $D_{cc}(D) \leq D_{min}$;
the collisional lifetime of particles in a disk with this distribution
is a minimum for particles for which $D_{cc}(D) = D_{min}$.
The size distribution of particles in a real disk is more complicated than
equation (\ref{eq:sigd}), however, equation (\ref{eq:fcc2}) can be used to
give a crude approximation for the collisional lifetime of a disk's large
particles:
\begin{equation}
  t_{coll}(D,r) \approx t_{coll}(D_{typ},r)*(D_{cc}(D)/D_{typ})^{3q-5}.
  \label{eq:tcol7}
\end{equation}

\subsection{Other Considerations for a Particle's Collisional Lifetime}
\label{app-tcoll3}
If the P-R drag lifetime, $t_{pr}$ (eq.~[\ref{eq:tpr}]), is comparable to, or
shorter than $t_{coll}$, then the effect of P-R drag on the collisional
lifetime must be accounted for;
e.g., on average, particles from a parent at $r_{parent}$, survive
until they reach $r_{coll}$ where $\int_{r_{parent}}^{r_{coll}}
\frac{800(M_\odot/M_\star)r}{t_{coll}(r)\beta} dr = 1$.
Also, if there is a significant change in $r$ along a particle's orbit due to
the eccentricity of its orbit, a similar analysis can be done to take this
into account when calculating its collisional lifetime.
By considering the washer-like disk of particles on orbits with $a$, $e$,
and random $\tilde{\omega}$ (\cite{syke90}):
$t_{coll}(a,e) = \pi [\int_{q}^{q'} (r/a)/[t_{coll}(r) \sqrt{(r-q)(q'-r})]dr]^{-1}$,
where $q=a(1-e)$ is this disk's inner edge, and $q'=a(1+e)$ is its outer
edge.

\section{Lobe Asymmetries in Disks where P-R Drag is Important}
\label{app-pr}

\subsection{Offset and Warp of a Disk's P-R Drag Affected Particles}
\label{app-spr}
To find the secular evolution of the orbital elements of a particle that
is affected by P-R drag, i.e., one with $\beta_{pr} < \beta < 0.1$,
the equations governing the evolution of its complex eccentricity,
$\dot{z} = \dot{z}_{sec} - 2.5(\alpha/a^2)z$
(eqs.~[\ref{eq:edot}] and [\ref{eq:zdot}]), and its complex inclination,
$\dot{y} = \dot{y}_{sec}$ (eq.~[\ref{eq:ydot}]),
must both be solved in conjunction with the P-R drag evolution of its
semimajor axis (eq.~[\ref{eq:adot}]).
While the solution given by equations (\ref{eq:z}) and (\ref{eq:y}) is
no longer applicable, the decomposition of the particle's complex
eccentricity and complex inclination into forced and proper elements,
and the physical meaning of these elements, is still valid;
however, each of these elements now depends on the particle's
dynamical history.

Consider the P-R drag affected particles created by the break-up of the
asteroid family group described in \S \ref{sssec-opsfam}.
Immediately after they are created, the orbital elements of these
particles are the same as those of the rest of the family;
i.e., they have semimajor axes $a$, and complex eccentricities
and complex inclinations that are uniformly distributed in these
planes around circles of radii equal to the proper elements of the
family, $e_p$ and $I_p$.
The dynamical evolution of a wave of these particles, i.e., those that
were created at the same time, can be followed by numerical
integration to ascertain how the orbital elements of the particles
in the wave vary as their semimajor axes decrease due to
P-R drag;
this is the "particles in a circle" method (\cite{dgdg92}).
It was found that the complex eccentricities and complex inclinations
of a wave of particles originating in the asteroid belt
remain on circles, and that as the wave's semimajor axis,
$a_{wave}$, decreases:
its effective proper eccentricity (the radius of the wave's circle
in the complex eccentricity plane) decreases
$\propto e_{p}*(a_{wave}/a)^{5/4}$;
its effective proper inclination (the radius of the wave's circle in
the complex inclination plane) remains constant at $I_p$;
the distributions of the particles' $\tilde{\omega}_p$ and $\Omega_p$
remain random;
while its effective forced elements (the centers of the circles in the
complex eccentricity and complex inclination planes) have a more
complicated variation (\cite{dgdg92}; \cite{liou93}).

Thus, the orbital element distributions, $n(z)$ and $n(y)$, of P-R drag
affected particles are like that of the large particles,
in that they are the addition of $z_f$ and $y_f$ to symmetrical proper
element distributions;
however, their forced and proper elements are different for particles
from different families, as well as being different for particles of
different sizes and with different orbital semimajor axes.
This means that their spatial distribution is subject to offset and
warp secular perturbation asymmetries.

\subsection{Origin of Inner Hole}
\label{app-oihpr}
For disks in which P-R drag is a significant physical process, the
existence of an inner hole implies not only that there must be no
large particles in the inner region, but also that there must be
some mechanism that prevents the particles that are in the outer disk
from evolving into the inner disk by P-R drag.
In HR 4796, and other disks in which P-R drag is insignificant, this
mechanism is collisions;
i.e., particles created in the outer disk are broken up by collisions
before they reach the inner region.
This section presents dynamical explanations of how a planet located
at the inner edge of the disk would help to prevent the emitting
particles from reaching the inner disk. 

This is not just important for disks in which its individual particles are
not affected significantly by P-R drag in their lifetime, since while the
inner edge of a disk's largest particles may be at $a_{min}$, the
cumulative effect of P-R drag on all of the stages of the collisional
cascade that a primordial particle goes through before the fragments are
small enough to be blown out of the system by radiation pressure
could mean that the inner edge of  these blow-out particles is further
in than $a_{min}$.
To assess whether the cumulative effect of P-R drag has an impact
on the inner edge of the emitting particles in the HR 4796 disk,
consider its size distribution to follow that assumed in equation
(\ref{eq:sigd}), where $D_{min} = D_{typ} = 2.5$ $\mu$m, and $q = 11/6$.
The lifetime of an intermediate particle can be found using
equations (\ref{eq:tcol5}) and (\ref{eq:fcc2}),
and the amount of P-R drag evolution in its lifetime is
(eqs.~[\ref{eq:adot}] and [\ref{eq:betah}]):
\begin{equation}
  da \approx -0.0125 t_{coll}(D,r)/Da, \label{eq:dapr}
\end{equation}
where $da$ and $a$ are measured in AU, $D$ in $\mu$m,
and $t_{coll}(D,r)$ in years.
Assuming that the largest fragment created in a collision has a
diameter half that of the original particle, a $\beta$ meteoroid particle
(e.g., a 2.5 $\mu$m particle) at 62 AU, assuming it to have originated from
a gravitationally bound particle (i.e., one with $\beta < 0.5$), can at the
very most be removed by 27 generations from its primordial ancestor, which
would at the very most have been originally at 62.1 AU.

One consequence of a planet at the inner edge of the disk is that it
accretes some of the particles that pass it on their way into the
inner region.
A simple estimate for the proportion of particles lost in such a way
can be obtained by considering the P-R drag evolution of a torus of particles
with orbital elements $a$, $e$, $I$, and random $\Omega$,
$\tilde{\omega}$ and $\lambda$;
the volume of this torus is $V_{tor} = 8\pi a^3 e \sin{I}$ (\cite{syke90}).
In the time it takes for the torus to pass the planet,
\hbox{$\Delta t = (1602/\beta)(M_\odot/M_\star)(a_{pl}/a_\oplus)^2e$
yrs}, the planet accretes a volume of dust given by (\cite{kess81}):
$V_{acc} = \sigma_{cap} v_{rel} \Delta t$, where
$\sigma_{cap} = \pi R_{pl}^2(1+v_e^2/v_{rel}^2)$ is the capture
cross-sectional area, $R_{pl}$ is the radius of the planet,
$v_e = \sqrt{2GM_{pl}/R_{pl}}$ is the escape velocity of the planet,
and $v_{rel}$ is the mean relative velocity of encounter between
the planet and the particles.
Thus, the proportion of dust accreted onto the planet,
$P = V_{acc}/V_{tor}$, is given by:
\begin{equation}
  P \approx \frac{12}{\beta \sqrt{1-\beta}}
            { \left( \frac{M_\odot}{M_\star}  \right) }^{3/2}
	    { \left( \frac{M_{pl}}{M_\odot}   \right) }^{4/3}
	    { \left( \frac{a_\oplus}{a_{pl}}  \right) }^{1/2}
	    { \left( \frac{\rho_J}{\rho_{pl}} \right) }^{1/3}
	    g(e,I),
\label{eq:paccrn}
\end{equation}
where $\rho_J = 1330$ kg/m$^3$ is the mean density of Jupiter,
$g(e,I) = [(v_{rel}/v)\sin{I}]^{-1}$, and $v$ is the velocity of the particle;
e.g., if $g(e,I) = 100$, then a Jupiter-like planet at the inner edge of the
disk would need $M_{pl} > 30M_J$ to accrete all of the particles passing
it (i.e., $P = 1$ for $\beta < 0.5$).
 
Another consequence of a planet at the inner edge of the disk is that it
traps some of the disk particles into its exterior mean motion resonances.
The resulting resonant ring has three consequences that may aid
with the formation of a clear inner region.
Firstly, the calculation of the probability of accretion onto the planet
given in equation (\ref{eq:paccrn}) does not apply to the particles that
are trapped in the ring.
It is thought that resonance trapping helps the accretion process,
since trapped particles may leave the resonance upon a close encounter
with the planet (\cite{kd98b}).
This means that the constraint on the mass of the planet derived in the
last paragraph is an upper limit for the accretion process to be the sole
removal mechanism.
Secondly, the ring increases the number of particles that are lost
by collisional break-up, since it both decreases the collisional lifetime of
disk particles (by increasing the number density of particles in the ring
region) and increases the amount of time it takes for particles that
are trapped in the ring to reach the inner region.
Finally, if the trapping timescales, $t_{res}$, are longer than the age
of the system, $t_{sys}$, then the trapped particles cannot have reached
the inner disk yet;
i.e., the ring causes a bottle-neck in the flow of particles to the inner disk.

\clearpage

\clearpage

\end{document}